\documentclass[final, 5p, times, twocolumn]{elsarticle}
\usepackage{hyperref}
\usepackage[multiple]{footmisc}
\usepackage[T1]{fontenc}
\usepackage[hang]{subfigure}
\usepackage{epstopdf}
\usepackage{graphicx}
\usepackage{placeins}
\usepackage{float}
\usepackage{nomencl}
\usepackage[table,xcdraw]{xcolor}
\usepackage{multirow}
\usepackage{comment}
\hypersetup{pdfauthor=author}
\usepackage[table,xcdraw]{xcolor}
\usepackage{longtable}
\usepackage[table,xcdraw]{xcolor}

\usepackage{hhline}
\usepackage{wasysym}
\usepackage{amssymb}
\usepackage{pifont}
\usepackage{graphics}
\usepackage{soul}
\makenomenclature

\journal{Journal of \LaTeX\ Templates}

\begin{document}

\begin{frontmatter}

\title{A Holistic Survey of Multipath Wireless Video Streaming}


\author[First]{Samira Afzal\corref{cor1}}
\ead{afzal.samira@usp.br}
\ead[url]{https://intrig.dca.fee.unicamp.br/}

\author[Second]{Vanessa~Testoni}
\author[Third]{Christian~Esteve~Rothenberg.}


\author[Fourth]{Prakash~Kolan}

\author[Fifth]{Imed~Bouazizi}

\cortext[cor1]{Corresponding author.}

\address[First]{Department of Electronic Systems Engineering Polytechnic School,
University of Sao Paulo~(USP), S\~{a}o Paulo, SP, Brazil. }

\address[Second]{Samsung Research Brazil~(SRBR), Campinas, S\~{a}o Paulo, Brazil.}

\address[Third]{School of Electrical and Computer Engineering~(FEEC),
University of Campinas~(Unicamp), Campinas, S\~{a}o Paulo, Brazil. }

\address[Fourth]{Samsung Research America~(SRA), Dallas, Texas, USA.}

\address[Fifth]{Qualcomm, San Diego, CA, USA}
\begin{abstract}

Most of today's mobile devices are equipped with multiple network interfaces and one of the main bandwidth-hungry applications that would benefit from multipath communications is wireless video streaming. However, most of current transport protocols do not match the requirements of video streaming applications or are not designed to address relevant issues, such as delay constraints, networks heterogeneity, and head-of-line blocking issues.
This survey provides a holistic literature review of multipath wireless video streaming, shedding light on the different alternatives from an end-to-end layered stack perspective, unveiling trade-offs of each approach, and presenting a suitable taxonomy to classify the state-of-the-art. Finally, we discuss open issues and avenues for future work.
\end{abstract}


\end{frontmatter}


\section{Introduction}
{\label{sec: Introduction}}

Multimedia  services~(e.g., Skype, FaceTime) and on-demand mobile video content~(e.g., Hulu, YouTube, Netflix) have become part of daily use. Likewise, online cloud gaming is a very popular entertainment{\color{black}~\cite{jarschel2011evaluation}}. Such applications require {\color{black}high-quality} video streaming capabilities to meet the  {\color{black}end-user} expectations.
The annual Cisco report~\cite{index2019cisco} shows that, since 2012, mobile video has represented more than half of global mobile data traffic and will keep being responsible for the {\color{black} most significant} traffic growth upfront. The increase of on-demand video is expected to affect mobile networks as much as fixed networks. Another trend is that 4K / Ultra HD~(UHD)\nomenclature{UHD}{Ultra HD} video will be more prevalent in the network, as well as Multi-View Video~(MVV)\nomenclature{MVV}{ Multi-View Video} and even 8K, in the short-mid term.

{\color{black}Delivering high-quality video streaming services makes the task of providing real-time wireless transmission of multimedia while ensuring Quality of  Experience~(QoE) quite challenging due to bandwidth and time {\color{black}constraints}~\cite{sani2017adaptive}.} One of the approaches to tackle this challenging scenario is to add multipath transmission where video streaming can be delivered over IP broadcast and/or broadband with bidirectional connectivity between video sources and users. {\color{black} Table~\ref{table:multipath-compared-singlepath} presents published results on the potential performance gains {\color{black}of} wireless video streaming when exploiting multiple network paths}.

{\color{black}Several surveys in the literature have covered different aspects of multipath data communications in general, such as~\cite{qadir2015exploiting,singh2015survey,li2016multipath,domzal2015survey,addepalli2013heterogeneous,habak2015bandwidth}. However, this survey focuses {\color{black}mainly} on the multipath transmission of wireless video.  {\color{black}More specifically, we focus on the data plane problem of how to schedule data on multiple paths. Out of the scope of our survey remain~(\textit{i}) control plane aspects of how to compute routes such as multipath proposals~\cite{barakabitze2018qualitysdn, herguner2017towards}
 based on Software-Defined Networking~(SDN)~\cite{kreutz2015software},~(\textit{ii}) Wireless Sensor Networks~(WSN) which we refer the reader to surveys~\cite{hasan2017survey,sha2013multipath} on multipath video streaming in this type of networks,~(\textit{iii}) Peer-to-peer~(P2P)\nomenclature{P2P}{peer-to-peer} video streaming applications surveyed in-depth in~\cite{Zhang:2012:SPL:2365364.2365643,liu2008survey}, and~(\textit{iv}) Internet of Things~(IoT)\nomenclature{IoT}{Internet of Thing}, a focused application scenario where also multipath connectivity has been surveyed~\cite{hodroj2021survey}.}


{\color{black}{
\subsection{Review of related surveys}}}

In the following, we provide a brief overview of the most related and recently published surveys on multipath data communications, {\color{black}{ as presented in}} Table~\ref{table:related surveys}.

\definecolor{Gray}{gray}{0.95}
\begin{table*}
\begin{center}

\caption{\color{black}{SELECTED PUBLISHED RESULTS ON MULTIPATH WIRELESS VIDEO STREAMING}}
\label{table:multipath-compared-singlepath}
\scalebox{0.83}{
    \begin{tabular}{| p{2cm} | p{3cm} | p{1.5cm} | p{8.5cm} |}
    \rowcolor[HTML]{C0C0C0}
    \hline
    \rowcolor[HTML]{C0C0C0}
\textbf{Publication}   & \textbf{Network environment}     & \textbf{Protocol/ feature} & \textbf{Performance improvements compared to single path}                                                                                \\ \hline
\rowcolor{Gray}{\cellcolor[HTML]{EFEFEF}}    MRTP~\cite{mao2006mrtp} &  Mesh ad hoc network with high burst loss & RTP & PSNR gains of 1.26 dB more in multipath than single path,  64.14\% loss rate reduction together with making packet losses more random. 
    \\  MPRTP~\cite{singh2013mprtp}                                                                & Two 3G links  with bandwidth variations                                                                                          & RTP                                     &  In the
quick bandwidth change scenario, PSNR  is better than  the single path with 0.5\%
and 1.0\% loss rate. In the slow bandwidth change scenario, it is comparable to the single path with 1.0\% loss rate.                                                                                                                                                                                                          \\
\rowcolor{Gray} RTRA~\cite{xing2014real}                                                                                                          & WiFi and blacktooth networks with bandwidth variations                                                                                          & DASH                                    &RTRA shows better results for both slow and rapid changing bandwidth scenarios in terms of startup delay~(reduced up to half),  playback fluency average~(no segment missing in multipath but high misses in some single path scenarios),  playback quality~(PSNR improved 1 to 3 dB),  quality switch~(up to 4 times reduction),
 and bandwidth utilization.                                                                      \\
 MPLOT~\cite{sharma2008mplot}                                                                                                              & Wireless mesh network with burst loss rate of 50\% & TCP                                     & MPLOT achieves more than 50\% goodput improvement compared to the single path.                                 \\

\rowcolor{Gray} {\cellcolor[HTML]{EFEFEF}} \begin{tabular}[c]{@{}l@{}}Apostolopoulos\\ et al. ~\cite{apostolopoulos2000reliable}\end{tabular} &  Burst lossy wireless network & IP source routing/relay&  While the proposed approach results in initial PSNR drops of only 1.5 to 7 dB, but single path  results in initial PSNR drops of 12 to 15 dB. \\ \hline
    \end{tabular}}
\end{center}
\end{table*}

\begin{table*}[!htb]
\centering
\caption{{\color{black}SELECTED SURVEYS ON MULTIPATH COMMUNICATIONS}}
\label{table:related surveys}
\scalebox{0.83}{
\begin{tabular}{| p{2cm} | p{2cm} | p{3cm} | p{8cm} |}
\hline
\rowcolor[HTML]{C0C0C0}
\textbf{Reference}                                                         &  \textbf{Year} &     \textbf{Scope}                                           & \textbf{Comments}                                                                                                                                                                                                                                                                       \\ \hline
\rowcolor{Gray}  Qadir et al.~\cite{qadir2015exploiting}     & 2015                        & Control and data plane & \begin{tabular}[c]{@{}l@{}}Multipath  for data in general,\\ Focus on network-layer multipath solutions\end{tabular}                                                                                                                                                 \\
Singh et al.~\cite{singh2015survey}         & 2015                        & Control and data plane                         & \begin{tabular}[c]{@{}l@{}}Multipath  for data in general,\\ Limited research on video streaming services\end{tabular}                                                                                                                                                \\

\rowcolor{Gray} {\cellcolor[HTML]{EFEFEF}} Li et al.~\cite{li2016multipath}            & 2016                        & Data plane                                     & \begin{tabular}[c]{@{}l@{}}Multipath  for data in general. \\ Regarding video streaming, relevant aspects not covered \end{tabular}

\\

 Trestian et al.~\cite{chaudhari2021survey} & 2018                        & Data plane                                     & \begin{tabular}[c]{@{}l@{}}Multimedia delivery solutions following three key\\ directions:  adaptation, energy efficiency and multipath\\ Multipath is limited  to MPTCP and SCTP/CMT \end{tabular}\\

\rowcolor{Gray} {\cellcolor[HTML]{EFEFEF}} {\color{black}Kaur et al.~\cite{kaur2020survey} }           & 2020                        & Control plane                                     & \begin{tabular}[c]{@{}l@{}} Focus on  QoS mechanisms and routing protocols in WSN,
\\   Limited research on multiathp routing
protocols in \\ relation to video streaming\end{tabular}

\\

 {\color{black}More et al.~\cite{more2020analytical} }           & 2020                        & Control plane                                     & \begin{tabular}[c]{@{}l@{}} Focus on multipath routing protocols in VANET in\\ general, \\ Limited research on routing
protocols in relation to \\video streaming\end{tabular}

\\

\rowcolor{Gray}{\cellcolor[HTML]{EFEFEF}}{\color{black} Hodroj et al.~\cite{hodroj2021survey}} & 2021                       & Control and Data plane                                     & \begin{tabular}[c]{@{}l@{}}Video streaming in multipath
and multihomed overlay \\networks, \\
Techniques in the survey are not explained in details,\\ Challenges and related aspects are not discussed, \\ Some important relevant protocols and techniques are \\not considered

\end{tabular} \\

{\color{black}This Survey}                                                    & 2021                        & Data plane                                     & \begin{tabular}[c]{@{}l@{}}Multipath wireless with the focus on  video streaming\\  In-depth discussion of techniques following different \\ aspects: \\- protocol layer perspective~(Application, Transport,\\ Network and Cross layer)\\ - scheduling functions~(packet selection, packet protection, \\path selection)\\- effected
features and related methods~(e.g., packet loss \\differentiation, fairness consideration, video codecs, \\the experimental environment, performance \\metrics, and video services)
\end{tabular}                                                                                                                                                                                                                              \\ \hline
\end{tabular}}
\end{table*}

Qadir et al.~\cite{qadir2015exploiting}  investigated multipathing for data in general, mainly on the network layer. Besides that, they have also investigated multipath transmission on the transport layer. Their investigation is organized by discussing key aspects of network-layer multipathing: 1) route computation~(source routing, hop-by-hop routing, overlay routing, and SDN-based routing); 2) routing metrics~(e.g., delay, bandwidth); 3) load balancing techniques~(static or dynamic); 4) number of paths to use; 5) how to use multiple paths together. 

Singh et al.~\cite{singh2015survey} covered multipathing for data communications in general, covering fundamentals of multipath routing, multipath computation algorithms, multipath forwarding algorithms, and traffic splitting algorithms. The work also reviews various multipath protocols following a layer-based structure, from the application layer to the physical layer.

Li et al.~\cite{li2016multipath}
investigated multipath solutions for data in general and presented research problems at various protocol layers{\color{black}, including cross-layer} approaches. {\color{black}Although multiple relevant video streaming multipath solutions are discussed in this survey, key aspects specific to video streaming  are not considered~(e.g., importance and influence of video content).}  In addition, the work does not cover multipath approaches based on fundamental video streaming protocols, e.g., Dynamic Adaptive Streaming over HTTP~(DASH), and MPEG Media Transport~(MMT).

As a related survey, we should also consider Trestian et al.~\cite{trestian2018seamless}, which is a survey on seamless multimedia delivery within a heterogeneous wireless networks environment. The authors evaluated three {\color{black}critical}  aspects of multimedia delivery: adaptation, energy efficiency, and multipath delivery. Regarding the latter, only proposals based on the Multipath TCP~(MPTCP) and Stream Control Transmission Protocol~(SCTP)/Concurrent Multipath Transfer~(CMT) are studied.

{\color{black} Kaur et al.~\cite{kaur2020survey} and More et al.~\cite{more2020analytical} point out various multipath routing protocols in WSN and VANET scenarios, respectively, providing limited insights on video streaming.

Hodroj et al.~\cite{hodroj2021survey} study the research works at different layers of overlay networks from transport
to the application. The work also covers the technologies like machine learning, Fog, Mobile Edge computing, VR 360 video, and the Internet of Multimedia Things~(IoMT). However, the survey provides only a high-lvel overview of different approaches without an in-depth analysis or discussion of multipath video streaming challenges. Furthermore, relevant  key video streaming protocols~(e.g., MMT), techniques~(e.g., MPRTP~\cite{singhmultipath2012}, MPQUIC~\cite{deconinck-quic-multipath-07}) and multipath scheduling functions~(e.g., content awareness, packet protection, and network characteristics) are not covered.}\\



{\color{black}{ \subsection{ Contributions of this survey }}

Differently from existing  surveys, this survey is centered on video streaming applications exploiting multipath wireless communications. 
Relevant approaches and new techniques in the field are covered in-depth by surveying existing works along two main strands.}

{\color{black}The first strand} relates to the protocol layer  perspective of each work: application layer, transport layer and/or network layer. We sub-group each layer approaches based on which standard protocol/feature is used in the schemes proposed by the authors. Such classification is beneficial to understand the advantages, drawbacks, and trade-offs of each layer and protocol/feature. We also indicate which part of the network~(server and/or client) requires adjustment {\color{black} to} become compatible with the multipath transmission approach.

{\color{black}On a second strand, we analyze} the approaches based on the specific scheduling functions  to transmit video data over {\color{black}wireless} link technologies. The works are  classified according to the following scheduling functions: packet selection, packet protection, and path selection.

{\color{black}In addition, we also discuss relevant topics such as packet loss differentiation,  fairness consideration, video codecs, the experimental environment, performance metrics, and video services.} Finally, we also cover fundamental research problems related to multipath video transmission, such as network heterogeneity, out-of-order packets, Head-of-Line~(HOL) blocking, end-to-end delay, overdue packets, implementation aspects, and pros and cons of each approach.\\

\begin{figure*}
\centering
\includegraphics[width=0.75\textwidth]{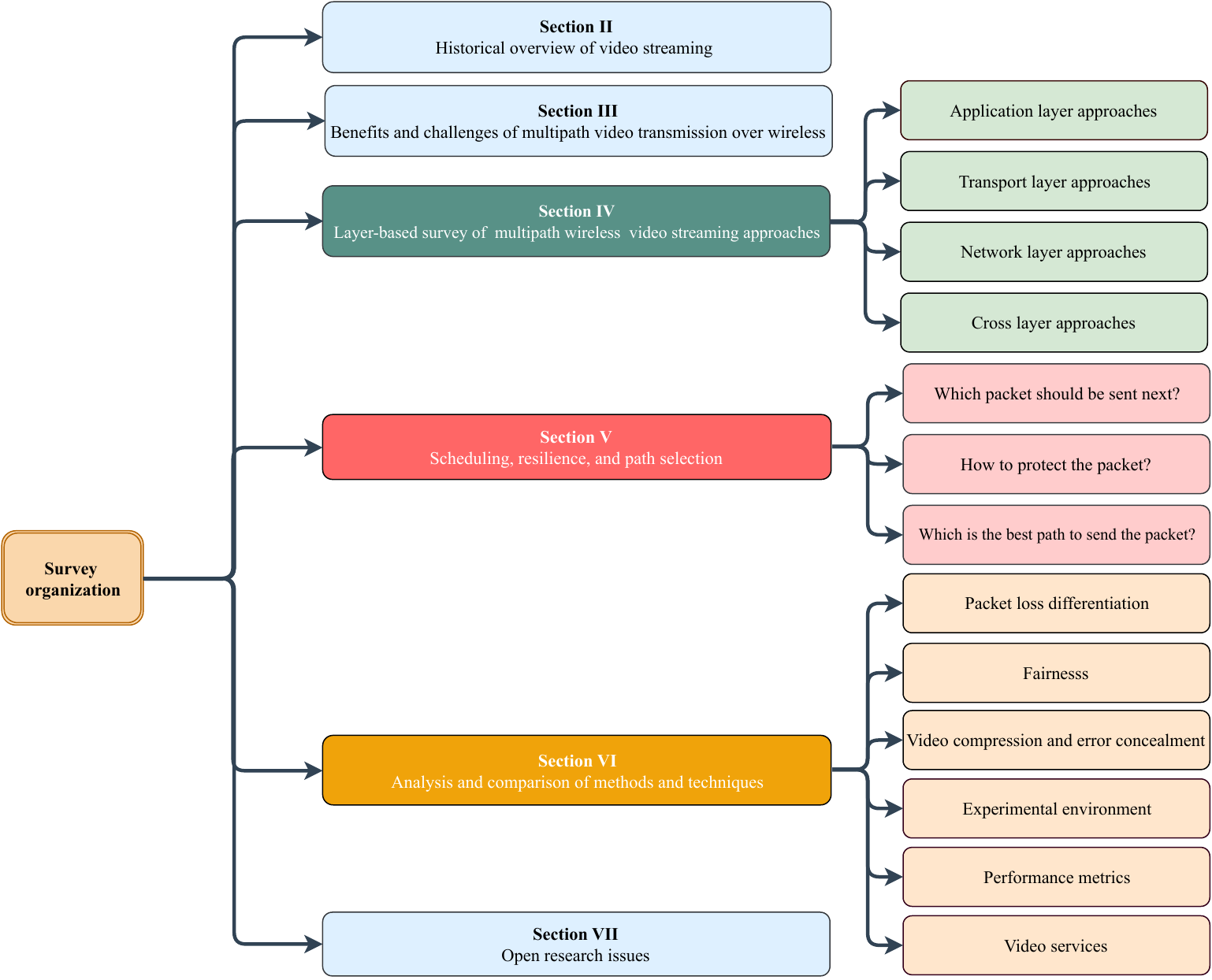}
\caption{ {\color{black}Visual representation of the organization of the survey.}}
\label{fig:organization}
\end{figure*}

{\color{black}
Altogether, this work contributes with} a holistic survey on the challenges, solutions, and open research problems for multipath wireless video streaming. The survey attempts to embrace  nearly all the key techniques and relevant scientific publications. {\color{black}The specific contributions of the survey are as follows:}

\begin{itemize}
 \item Historical overview of {\color{black} video streaming standard and technology developments}.

\item {\color{black}Comprehensive study} regarding the benefits of  multipath video transmission over wireless to improve  QoE.

 \item In-depth discussion regarding the challenges of deploying a wireless multipath solution for video delivery considering the type of video streaming service.

 \item Taxonomies from different aspects of stack layers and scheduling functionalities to classify the state-of-the-art.

  \item
The survey provides a comprehensive literature review of multipath wireless video streaming covering relevant approaches and new techniques in the field~(close to fifty research works), classifying them accordingly, assessing each approach from an end-to-end layered stack perspective, scheduling functions, methods and features, also, outlining their trade-offs.



 \item {\color{black}Low-level multipath wireless scheduling functions are discussed to improve the video streaming QoE;~(\textit{i}) key features to select the proper packets to be transmitted through êach network interface,~(\textit{ii}) relevant packet protection techniques~(channel-level and source-level) to achieve better video streaming goodput and QoE, and~(\textit{iii})  critical path characteristics to select the most qualified path to transfer the packet.}

 \item
 Relevant aspects of multipath video streaming approaches are considered, including packet loss differentiation, fairness, video compression, error concealment, experimental environment, performance metrics, and video services.

\item Several open issues and trends are outlined for future research.

\end{itemize}}

{\color{black}{
\subsection{Roadmap of the survey}}}

The high-level organization of this survey is illustrated in Figure~\ref{fig:organization}. An overview of video streaming protocols is provided in Section~\ref{sec: Historical overview of Video Streaming Protocols}. The benefits and challenges of adding multipath transmission to video streaming scenarios are presented in Section~\ref{sec: Benefits and Challenges}. Surveyed works are then introduced in Section~\ref{sec: MultiPath Mobile Video Streaming Approaches} and classified based on the protocol layer and on the used protocol/feature. In Section~\ref{sec: Scheduler, Resilience, and Routing Functions in Multipath Video}, the works are then investigated based on the scheduling functions: choice of the next packet to be transmitted~(packet selection), data packet protection method~(packet protection), and selection of the proper network channel~(path selection).
Section~\ref{sec: Analysis and Comparison of Candidate Methods and Techniques} provides additional information about the surveyed works that may also be of interest for the reader, such as packet loss differentiation, fairness consideration, video codecs, the experimental environment, performance metrics, and video services. {\color{black} Section~\ref{sec: Open Research Issues} presents research issues and directions. Finally, Section~\ref{sec: Conclusions} provides
concluding remarks.} {\color{black}A list of abbreviations can be found in the appendix to help readers handling the myriad of acronyms.}

\section{Historical Overview of Video Streaming}{\label{sec: Historical overview of Video Streaming Protocols}}

\begin{figure*}[ht]
\centering
\includegraphics[width=0.7\textwidth]{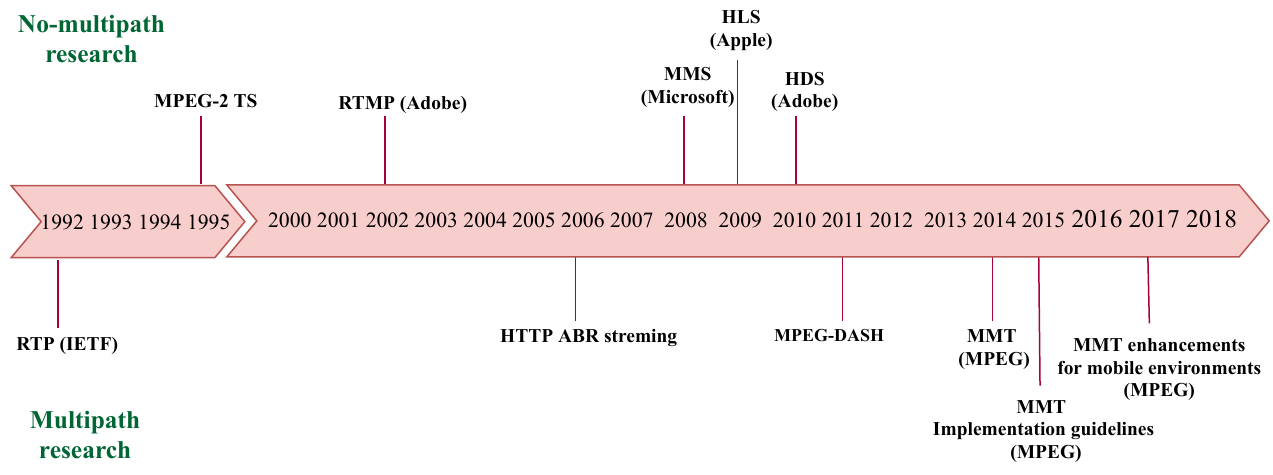}
\caption{{\color{black}Historical overview of video streaming protocols.}}
\label{fig:videoStreamingProtocols}
\end{figure*}

This section provides a general picture of video streaming development as presented by the timeline and milestones in Figure~\ref{fig:videoStreamingProtocols}. Interested readers are referred  to~\cite{yuste2015understanding} for a deeper review of different MPEG standards.

The first widely used video streaming protocol is the Real-time
Transport Protocol~(RTP)~\cite{Schulzrinne1992}, which was initially released in 1992 by IETF. It is a UDP-based protocol used for unidirectional real-time video streaming. RTP has very low overhead and works well in managing IP networks. However, it requires a payload format for each media type or codec~\cite{lim2014new}, suffers from lack of multiplexing and has limited support for non-real-time video. Another disadvantage of RTP is that many Content Delivery Networks~(CDNs)\nomenclature{CDN}{Content Delivery Networks} do not support it because the server must manage a separate streaming session for each client,  turning large-scale deployment more resource intensive. Moreover, RTP cannot traverse firewalls and is connectionless. Therefore, RTP is generally employed for private managed networks where the number of packet losses is small, such as pay-TV cable networks. More technical details on RTP will be provided in Section~\ref{sec: Application Layer Approaches}.

The next widely known and adopted video streaming protocol
shown in Figure~\ref{fig:videoStreamingProtocols} is the MPEG-2 Transport System
(MPEG-2 TS)~\cite{Information2007Part1}\nomenclature{MPEG-2 TS}{MPEG-2 Transport System}.  It has been widely used since 1995 in digital broadcasting, mobile broadcasting systems and streaming over the Internet. Several standards have also adopted this protocol, such as the Terrestrial Digital Multimedia Broadcasting~(T-DMB),~\nomenclature{T-DMB}{Terrestrial Digital Multimedia Broadcasting} the Digital Video Broadcasting Handheld~(DVB-H),\nomenclature{DVB-H}{Digital Video Broadcasting Handheld} the Advanced Television Systems Committee~(ATSC)~\nomenclature{ATSC}{Advanced Television Systems Committee} and the Internet Protocol TeleVision~(IPTV)\nomenclature{IPTV}{Internet Protocol TeleVision}~\cite{yie2016method}. MPEG-2 TS is not only a format for fast and reliable packetized streaming delivery but also a format for storage. In addition, MPEG-2 TS is fully codec agnostic. Since the requirements for on-demand and personalized video delivery over the Internet have dramatically increased, it became challenging for MPEG-2 TS to achieve the high requirements of broadcasting over IP~\cite{lim2014new, lim2013mmt}.
For example, MPEG-2 TS is {\color{black}inappropriate} for UHD delivery over packet networks due to the pre-multiplexing mechanism, not flexible packetization and small-fixed packet size~(188 bytes).

The next protocol in the timeline of Figure~\ref{fig:videoStreamingProtocols} is the {\color{black}Real-Time Messaging Protocol~(RTMP)~\cite{parmar2012real}}.\nomenclature{RTMP}{Real Time Messaging Protocol} It is an Adobe proprietary protocol standardized in 2002 that was initially developed by Macromedia. RTMP is typically over  TCP and used for bidirectional video streaming. This protocol provides the advantage of multiplexing capability but requires flash player plugin.   Another disadvantage is that RTMP suffers from not being codec agnostic and not supporting some newer video codecs, such as High Efficiency Video Coding~(HEVC)~\cite{hevc}. Yet another disadvantage is that it is blocked by firewalls and not supported by all CDNs.

The next highlighted point in the timeline of Figure~\ref{fig:videoStreamingProtocols} is not a protocol, but a video streaming technique introduced in 2006 that become highly adopted in the subsequent streaming protocols. Adaptive Bit Rate~(ABR)~\nomenclature{ABR}{Adaptive Bit Rate} streaming was introduced by Move Networks~\cite{BrueckMark2010Apparatus} and is over HTTP\nomenclature{HTTP}{Hypertext Transfer Protocol}~\cite{fielding1999hypertext} with some rate adaptation techniques considering different parameters, such as bandwidth availability or media playout situations~\cite{seufert2015survey}.

HTTP-based video streaming solutions are easy to deploy in the current Internet architecture and can traverse firewalls. Moreover, the client can manage the streaming without the need to maintain a session state on the server, thus it improves scalability~\cite{sodagar2011mpeg}. Therefore, HTTP is supported by most of CDNs~\cite{james2016multipath}, contributing to the growing interest as a video streaming protocol. In 2015, a new version of HTTP, namely HTTP/2, was standardized~\cite{rfc7540}  and {\color{black}showed} improvement in video quality and performance. More technical details on HTTP/2 will be provided in Section~\ref{sec: Application Layer Approaches}.

The initial HTTP Adaptive Streaming~(HAS)~\nomenclature{HAS}{HTTP Adaptive Streaming} commercially successful protocols~\cite{seufert2015survey} were {\color{black}the Microsoft Silverlight Smooth Streaming~(MSS)\footnote{\color{black}https://docs.microsoft.com/en-us/iis/media/smooth-streaming/smooth-streaming-transport-protocol}} developed by Microsoft in 2008, the HTTP Live Streaming~(HLS)~\cite{rfc8216} developed by {\color{black} Apple in 2009 and the Adobe HTTP Dynamic Streaming~(HDS)}\footnote{\color{black}http://www.adobe.com/products/hds-dynamic-streaming.html} developed by Adobe in 2010. Since all these protocols were proprietary and incompatible, in 2011, the Dynamic Adaptive Streaming over HTTP~(MPEG-DASH)~\cite{DASH2011Part6} protocol was developed to become a unified codec agnostic standard. MPEG-DASH flexible delivery and codec agnostic properties have turned it into a successful protocol widely adopted by content providers~\cite{MPEGDASHYouTubeNetflix}, such as Netflix and YouTube. Another advantage is that DASH supports both multiplexed and unmultiplexed encoded content. However, the protocol has some performance limitations, for instance, low latency delivery~\cite{lim2014new}, several switches, freezes, and poor QoE~\cite{yahiahttp}. More technical details on DASH are provided in Section~\ref{sec: Application Layer Approaches}. All ABR-based protocols support live and video on demand~(VoD)~\nomenclature{VoD}{Video on Demand} delivery. More details  on  HAS protocols are discussed  in~\cite{seufert2015survey}.

The next protocol in the timeline of Figure~\ref{fig:videoStreamingProtocols} is the MPEG Media Transport~(MMT)~\cite{MPEG-H2014part1}. It was standardized by MPEG in 2014 considering recent changes in multimedia delivery and requirements for Internet technologies, such as IP and HTML for Internet-based video streaming solutions~\cite{lim2014new}. This protocol also supports UHD resolution and HEVC video codec.
MMT was designed to inherit some MPEG-2 TS features, such as {\color{black}content-agnostic} media delivery, easy conversion between storage and delivery format and {\color{black}multiplexing support}. In addition, MMT was developed due to a need for an international standard to support hybrid delivery in various heterogeneous network environments. Then, in 2015, implementation guidelines {\color{black} were standardized} to provide technical guidelines for implementing and deploying MMT systems.

{\color{black}
The last highlight point in the timeline is MMT enhancement for mobile environment specifying multipath support which has already been added to the protocol and standardized~\cite{Coding2017}.  }
MMT was adopted by some recent standards, such as the ATSC 3.0~\cite{ye2016shvc}, which is a recent standard with a hybrid delivery model which includes MMT and DASH. Especially, in ATSC 3.0, MMT protocol~(MMTP)~\nomenclature{MMTP}{MPEG Media Transport Protocol} is proposed for broadcasting, and DASH over HTTP is proposed for broadband service.

One important difference between DASH and MMT is that, typically, DASH supports a client driven Quality of Service~(QoS)~\nomenclature{QoS}{Quality of Service} control standard, while MMT supports a server driven QoS control services~\cite{ye2016shvc}. More technical details on MMT are provided in Section~\ref{sec: Application Layer Approaches}.

Adding multipath capabilities has been investigated in some of the {\color{black}above-mentioned} video streaming protocols but not in all of them, especially {\color{black}it has not been investigated} for the proprietary protocols due to their closed and incompatible design~(Figure~\ref{fig:videoStreamingProtocols}).

\noindent \textbf{Commercial services.}~We already mentioned some companies using their own developed proprietary protocols, such as Move Networks, Microsoft~(MMS), Apple~(HLS) and Adobe~(RTMP, HDS). Besides them, there are  other company services adopting or in the process of developing streaming solutions. For example,  Skype and WhatsApp are mobile application platforms providing video calls or video conferences for their users. These services use RTP for video streaming~\cite{karya2018rtp}. Hulu is an online video service providing on-demand shows, movies, documentaries, and more. Hulu requires flash player for video streaming through the RTMP protocol~\cite{trestian2018seamless}.


A number of video service providers use DASH. Among the most famous ones are~YouTube, Netflix, Twitch and Vimeo. 
Another commercial streaming service is  Bitmovin, which  provides adaptive streaming supporting MPEG-DASH and HLS.
Generally, DASH has gotten broad support from commercial companies -- {\color{black}see DASH Industry Forum member list}\footnote{{\color{black}https://dashif.org/members/}}. In addition, browsers, such as Chrome and Firefox, also support DASH~\cite{MPEGDASHYouTubeNetflix}. {\color{black}NHK, Nippon Hoso Kyokai}\footnote{\color{black}https://www.nhk.or.jp/corporateinfo/}, is a Japan's telecommunication company~(public service broadcaster) uses MMT as the protocol of choice for 4K/8K Super Hi-Vision.

\section{Multipath Video Transmission over Wireless: Benefits and Challenges}{\label{sec: Benefits and Challenges}}

\begin{figure}[!t]
\centering
\includegraphics[width=0.5\textwidth]{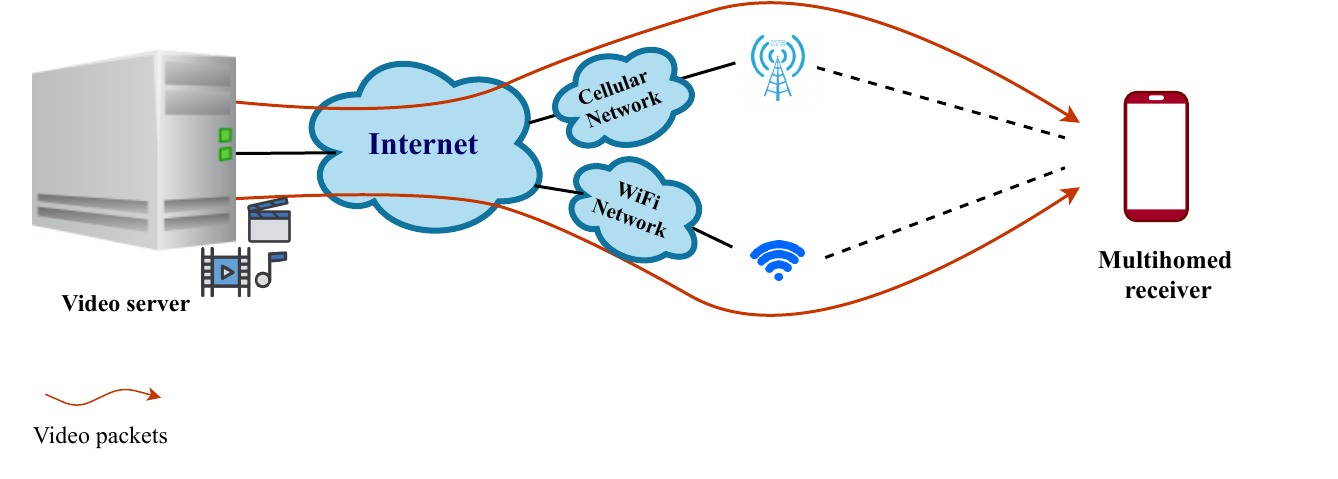}
\caption{{\color{black} Multipath wireless video streaming over LTE and WiFi networks.}}
\label{fig:multipath}
\end{figure}

As today, most of current portable  devices are already equipped with both cellular and WiFi interfaces. These multiple interface devices{\color{black}, which could be equipped with two or more than two interfaces, having} the ability to connect simultaneously to multiple network paths are known as multihomed devices, as illustrated in Figure~\ref{fig:multipath}. Multihomed devices can utilize multipath communication by aggregating the available bandwidth from multiple Radio Access Technologies~(RATs)\nomenclature{RAT}{Radio Access Technologies}. With multiple interfaces, users can receive data through parallel paths with multiple IP addresses.\\
As previously stated, providing high/optimal QoE for the final user in the {\color{black}wireless} video streaming scenario requires high bandwidth and low transmission delay. This is a challenging task, considering the several aspects involved in {\color{black}wireless} transmission, such as bandwidth constraints, lossy wireless channels, delay, lack of coverage and congested networks.

 {\color{black}
\subsection{Benefits} }
Adding multipath transmission capabilities can provide the benefits discussed next.

\noindent  \textbf{Throughput increase.}~By aggregating bandwidth and distributing video traffic over multiple network paths, faster transmission can be achieved, which is essential for real-time video streaming applications~\cite{yap2012making}.\\

\noindent  \textbf{Load balancing.}~Efficiently distributing video traffic through the available network paths relieves congestion~\cite{qadir2015exploiting}. In addition, load balancing improves stability by achieving lower variability and inter-packet delay~(jitter);
\\

\noindent \textbf{Reliability and seamless connectivity.}~With multipath, users can simultaneously utilize multiple available network connections. Better service continuity can be  achieved by increasing the probability of keeping end-to-end connections alive. In the case of failure or congestion in  one network path, multipathing  provides a resilient alternative, resulting in improved  user video  experience.\\

\noindent  \textbf{Reduction of burst loss length.}~Continuous packet losses are harmful to the perceived video streaming quality~\cite{apostolopoulos2000reliable}. Although decoders
can recover from the loss of a small number of video packets by exploiting correlations in the previously received video sequences, its effectiveness decreases dramatically in case of losing large number of continuous video packets.
Using multipath streaming benefits to convert burst losses to isolated losses, and consequently, increase the probability of recovering from lost packets\\

\noindent \textbf{Delay decrease.}~Using multiple paths contributes to having video data ready at the receiver faster, thus, decreasing the effective delay especially from an application perspective. Probing  multiples paths enables to get the data from the lowest delay path,  reducing the Time To First Byte~(TTFB)\nomenclature{TTFB}{Time To First Byte}, i.e., the time between the video request being sent and the first packet received after the request~\cite{gutterman2019requet}.\\

\noindent  \textbf{Security.}~Splitting video streams over multiple paths improves protection to some  security threats~\cite{singh2015survey} once that each network path only carries parts of the whole video stream.

 {\color{black}

\subsection{Challenges} }
Despite all of its benefits, attempting to deploy a multipath solution for video delivery may  {\color{black} entangle a series of potential} roadblocks from different  perspectives.\\

\noindent \textbf{Compatibility.}~Implementation of a general multipath solution usually requires changing one or both of server and {\color{black}client} sides, modifying standardized protocols, improving operating systems kernel and/or changing third-party network equipment. \\

\noindent \textbf{Networks heterogeneity.}~Heterogeneous wireless networks vary based on different bandwidth constraints, delays, jitters and packet loss rates. These different physical properties cause asymmetric communication for video transmission, and consequently, may decrease the overall streaming quality. For instance, a large difference between LTE and WiFi bandwidth decreases the bandwidth aggregation performance~\cite{bui2013greenbag}. Second and third generation~(2G and 3G) of cellular networks do not provide enough bandwidth to support live video streaming due to high data rate~\cite{han2011end,luo2011digital,chen2013measurement}. In addition, {\color{black}the retransmission mechanism in 3G~\cite{chen2013measurement}} may increase the Round-Trip Time~(RTT) and the rate variability. 
4G LTE offers higher data transmission rate and signal coverage than 2G/3G~\cite{yoon2012muvi,chen2013measurement}. When comparing 4G to WiFi, there relevant differences in terms of bandwidth, packet loss, and round-trip time can be observed~\cite{xu2013cmt,chen2013measurement}.
These aspects, in addition to wireless losses being recognized as congestion by some protocols~(e.g., TCP) resulting in decreased  network throughput~\cite{chan2005tcp}, turn multipath communications over heterogeneous wireless networks a truly challenging task.  \\



\noindent \textbf{Out-of-order packets.}~Spreading data over heterogeneous paths with different RTTs, throughput fluctuations, and jitter existence introduces the out-of-order packets problem. This phenomenon causes unnecessary packets retransmissions, wasted bandwidth, and consequently, network congestion. In addition, more time is required to recover the ordered data. A robust multipath transmission solution is required to cope with packet reordering in heterogeneous wireless networks~\cite{li2016multipath} to avoid video quality degradation. \\

\begin{figure}[!t]
\centering
\includegraphics[width=0.5\textwidth]{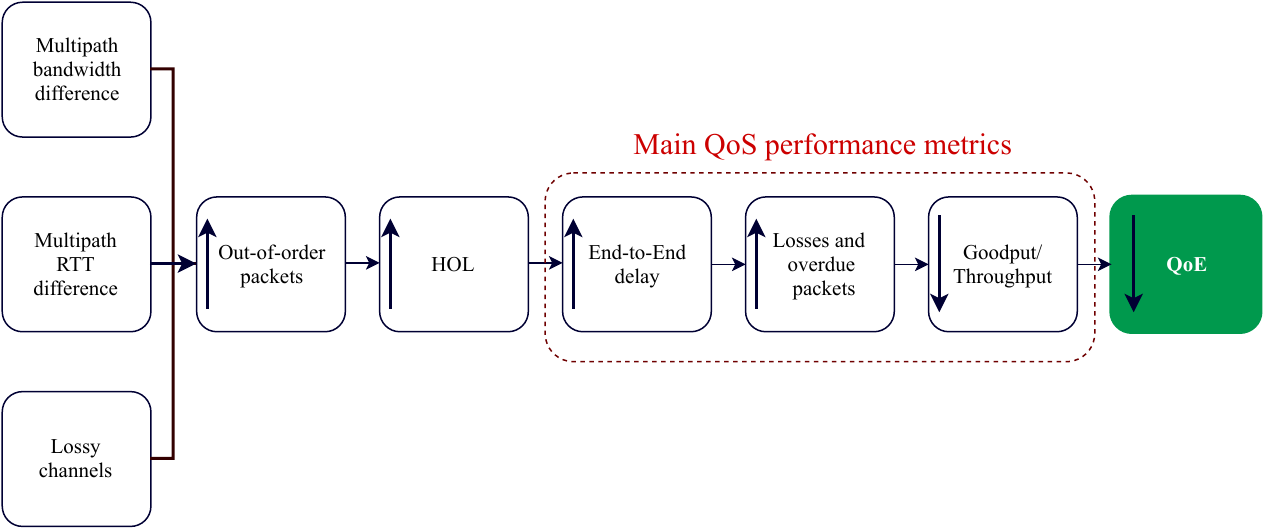}
\caption{{\color{black}Challenges of employing multipath transmission in wireless video streaming applications and possible adverse effects to be avoided.}}
\label{fig:Challenges causes and effects}
\end{figure}

\noindent \textbf{Head-of-Line~(HOL) blocking.}~When many packets are stored in the destination buffer waiting for delayed packets, the buffer may become full and blocked. This issue is referred as Head-of-Line blocking\nomenclature{HOL}{Head-of-Line}~\cite{corbillon2016cross,ferlin2014tackling}. Generally, buffer blocking occurs with reliable protocols that guarantee in-order packet delivery, such as TCP, and it may become worse in case of multipath delivery. The Bufferbloat phenomenon is the main reason {\color{black}for} HOL blocking, contributing to high latency, especially in 3G/4G cellular networks~\cite{chen2014bufferbloat,ferlin2014tackling}. Bufferbloat  occurs because of significantly large network buffers ~(e.g., large router queues) that avoid packet loss at the cost of adding high latencies under  congestion. The problem can become worse in case of multipath delivery
because if  bufferbloat occurs in one of the paths, those  packets  arrive at the destination with high delay and out-of-order, resulting in HOL blocking.
Consequently,  HOL blocking  not only increases end-to-end delay and jitter but successfully arrived packets may become obsolete due to the long waiting time  in the destination buffer.  \\

\noindent \textbf{End-to-end delay.}~Real-time video streaming requires a bounded end-to-end delay~\cite{sun2016mars}, which refers to the measured delay from the generation of a video frame to the moment when it can be decoded. {\color{black}End-to-end} delay includes holding time of a video frame at both sender and receiver sides and the transmission delay. It could also include the queuing delay, propagation delay, access delay, and reordering delay. The queuing delay refers to packet buffering in the sender, receiver and other nodes in the network during packet transmission. The transmission delay and radio access delay occur in the physical transmitter to map the data from packets to bits on physical radio interface's hardware. The distance between entities causes the propagation delay. The access points introduce transfer and propagation delay. In the case of video streaming on multipath networks, reordering delay can be increased~\cite{brosh2010delay}.   \\

\noindent \textbf{Overdue packets.}~Video data packets arriving at the destination after decoding deadlines are expired and known as overdue packets. {\color{black}While overdue packets for UDP-like transmissions may cause video distortions~(i.e., degradation of the visual video fidelity~\cite{sani2017adaptive}) similar to lost packets, in reliable transport protocols like TCP  the effects surface as stalling~(i.e., video freezes) or rebuffering. Avoiding stalls becomes most critical in live streaming scenarios. Thus, this kind of real-time applications, even when based on TCP-like solutions, consider the overdue packets as lost packets since they are discarded. This concept is called liveness~\cite{sani2017adaptive}. Therefore, suitable multipath streaming strategies need to consider potential decoding deadlines of the  receivers.  \\

\noindent \textbf{Wrapping up the Challenges.}~Figure~\ref{fig:Challenges causes and effects} aims at putting together all key issues and  possible adverse effects of multipath wireless video streaming. {\color{black}It is important to avoid or minimize such effects to design any multipath wireless video streaming solution.} In other words, optimization of QoS-related parameters leads to
 improved QoE\nomenclature{QoE}{Quality of Experience}~\cite{barakovic2013survey}.
%
QoS measurements may differ based on the type of video streaming service~\cite{van2011multimedia,sani2017adaptive}, such as VoD, live or {\color{black}real-time}. VoD is a video streaming service which encoded media is pre-stored at the server, and the user can select and watch it at any time~(e.g., Netflix movies). In contrast, in live and {\color{black}real-time} video streaming services {\color{black}(e.g., live sport streaming, real-time video  including interactive video call, gaming, etc.)} the video content is not pre-stored/available when the streaming starts.  In live streaming, the buffer is smaller compared to VoD streaming to avoid long delays and it also has stricter deadlines. {\color{black}Real-time} video streaming has even shorter delay constraint. For example, according to~\cite{recommendation20011010} and~\cite{wu2016bandwidth}, a large delay of 5 seconds may be acceptable for VoD and around 1~second delay is acceptable for live streaming, but in order to achieve excellent {\color{black}real-time} streaming quality, the solution should provide the end-to-end delay not exceed 150 ms. Besides, packet loss rates higher than 1\% are not acceptable for live video streaming solutions~\cite{austerberry2005technology, chow2009ems}. In some applications with high scenes variability, such as football, it was reported~\cite{mwela2010impact} that subjects already become uncomfortable for packet loss rate slightly above 0.3\%.
Finally, meeting all QoS requirements does not necessarily guarantee high(est) user QoE. Devices' operating system, hardware, battery, operator pricing, light, people around the user and emotion are some examples of factors that impact the users’  experience~\cite{trestian2018seamless,barakovic2013survey}.}\\

\section{{\color{black}Layer-based survey of {\color{black} multipath wireless}  video streaming approaches}}{\label{sec: MultiPath Mobile Video Streaming Approaches}}

In this section, the surveyed multipath wireless video streaming works are initially introduced and classified in Table~\ref{layered classification} according to protocol stack layers and protocol/features. The table also indicates which parts of the network equipment~(whether client, server,  network or a  combination of them) need to be adjusted to become compatible with multipath transmission schemes.
Most flexible solutions require only client side modification because they are compatible with the current network infrastructure and {\color{black} do not need any change on the server or network infrastructure.} On the other hand, some other approaches  require server side modification or even both server and client together.  Most difficulties are with solutions that they need to adjust network infrastructure.

\subsection{Application Layer Approaches}{\label{sec: Application Layer Approaches}}

Video streaming approaches focused on the application layer have the advantage of accessing player buffer status and relevant video content information, such as frames priorities and coding dependencies. {\color{black}Application-specific information provides the multipath approach with rich inputs to define the video streaming scheduling strategies. One key advantage is that there is no need to change lower layer protocols. {\color{black}However, a significant drawback of these solutions is that they commonly require modifications of the video software. An application-level sequence number is generally used for loss detection in application layer approaches, which often increases the overall protocol overhead. In addition, to }perform knowledgeable packet scheduling decisions, the application requires a mechanism to estimate the network paths' performance, e.g., through application-specific probes or from TCP congestion control information~\cite{li2016multipath}.}

\begin{figure}[!t]
\centering
\includegraphics[width=0.3\textwidth]{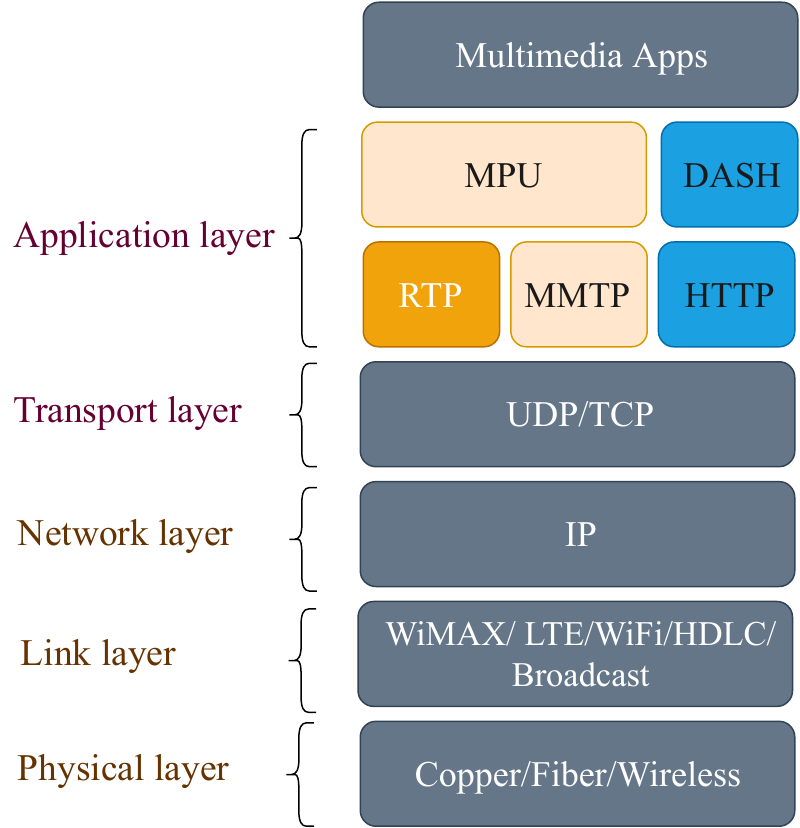}
\caption{{\color{black}Application layer protocols in a network protocol stack.}}
\label{fig:ApplicationLayerProtocolStack}
\end{figure}

In this subsection, we discuss relevant works that are based on RTP, DASH, MMT, {\color{black}QUIC} and other adaptive streaming approaches. {\color{black} Most of these protocols} were  previously introduced in  Section~\ref{sec: Historical overview of Video Streaming Protocols} and will be further detailed here. Figure~\ref{fig:ApplicationLayerProtocolStack} illustrates the protocol stack position of these protocols and Table~\ref{layered classification} presents each category.\\

\definecolor{ForestGreen}{RGB}{34,139,34}
\definecolor{skyblack}{rgb}{0,0.55,1}
\definecolor{LightCyan}{rgb}{0.92,1,1}
\definecolor{Gray}{gray}{0.95}

\vspace{-1.0em}

\begin{table*}[!t]
\centering
\caption{{\color{black}CLASSIFICATION OF  SURVEYED WORKS ACCORDING TO THE PROTOCOL LAYER  IMPLEMENTATION}}
\label{layered classification}
\scalebox{0.93}{
\begin{tabular}{llll}
\hline
\rowcolor[HTML]{C0C0C0}
\multicolumn{1}{|l|}{\cellcolor[HTML]{C0C0C0}\textbf{Protocol layer}}                       & \multicolumn{1}{l|}{\cellcolor[HTML]{C0C0C0}\textbf{Applied protocols/features}}                           & \multicolumn{1}{l|}{\cellcolor[HTML]{C0C0C0}\textbf{Works}}                      & \multicolumn{1}{l|}{\cellcolor[HTML]{C0C0C0}\textbf{Compatibility}}                                                                            \\ \hline
 \multicolumn{1}{|l|}{\cellcolor[HTML]{EFEFEF}}                                            & \multicolumn{1}{l|}{{   }}                                                      & \multicolumn{1}{l|}{{\cellcolor[HTML]{EFEFEF}}MRTP~\cite{mao2006mrtp}}                  & \multicolumn{1}{l|}{ {\cellcolor[HTML]{EFEFEF}}Server and Client}                                                                                  \\
\multicolumn{1}{|l|}{\cellcolor[HTML]{EFEFEF}}                                            & \multicolumn{1}{l|}{\multirow{-2}{*}{{   RTP}}}                                 & \multicolumn{1}{l|}{{MPRTP~\cite{singh2013mprtp}}}          & \multicolumn{1}{l|}{{Server and Client}}                                                                                  \\

\multicolumn{1}{|l|}{\cellcolor[HTML]{EFEFEF}}                                            & \multicolumn{1}{l|}{{   }}                                                      & \multicolumn{1}{l|}{{\cellcolor[HTML]{EFEFEF}MRTP-AR~\cite{leiwm-avtcore-mprtp-ar-09}}}                  & \multicolumn{1}{l|}{{{\cellcolor[HTML]{EFEFEF}}Server and Client}}                                                                                  \\   \cline{2-4}

\multicolumn{1}{|l|}{\cellcolor[HTML]{EFEFEF}}                                            & \multicolumn{1}{l|}{{   }}                                                      & \multicolumn{1}{l|}{{Xing et al.~\cite{xing2012rate}}}                     & \multicolumn{1}{l|}{{Client}}                                                                                             \\
\multicolumn{1}{|l|}{\cellcolor[HTML]{EFEFEF}}                                            & \multicolumn{1}{l|}{{   }}                                                      & \multicolumn{1}{l|}{{{\cellcolor[HTML]{EFEFEF}}Chowrikoppalu et al.~\cite{chowrikoppalu2013multipath}}} & \multicolumn{1}{l|}{{{\cellcolor[HTML]{EFEFEF}}Client}}                                                                                             \\
\multicolumn{1}{|l|}{\cellcolor[HTML]{EFEFEF}}                                            & \multicolumn{1}{l|}{\multirow{-4}{*}{{   }}}                                & \multicolumn{1}{l|}{{RTRA~\cite{xing2014real}}}                 & \multicolumn{1}{l|}{{Client}}                                                                                             \\
\multicolumn{1}{|l|}{\cellcolor[HTML]{EFEFEF}}                                            & \multicolumn{1}{l|}{{   }}                                                      & \multicolumn{1}{l|}{{{\cellcolor[HTML]{EFEFEF}}Houz{\'e} et al.~\cite{houze2016applicative}}}             & \multicolumn{1}{l|}{{{\cellcolor[HTML]{EFEFEF}}Client}}                                                                                             \\
\multicolumn{1}{|l|}{\cellcolor[HTML]{EFEFEF}}                                            & \multicolumn{1}{l|}{\multirow{-4}{*}{{   DASH}}}                                & \multicolumn{1}{l|}{{{\color{black}Go et al.}~\cite{go2019hybrid}}}                 & \multicolumn{1}{l|}{{Server and Client}}                                                                                           \\ \cline{2-4}
\multicolumn{1}{|l|}{\cellcolor[HTML]{EFEFEF}}                                            & \multicolumn{1}{l|}{}                                                                          & \multicolumn{1}{l|}{{\cellcolor[HTML]{EFEFEF}}Kolan et al.~\cite{kolan2016method}}                                 & \multicolumn{1}{l|}{{\cellcolor[HTML]{EFEFEF}}Server and Client}                                                                                                         \\
\multicolumn{1}{|l|}{\cellcolor[HTML]{EFEFEF}}                                            & \multicolumn{1}{l|}{MMT}                                                                          & \multicolumn{1}{l|}{{\color{black}Afzal et al.~\cite{afzal2018novel,afzal2021multipath}}}                                 & \multicolumn{1}{l|}{Server and Client}                                                                                                         \\
\multicolumn{1}{|l|}{\cellcolor[HTML]{EFEFEF}}                                            & \multicolumn{1}{l|}{}                                                                          & \multicolumn{1}{l|}{{\cellcolor[HTML]{EFEFEF}}Sohn et al.~\cite{sohn2015synchronization}}                                 & \multicolumn{1}{l|}{{\cellcolor[HTML]{EFEFEF}}Server and Client}                                                                                                         \\ \cline{2-4}

\multicolumn{1}{|l|}{\cellcolor[HTML]{EFEFEF}}                                            & \multicolumn{1}{l|}{\multirow{1}{*}{{\color[HTML]{333333}{\color{black} QUIC}}}}                       & \multicolumn{1}{l|}{{{\color{black}Michel et al.}~\cite{michel2018adding}}}               & \multicolumn{1}{l|}{{Server and Client}}                                                                                 \\\cline{2-4}
\multicolumn{1}{|l|}{\cellcolor[HTML]{EFEFEF}}                                            & \multicolumn{1}{l|}{{   }}                                                      & \multicolumn{1}{l|}{{\cellcolor[HTML]{EFEFEF}Evensen et al.~\cite{evensen2010quality}}}                                 & \multicolumn{1}{l|}{{\cellcolor[HTML]{EFEFEF}Client}}                                                                                             \\
\multicolumn{1}{|l|}{\cellcolor[HTML]{EFEFEF}}                                            & \multicolumn{1}{l|}{{   }}                                                      & \multicolumn{1}{l|}{{ Evensen et al.~\cite{evensen2011improving}}}             & \multicolumn{1}{l|}{Client}                                                                                             \\
\multicolumn{1}{|l|}{\cellcolor[HTML]{EFEFEF}}                                            & \multicolumn{1}{l|}{{   }}                                                      & \multicolumn{1}{l|}{{\cellcolor[HTML]{EFEFEF}Evensen et al.~\cite{evensen2012using}}}                 & \multicolumn{1}{l|}{{\cellcolor[HTML]{EFEFEF}Client}}                                                                                             \\
\multicolumn{1}{|l|}{\multirow{-11}{*}{\cellcolor[HTML]{EFEFEF}\textbf{Application Layer}}} & \multicolumn{1}{l|}{\multirow{-4}{*}{{   Other Adaptive Streaming Approaches}}} & \multicolumn{1}{l|}{{GreenBag~\cite{bui2013greenbag}}}          & \multicolumn{1}{l|}{{Client}}                                                                                            \\  \multicolumn{1}{|l|}{\multirow{-11}{*}{\cellcolor[HTML]{EFEFEF}}} & \multicolumn{1}{l|}{\multirow{-4}{*}{{   }}} & \multicolumn{1}{l|}{{\cellcolor[HTML]{EFEFEF} {\color{black}MP-H2~\cite{nikravesh2019MP-H2}}}}          & \multicolumn{1}{l|}{{\cellcolor[HTML]{EFEFEF}Client}} \\ \hline
\multicolumn{1}{|l|}{\cellcolor[HTML]{EFEFEF}}                                            & \multicolumn{1}{l|}{{\color[HTML]{333333} }}                                                      & \multicolumn{1}{l|}{{\color[HTML]{333333} BEMA* ~\cite{wu2016bandwidth}}}            & \multicolumn{1}{l|}{{\color[HTML]{333333} Server and Client}}                                                                                  \\
\multicolumn{1}{|l|}{\cellcolor[HTML]{EFEFEF}}                                            & \multicolumn{1}{l|}{{\color[HTML]{333333} }}                                                      & \multicolumn{1}{l|}{{\cellcolor[HTML]{EFEFEF}Freris at al.~\cite{freris2013distortion}}}             & \multicolumn{1}{l|}{{\cellcolor[HTML]{EFEFEF}Server and Client}}                                                                                  \\
\multicolumn{1}{|l|}{\cellcolor[HTML]{EFEFEF}}                                            & \multicolumn{1}{l|}{\multirow{-3}{*}{{\color[HTML]{333333} Multipath UDP}}}                       & \multicolumn{1}{l|}{{Correia at al.~\cite{correia2012optimal}}}               & \multicolumn{1}{l|}{{Server and Client}}                                                                                  \\ \cline{2-4}

 \multicolumn{1}{|l|}{\cellcolor[HTML]{EFEFEF}}                                            & \multicolumn{1}{l|}{{Multipath TCP}}                                         & \multicolumn{1}{l|}{{\cellcolor[HTML]{EFEFEF}MPLOT~\cite{sharma2008mplot}}}            & \multicolumn{1}{l|}{{\cellcolor[HTML]{EFEFEF}Server and Client}}                                                                                  \\ \cline{2-4}
 \multicolumn{1}{|l|}{\cellcolor[HTML]{EFEFEF}}                                            & \multicolumn{1}{l|}{{Multipath DCCP}}                                         & \multicolumn{1}{l|}{{MP-DCCP~\cite{huang2012qos}}}            & \multicolumn{1}{l|}{{Server}}                                                                                  \\ \cline{2-4}
\multicolumn{1}{|l|}{\cellcolor[HTML]{EFEFEF}}                                            & \multicolumn{1}{l|}{{   }}                                                      & \multicolumn{1}{l|}{{\cellcolor[HTML]{EFEFEF}ADMIT~\cite{wu2016streaming}}}             & \multicolumn{1}{l|}{{\cellcolor[HTML]{EFEFEF}Server and Client}}                                                                                  \\
\multicolumn{1}{|l|}{\cellcolor[HTML]{EFEFEF}}                                            & \multicolumn{1}{l|}{{   }}                                                      & \multicolumn{1}{l|}{{{\color{black}DEAM}~\cite{Wu2019Energy}}}             & \multicolumn{1}{l|}{{Server and Client}}                                                                                  \\

\multicolumn{1}{|l|}{\cellcolor[HTML]{EFEFEF}}                                            & \multicolumn{1}{l|}{{   }}                                                      & \multicolumn{1}{l|}{{\cellcolor[HTML]{EFEFEF}{\color{black}EDAM~\cite{7937943}}}}            & \multicolumn{1}{l|}{{\cellcolor[HTML]{EFEFEF}Server and Client}}                                                                                             \\
\multicolumn{1}{|l|}{\cellcolor[HTML]{EFEFEF}}                                            & \multicolumn{1}{l|}{{   }}                                                      & \multicolumn{1}{l|}{{MPTCP-SD~\cite{diop2012qos}}}            & \multicolumn{1}{l|}{{Server}}                                                                                             \\
\multicolumn{1}{|l|}{\cellcolor[HTML]{EFEFEF}}
& \multicolumn{1}{l|}{\multirow{-3}{*}{{  MPTCP }}}                               & \multicolumn{1}{l|}{\cellcolor[HTML]{EFEFEF}{MPTCP-PR~\cite{diop2012qos}}}             & \multicolumn{1}{l|}{\cellcolor[HTML]{EFEFEF}{Client}}                                                                                             \\
\multicolumn{1}{|l|}{}                                            & \multicolumn{1}{l|}{{   }}                                                      & \multicolumn{1}{l|}{{Xu et al.~\cite{zhang2015multipath}}}            & \multicolumn{1}{l|}{{Server and Client}}                                                                                             \\
\multicolumn{1}{|l|}{\cellcolor[HTML]{EFEFEF}}                                            & \multicolumn{1}{l|}{{   }}                                                      & \multicolumn{1}{l|}{\cellcolor[HTML]{EFEFEF}{PR-MPTCP$^+$~\cite{cao2016pr}}}            & \multicolumn{1}{l|}{\cellcolor[HTML]{EFEFEF}{Server}}                                                                                             \\  \cline{2-4}
\multicolumn{1}{|l|}{\cellcolor[HTML]{EFEFEF}}                                            & \multicolumn{1}{l|}{{   }}
& \multicolumn{1}{l|}{{\cellcolor[HTML]{EFEFEF}Kelly et al.~\cite{kelly2004delay}}}                   & \multicolumn{1}{l|}{{\cellcolor[HTML]{EFEFEF}Not defined}}                                                                                             \\

\multicolumn{1}{|l|}{\cellcolor[HTML]{EFEFEF}}                                            & \multicolumn{1}{l|}{{   }}                                                      & \multicolumn{1}{l|}{{Okamoto et al.~\cite{okamoto2014performance}}}           & \multicolumn{1}{l|}{ {Server}}                                                                                             \\
\multicolumn{1}{|l|}{\cellcolor[HTML]{EFEFEF}}                                            & \multicolumn{1}{l|}{{   }}                                                      & \multicolumn{1}{l|}{{\cellcolor[HTML]{EFEFEF}SRMT~\cite{da2016preventing}}}             & \multicolumn{1}{l|}{{\cellcolor[HTML]{EFEFEF}Server}}                                                                                             \\
\multicolumn{1}{|l|}{\cellcolor[HTML]{EFEFEF}}                                            & \multicolumn{1}{l|}{{   }}                                                      & \multicolumn{1}{l|}{PR-SCTP~\cite{sanson2010pr}}                                            & \multicolumn{1}{l|}{Server}                                                                                                                    \\
\multicolumn{1}{|l|}{\cellcolor[HTML]{EFEFEF}}                                            & \multicolumn{1}{l|}{{   }}                                                      & \multicolumn{1}{l|}{{\cellcolor[HTML]{EFEFEF}CMT-QA~\cite{xu2013cmt}}}           & \multicolumn{1}{l|}{{\cellcolor[HTML]{EFEFEF}Server and Client}}                                                                                  \\
\multicolumn{1}{|l|}{\cellcolor[HTML]{EFEFEF}}                                            & \multicolumn{1}{l|}{{   }}                                                      & \multicolumn{1}{l|}{{CMT-DA~\cite{wu2015distortion}}}             & \multicolumn{1}{l|}{{Server and Client}}                                                                                  \\
\multicolumn{1}{|l|}{\multirow{-14}{*}{\cellcolor[HTML]{EFEFEF}\textbf{Transport Layer}}}   & \multicolumn{1}{l|}{\multirow{-7}{*}{{   SCTP and CMT ~(extension of SCTP)}}}   & \multicolumn{1}{l|}{{\cellcolor[HTML]{EFEFEF}CMT-CA~\cite{wu2016content}}}                  & \multicolumn{1}{l|}{{\cellcolor[HTML]{EFEFEF}Server and Client}}                                                                                   \\ \hline

\multicolumn{1}{|l|}{\cellcolor[HTML]{EFEFEF}}                                            & \multicolumn{1}{l|}{{   }}                                                      & \multicolumn{1}{l|}{{Yap at al.~\cite{yap2012making}}}                    & \multicolumn{1}{l|}{{\begin{tabular}[c]{@{}l@{}}Server~(depends on the application), \\ Client and Network\end{tabular}}} \\
\multicolumn{1}{|l|}{\cellcolor[HTML]{EFEFEF}}                                            & \multicolumn{1}{l|}{\multirow{-2}{*}{{   SDN }}}                            & \multicolumn{1}{l|}{{\cellcolor[HTML]{EFEFEF}MARS~\cite{sun2016mars}}}                  & \multicolumn{1}{l|}{{\cellcolor[HTML]{EFEFEF}Network}}                                                                                            \\ \cline{2-4}
\multicolumn{1}{|l|}{\multirow{-3}{*}{\cellcolor[HTML]{EFEFEF}\textbf{Network Layer}}}      & \multicolumn{1}{l|}{{Proxy}}                                            & \multicolumn{1}{l|}{{BAG~\cite{chebrolu2006bandwidth}}}         & \multicolumn{1}{l|}{{Client and Network}}                                                                                 \\ \hline
\multicolumn{1}{|l|}{\cellcolor[HTML]{EFEFEF}}                                            & \multicolumn{1}{l|}{{   }}                                                      & \multicolumn{1}{l|}{{\cellcolor[HTML]{EFEFEF}Corbillon et al.~\cite{corbillon2016cross}}}               & \multicolumn{1}{l|}{{\cellcolor[HTML]{EFEFEF}Server}}                                                                                             \\
\multicolumn{1}{|l|}{\cellcolor[HTML]{EFEFEF}}                                            & \multicolumn{1}{l|}{{   }}                                                      & \multicolumn{1}{l|}{{Ojanper{\"a} et al.~\cite{ojanpera2016network}}}              & \multicolumn{1}{l|}{{Server, Client and Network}}                                                                         \\
\multicolumn{1}{|l|}{\cellcolor[HTML]{EFEFEF}}                                            & \multicolumn{1}{l|}{{   }}                                                      & \multicolumn{1}{l|}{{\cellcolor[HTML]{EFEFEF}GALTON~\cite{wu2015goodput}}}              & \multicolumn{1}{l|}{{\cellcolor[HTML]{EFEFEF}Server and Client}}                                                                                  \\
\multicolumn{1}{|l|}{\cellcolor[HTML]{EFEFEF}}                                            & \multicolumn{1}{l|}{\multirow{-4}{*}{{  }}}              & \multicolumn{1}{l|}{{FRA-JSCC~\cite{wu2013joint}}}              & \multicolumn{1}{l|}{{Server and Client}}                                                                                  \\

\multicolumn{1}{|l|}{\cellcolor[HTML]{EFEFEF}}                                            & \multicolumn{1}{l|}{\multirow{-4}{*}{{   Application Layer Decision}}}              & \multicolumn{1}{l|}{{\cellcolor[HTML]{EFEFEF}{\color{black}Deng et al.~\cite{deng2021cross}}}}              & \multicolumn{1}{l|}{{\cellcolor[HTML]{EFEFEF}Client}}                                                                                  \\ \cline{2-4}
\multicolumn{1}{|l|}{\cellcolor[HTML]{EFEFEF}}                                            & \multicolumn{1}{l|}{}                                                                             & \multicolumn{1}{l|}{MP-DASH~\cite{han2016mp}}                                        & \multicolumn{1}{l|}{Server and Client}                                                                                                         \\
\multicolumn{1}{|l|}{\cellcolor[HTML]{EFEFEF}}                                            & \multicolumn{1}{l|}{}                                                                             & \multicolumn{1}{l|}{{\cellcolor[HTML]{EFEFEF}Nam et al.~\cite{nam2016towards}}}                   & \multicolumn{1}{l|}{{\cellcolor[HTML]{EFEFEF}\begin{tabular}[c]{@{}l@{}}Server~(depends on the application), \\ Client and Network\end{tabular}}} \\
\multicolumn{1}{|l|}{\multirow{-7}{*}{\cellcolor[HTML]{EFEFEF}\textbf{Cross Layer}}}        & \multicolumn{1}{l|}{\multirow{-3}{*}{Transport Layer Decision}}                                      & \multicolumn{1}{l|}{{CMT-CL/FD~\cite{xu2015cross}}}             & \multicolumn{1}{l|}{{Server}}  \\

\multicolumn{1}{|l|}{\multirow{-7}{*}{\cellcolor[HTML]{EFEFEF}\textbf{}}}        & \multicolumn{1}{l|}{\multirow{-3}{*}{}}                                      & \multicolumn{1}{l|}{{\cellcolor[HTML]{EFEFEF}{\color{black}OLS~\cite{xing2021low}}}}             & \multicolumn{1}{l|}{{\cellcolor[HTML]{EFEFEF}Server and Client}}  \\ \hline

\multicolumn{4}{l}{*BEMA: UDP~(for video data transmission) and TCP~(for connection establishment and feedback information).}
\end{tabular}}
\end{table*}

\subsubsection{\textbf{RTP}}{\label{sec: RTP Based Protocols}} {\color{black}The Real-time Transport Protocol~(RTP)~\nomenclature{RTP}{Real-time Transport Protocol} is an application layer transport protocol to support live, on-demand, and interactive multimedia applications. Next, we highlight more properties of the protocol, and then we survey multipath works based on RTP.\\

\noindent \textbf{RTP properties.}~Although RTP is designed to run over UDP, it could also carry data over other transport protocols, such as TCP or SCTP.} Another property of RTP is that it can be used in conjunction with the RTP Control Protocol~(RTCP) to send monitored information and QoS parameters periodically.  RTP also can be used in conjunction with other protocols, such as Real-time Streaming Protocol~(RTSP)~\nomenclature{RTSP}{Real-time Streaming Protocol}~\cite{schulzrinne1998real}, which is used to control multimedia playback.
A big problem of RTP, running over UDP, is that it lacks congestion control and it is unfair to give room to other flows. There is also no guarantee of reliable delivery and it needs a method to protect high priority frames~(I-frames). Furthermore, a challenge to improve RTP to support multipath streaming is that RTP establishes at the media session level and receiver reports per media~(video or audio) flow~\cite{singh2013mprtp}. \\

\noindent \textbf{{\color{black}Multipath support.}}~Multiflow Real-Time Transport Protocol
(MRTP)~\cite{mao2006mrtp}, Multipath RTP~(MPRTP)~\cite{singh2013mprtp} and Multipath Real-Time Transport Protocol Based on Application-Level Relay~(MPRTP-AR){\color{black}~\cite{leiwm-avtcore-mprtp-ar-09}} improved RTP to support multipath video streaming. 

The works  MRTP and MPRTP are  Constant  Bit  Rate~(CBR) approaches. Since RTP  lacks  congestion  control,
a  considerable  receiver  buffer  is  required  to  compensate {\color{black} for}
the  different  path  latencies  of  RTP  streams  when  playing
a  CBR video~\cite{singh2015survey}. Both MRTP and MPRTP use QoS reports, similar to RTCP reporting in RTP, to carry periodic per flow and session statistics.

Multiflow Real-time Transport Protocol~(MRTP)~\cite{mao2006mrtp} aims to remedy the failure and congestion in mobile wireless ad hoc networks and claimed by its authors that the approach is also applicable to the Internet. In MRTP, media divides into flows,
and each flow is for one path~(in MRTP, the concept of flow is used for series of video packets {\color{black}that} are transmitted through an individual path). Then, there is a reassembly buffer at the receiver side to compensate {\color{black} for} jitter, reorder and reassemble packets by utilizing session ID, flow ID and flow sequence number.
MRTP/MRTCP~(Multi-flow Realtime Transport Control Protocol) is an extension of the RTP/RTCP. MRTP dynamically adds or removes paths based on the QoS reports. QoS reports are also used for the sender to adapt to transmission errors. For example, by adding redundancy to increase error resilience and by assigning data to more proper paths. Therefore, these reports are transferred through the best path or multiple paths to guarantee reliable delivery.

Different error control schemes, including Forward Error Correction~(FEC), Multiple Description Coding~(MDC) or Automatic Repeat reQuest~(ARQ)  could incorporate with MRTP. Finally, the results of the surveyed work show that MRTP outperforms single path RTP on received video quality.

In MRTP, it is possible to choose the data distribution method. For example, it could be just a simple Round Robin, striping~(over multiple servers), layered coding, multiple description coding or object-oriented coding~(video or audio objects encode individually).

Another RTP extension effort with multipath transmission capability for real-time media is Multipath RTP~(MPRTP) protocol~\cite{singh2013mprtp}  {\color{black}with the aim of} minimizing the latency. MRTP uses RTCP to monitor and control information~(e.g., jitter and packet loss).
As a result, paths are categorized as congested, mildly congested, and non-congested conditions based on the packet loss information.
The scheduler, which is responsible for packet distribution over different paths, assigns more media data on the non-congested path and fewer media data on congested ones. I-frames have the highest priority and are transferred over the path with the highest bandwidth, the least delay and packet losses. The sender is informed to retransmit packets by NACK and also retransmits packets on the path with the highest bandwidth, least delay and packet losses.

The approach is not integrated with congestion control but tries to keep the load balancing by using network characteristics.
The authors developed a de-jitter algorithm at the receiver side to overcome the variation of RTT and packet reordering with an adaptive playback buffer.
An MPRTP sender assigns a subflow ID to each path~(in MPRTP, the concept of subflow is used for series of video packets transmitted over a single path) and subflow-specific sequence numbers to determine subflow-related packet jitter, packet loss, and packet discards at the receiver side.
The approach is less unfair than RTP with the aim of system balancing and spreading data over paths.

\begin{figure}[!t]
\centering
\includegraphics[width=0.25\textwidth]{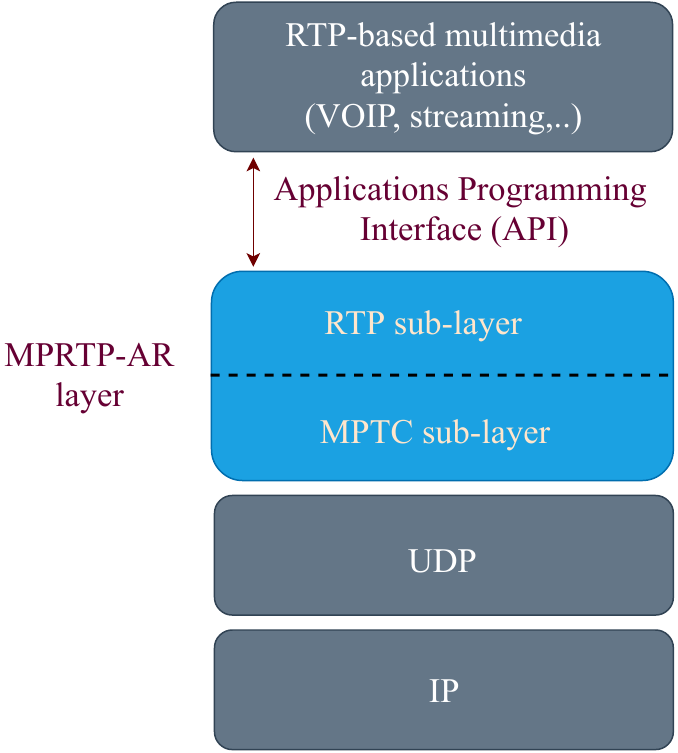}
\caption{{\color{black}MPRTP-AR protocol stack~(source: adapted from \cite{leiwm-avtcore-mprtp-ar-09}).}}
\label{fig:MPRTP-AR}
\end{figure}

{\color{black}Recently}, Multipath Real-Time Transport Protocol Based on Application-Level Relay~(MPRTP-AR)~\cite{leiwm-avtcore-mprtp-ar-09} was defined by IETF. As shown in Figure~\ref{fig:MPRTP-AR}, the proposed MPRTP-AR protocol stack has two sub-layers: RTP sub-layer and multipath
transport control~(MPTC) sub-layer. The RTP sub-layer helps this protocol to be fully compatible with existing RTP applications. Therefore, there is no need to change the Application Programming Interface~(API)\nomenclature{API}{Application Programming Interface}. The MPTC sub-layer is responsible for functions such as flow partitioning, subflow packaging and recombination, and also subflow reporting.

At the sender side, data from the application layer are formatted in RTP packets which are sent to the MPTC sub-layer. Then, MPTC  formats them into MPRTP-AR data packets. At the receiver side, MPTC extracts the fixed header of MPRTP-AR data packets and sends them to the RTP sub-layer. RTCP packets could also be generated by the RTP sub-layer for generating media transport statistics. RTCP data could be packaged in MPRTP-AR data packets which would be distributed over multiple paths by MPTC sub-layer.

In addition to MPRTP-AR data packets, MPRTP-AR control packets are defined for providing keep-alive packets and MPRTP-AR reports. MPRTP-AR reports~(MPRTP-AR Subflow Receiver Report~(SRR) and MPRTP-AR Flow Recombination Report~(FRR)) contain transport qualities of active paths~(e.g., packet loss rate and jitter) and effects on scheduling and flow partitioning. Flow partitioning methods are categorized into two groups that are named coding-aware methods and coding-unaware methods.  Coding-aware methods are used for layering coding, multiple description coding or object-oriented coding, and are on  RTP sub-layer. In this method, each coding flow is assigned to a subflow, or several coding flows are multiplexed into one subflow. Coding-unaware methods are on MPTC sub-layer, and the RTP/RTCP that are passed from upper layer would evenly spread based on the quality of the associated active paths. Flow reporting is also optionally available for the whole recombined flows.\\

\vspace{-1.0em}


\subsubsection{\textbf{DASH}}{\label{sec: DASH Based Approaches}}
Dynamic Adaptive Streaming over HTTP~(MPEG-DASH)~\cite{DASH2011Part6}~\nomenclature{DASH}{Dynamic Adaptive Streaming over HTTP} is an application layer protocol to support both VoD and live video delivery. We first detail the DASH system and its main performance limitations. Then, we explain the rate  adaptation methods. Finally, we discuss relevant works based on this protocol suite.\\

\begin{figure}[!t]
\centering
\includegraphics[width=0.5\textwidth]{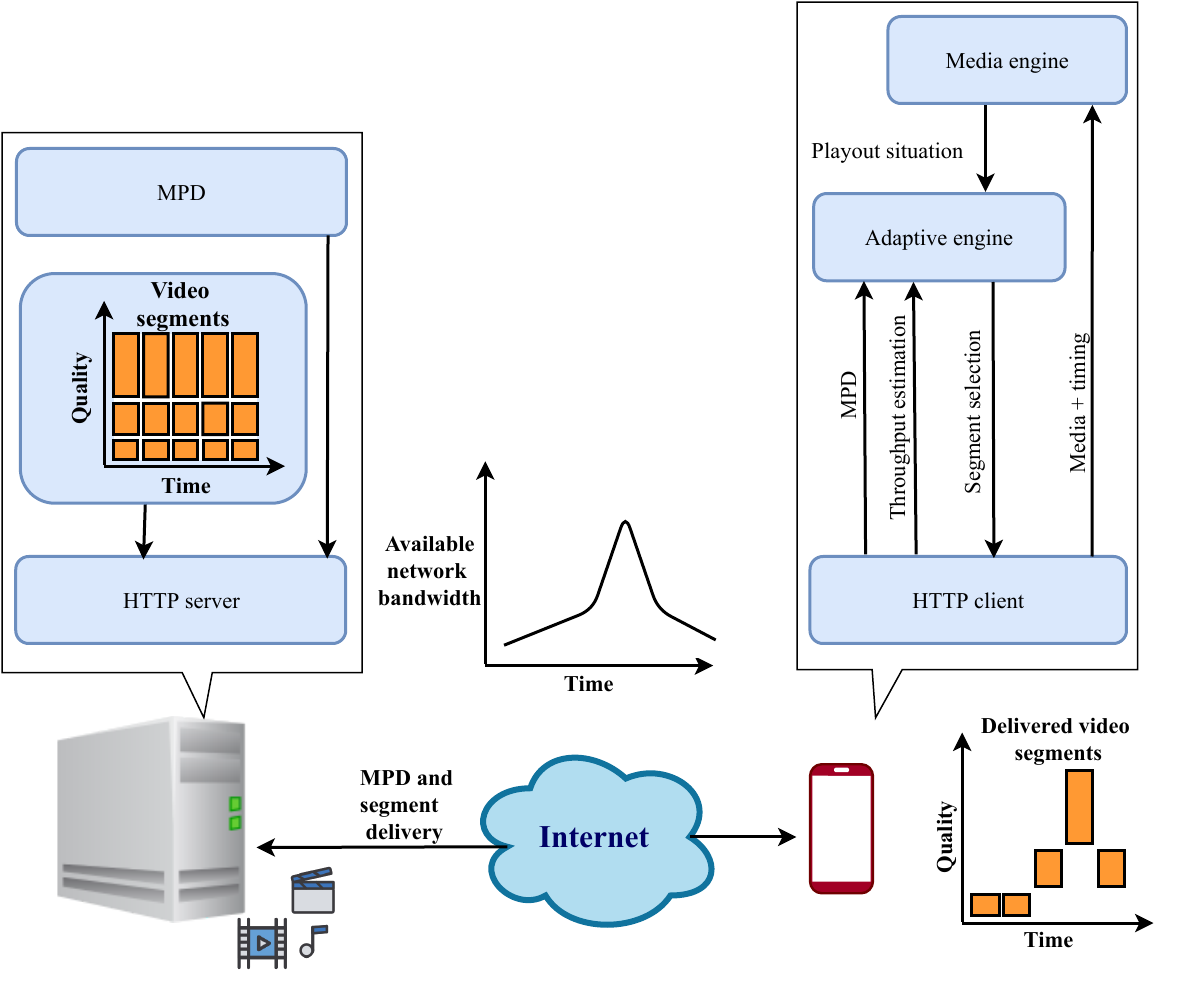}
\caption{{\color{black}DASH system~(source: adapted from~\cite{seufert2015survey}).}}
\label{fig:DASHSystem}
\end{figure}

\noindent \textbf{{\color{black}DASH system.}}~As explained in Section~\ref{sec: Historical overview of Video Streaming Protocols}, DASH has the same background technology  {\color{black}as} HTTP adaptive streaming. As shown in Figure~\ref{fig:DASHSystem},  the video sequences are stored in various resolutions~(bit rates), called  representation,
and are fragmented into small segments at  DASH server side. DASH component characteristics~(text, video, audio, etc.) are described in  {\color{black}an} XML document named Media Presentation Description~(MPD)\nomenclature{MPD}{Media Presentation Description}. {\color{black}Typically,} DASH clients are responsible for choosing the next media segments and requesting the related HTTP URL. Therefore, a rate adaptation method, named adaptation engine in Figure~\ref{fig:DASHSystem}, is required to select the proper segments' bit rate by considering the segment availability indicated by the MPD, the network conditions and the media playout situation~(e.g., playout buffer level)~\cite{seufert2015survey}.\\

 \vspace{-1.0em}

\noindent \textbf{Performance limitations.}~The rate adaptation method is responsible for key issues that influence QoE, namely, startup delay, stall, and video quality switches. Startup delay refers to the time since the client  {\color{black}requests} a video until it starts to play, namely pre-buffering. This delay occurs because, generally,  one or more segments have to be downloaded completely before the video starts to play. Noting that while VoD applications can pre-buffer few seconds of video, live  and interactive applications can only pre-buffer few hundreds of ms of video~\cite{singh2013mprtp}.

Stall or interruption refers to the pauses during the video playback due to the playback buffer is emptied, and it needs to wait to re-buffer the video~\cite{singh2012performance}. {\color{black}Generally, this issue occurs because of insufficient bandwidth.} One approach to decrease latency, and consequently solve the stall issue, is applying subsegmentation transmission and sending subsegments over more than one link simultaneously, which means adding multipath transmission capability to increase the fetching segment speed. However, the subsegmentation transmission technique also increases HTTP request overhead~\cite{xing2012rate}. In particular, the overhead problem is caused by subsegmentation transmission because at the client side after each request, an average of RTT is required to receive the response from the server. In the case of a large file with small segments/subsegments, this overhead causes a high latency. The HTTP/2 protocol provides intra-connection to support parallelism through
multiplexing streams within the same connection, mitigating the overhead problem~\cite{nikravesh2019MP-H2,xiao2016evaluating}.

Switching between different video quality representations is also a problem that impacts the video quality on the user side and causes annoying of viewers. Video switching happens because of the network bandwidth changing or buffer occupancy status. Therefore, it is important to adapt a suitable rate adaptation method, which could identify the network resources and congestion on time in order to have an optimal user experience~\cite{ojanpera2016network}.\\

\noindent \textbf{{\color{black}Rate  adaptation  methods.}}~{\color{black}
Typically, rate adaptation methods use throughput monitoring, receiver buffer status, or power level in the process of video segment bit rate decision~\cite{sani2017adaptive}. Rate adaptation methods  perform more efficiently if they can access the network information~\cite{ojanpera2016wireless}. For example, SDN  is a technology to implement such a mechanism~\cite{kleinrouweler2016delivering,cofano2017design}. Another example is Server and Network-assisted DASH~(SAND)\nomenclature{SAND}{Server and Network-assisted DASH}~\cite{Information2014Part5,thomas2016applications}, which is a system standardized recently by MPEG to collect and propagate the network information for DASH bit rate adaptation decision.
The proposed architecture in~\cite{ojanpera2016network} is built upon the Distributed Decision Engine~(DDE)\nomenclature{DDE}{Distributed Decision Engine}~\cite{luoto2015distributed} framework to provide more network information~(e.g., available capacity, load, QoS) for better rate adaptation decision in multipath scenario.}\\


\begin{table}[t]
\centering
\caption{{\color{black}MULTIPATH SUPPORT FOR DASH}}
\label{DASH Table}
\scalebox{0.83}{
\begin{tabular}{|
>{\columncolor[HTML]{FFFFFF}}l |l|l|l|}
\hline
\cellcolor[HTML]{C0C0C0}\textbf{Works} & \cellcolor[HTML]{C0C0C0}\textbf{Year} & \cellcolor[HTML]{C0C0C0}\textbf{Multipath connection} \\ \hline

\rowcolor[HTML]{EFEFEF}Xing et al.~\cite{xing2012rate}              & 2012                         & Separate TCP connections                                               \\

RTRA~\cite{xing2014real}          & 2014                         & Separate TCP connections                                             \\
\rowcolor[HTML]{EFEFEF}Houz{\'e} et al.~\cite{houze2016applicative}      & 2016                         & Separate TCP connections                                              \\
{\color{black}Go et al.~\cite{go2019hybrid}}      & 2019                         & TCP and UDP connections                                             \\ \hline \hline

\rowcolor[HTML]{EFEFEF}Chowrikoppalu et al.~\cite{chowrikoppalu2013multipath}     & 2013                         & MPTCP                                               \\

Corbillon et al.~\cite{corbillon2016cross}        & 2016                         & MPTCP                                              \\
\rowcolor[HTML]{EFEFEF}Ojanper{\"a} et al.~\cite{ojanpera2016network}       & 2016                         & MPTCP                                             \\
MP-DASH~\cite{han2016mp}          & 2016                         & MPTCP                                              \\
 \rowcolor[HTML]{EFEFEF}Nam et al.~\cite{nam2016towards}          & 2016                         & MPTCP                                               \\ \hline
\end{tabular}}
\end{table}

\noindent \textbf{{\color{black}Multipath support.}}~Current DASH version lacks multipath support, but it is being promoted as its future. {\color{black}Table~\ref{DASH Table} presents relevant efforts to provide multipath delivery for DASH through separate TCP connections~(e.g.,~\cite{ xing2012rate,xing2014real,houze2016applicative}),  MPTCP~(e.g.,~\cite{corbillon2016cross},~\cite{ojanpera2016network}, MP-DASH~\cite{han2016mp},~\cite{nam2016towards} and~\cite{chowrikoppalu2013multipath}), or both TCP and UDP~(e.g.,~\cite{go2019hybrid}). We note an increasing in combining DASH with MPTCP.}
More technical details about MPTCP will be provided in Section~\ref{sec: MPTCP Based Approaches}. At a high level, the reasons behind MPTCP growing interest include the native aggregation of bandwidth and  mobility support, and the facts that  MPTCP is friendly to middleboxes and supported in the Linux kernel, factors  contributing to the industry's attention~\cite{li2016multipath},~\cite{habib2016past}.

James et al.~\cite{james2016multipath} explored ``Whether MPTCP would always benefit mobile video streaming?".
This research analyzed the performance of different scenarios for DASH over MPTCP.  The results show that having two paths with stable bandwidths is beneficial even with small bandwidth capacity on the secondary path. Another positive impact of an additional link is when the primary path has high bandwidth variability. However, there are some harmful cases too. For example, adding an unstable secondary path could harm the stable primary path or when the bandwidth of the secondary path is not enough to transmit higher video bit rates.  Therefore, MPTCP is significantly sensitive to bandwidth fluctuation. The results also show that unnecessary multipath causes more energy consumption, resource wasting or increase cost of the quality switch.

One note regarding provide multipath delivery for DASH is about which one of the client or the server is responsible for packet scheduling decisions. {\color{black}In Table~\ref{DASH Table}, all the surveyed works that spread data over separate TCP connections, and also the proposed work~\cite{go2019hybrid}, which uses both TCP and UDP connections,} the client is responsible for choosing the proper path and fetching the suitable segments/subsegments over that path due to the fact that DASH logic is on the client side. But, integration of DASH with MPTCP is challenging when DASH logic resides on the client side, and MPTCP scheduler is on the server side. Besides, MPTCP is transparent to the application layer. Therefore, in the surveyed works of Table~\ref{DASH Table}, which MPTCP is used as transport protocol, rate adaptation logic is kept at the client side. But, scheduling decisions related to packet selection and distributing them through the paths are placed at the server side or both client and server side.
The surveyed works~\cite{corbillon2016cross},~\cite{ojanpera2016network}, MP-DASH~\cite{han2016mp} and~\cite{nam2016towards} are more related to the {\color{black}cross-layer} protocol stack. So, we will discuss them with details about scheduling strategies in Section~\ref{sec: Cross Layer Approaches}. The other works are explained in more details below.

Xing et al. used Markov Decision Process~(MDP)~\cite{bokani2013http}~\nomenclature{MDP}{Markov Decision Process} to formulate video streaming process as a reinforcement learning task in their works~\cite{xing2012rate} and~\cite{xing2014real} for non-scalable and Scalable Video Coding~(SVC), respectively.  The works' goals are decreasing startup delay, improving video quality and achieving better smoothness. In each of these works, the implemented rate adaptation
method selects the next segment based on the current queue length and estimated available bandwidth. To estimate an accurate available bandwidth, Markov channel model is used.
This way, adaptation logic finds the transit probability of each link in real-time and determines the best action~(e.g., using both links, using only WiFi link, client wait or smoothing). There is also a reward function implemented to reward each action with concern of video QoS requirements~(by measuring interruption rate, video quality, video smoothness and search time cost). However, the major problem of using MDP is the high computational cost of solving the complex optimization problem, especially in online and high mobile speed users~\cite{bokani2013http}. In addition, the approach is not a content-aware solution.


Chowrikoppalu et al.~\cite{chowrikoppalu2013multipath} modified DASH in order to utilize multipath capability. In this work, the adaptation logic is fed with a proposed bandwidth estimation algorithm and some proposed parameters, including path stability, total path stability and buffer level. The bandwidth estimation algorithm is based on sniffing packets on the interface level. Path stability and total path stability are defined to show the variation of bandwidth on each path and on MPTCP connection, respectively. However, the main problem of this approach is that it does not access the video content information.

Houz{\'e} et al.~\cite{houze2016applicative} implemented a video player utilizing multipath capability over multiple TCP connections. The goal of this scheme is achieving low-latency in DASH video delivery~(below 100 ms).
In this approach, server encodes frames of each representation and put them in the related segment every \textit{x} ms~(\textit{x} depends on the frame rate, for example, \textit{x} is 40 ms for 25 fps). The client has to fetch each whole frame before the deadline~(play time of the frame) and in \textit{x} ms before a new frame becomes available to fetch. For this target, the authors utilized video delivery over multiple paths as a way to reduce latency.
Each frame divides to byte ranges to transfer over different paths and the approach has a mechanism to find the best byte range size in order to receive them with a small inter-arrival time.
The larger byte ranges are transferred over faster paths, this way,  the variation of transfer delay decreases, and consequently, HOL blocking problem mitigates.
Besides, another adapted mechanism is proposed to select the proper representation. In this mechanism, when a segment starts, the biggest frame of each representation is considered in making the decision. The biggest frame is commonly the first frame of each representation~(I-frame). Therefore, a representation would be selected that the biggest frame has high probability of reaching the destination on time.
The problem, however, is that the approach does not consider the video content information. In addition, while the work uses RTT to estimate each path speed, it needs to improve the scheduling strategy to manage the paths.

{\color{black}Go et al.~\cite{go2019hybrid} proposed a hybrid TCP/UDP-based enhanced
HTTP adaptive streaming and a MPEG-DASH-based enhanced system for multi-homed mobile devices. 
In the proposed approach, an HTTP client firstly requests MPD over the wireless network with the strongest signal strength. Then, the client analyses the MPD and also estimates network conditions~(e.g., throughput, RTT, and PLR) during the data is downloaded from the server.
The analyzed MPD information together with the estimated network condition and buffered video time are then used to determine the types of transport protocols, requested video quality, and the amount of data that should be fetched through each network. When UDP is selected to transfer data, FEC would be applied and Raptor codes parameters~(symbol size and amount of redundant data) are determined to provide reliable data transmission. If even with using FEC, many packets are lost and it is not possible to recover the lost packets at the client, then the proposed system employs the NACK-based retransmission mechanism. Besides, the approach also adopts TCP-Friendly Rate Control~(TFRC)~\cite{floyd2000equation}\nomenclature{TFRC}{TCP-Friendly Rate Control} to guarantee fairness toward TCP flows. TFRC is an equation-based congestion control algorithm which is designed for unicast multimedia traffic. TFRC estimates the loss rate at the receiver and informs it to the sender, which adapts its transmission rate based on the congestion estimation and on the equation that models TCP congestion control behavior. TFRC responds to the congestion with less fluctuation than standard TCP congestion control and over longer periods of time~\cite{cen2003end}. However, TFRC may cause unnecessary reduction of transmission rate during wireless losses. Therefore, in this proposed approach, Spike scheme~\cite{cen2003end} is used for packet loss differentiation. Transmission rate is only calculated based on congested PLR, and the amount of FEC redundant data is calculated based on wireless PLR. To improve energy efficiency, two methods are provided; the first one is a model to estimate energy consumption for wireless network interfaces. The experimental result for this model shows that when the network condition dynamically fluctuates and it is error-prone, UDP with Raptor codes provides better energy efficiency than TCP. 
The second one is a model to estimate the energy and delay for the Raptor decoding process.}

\subsubsection{\textbf{MMT}}
MPEG Media Transport~(MMT)\nomenclature{MMT}{MPEG Media Transport}~\cite{MPEG-H2014part1} is a part of the ISO/IEC 23008 High Efficiency Coding and Media Delivery in Heterogeneous Environments~(MPEG-H) standard~\cite{lim2014new}. {\color{black}This application layer transport protocol }supports VoD and live video streaming. MMT has been widely used for Virtual Reality~(VR) and Augmented Reality~(AR) technologies, three-dimensional~(3-D) scene communication, MVV, and for major advances in televisual technology worldwide~\cite{bae2013method,aoki2017Emerging}.
We previously explained some of the properties and behaviors of MMT in Section~\ref{sec: Historical overview of Video Streaming Protocols}. Here, firstly, we indicate more properties of the protocol. Then, we explain the related technologies, and data transmission details. Finally, the surveyed works that are based on the MMT protocol are discussed. \\

\noindent \textbf{{\color{black}MMT properties.}}~MMT could be used for all unidirectional, bidirectional, unicast, multicast, multisource and, even, multipath media delivery. Besides, MMT supports both broadcast and broadband video streaming~\cite{ITU-R-MMT},{~\color{black}~\cite{I-D.bouazizi-mmtp}}.  It also provides traditional IPTV broadcasting service and all-Internet Protocol~(All-IP) networks.

Capability of hybrid media delivery is one of the most important properties of MMT. Hybrid media delivery~\cite{jung2015overview} refers to the combination of delivered media components over different types of network. For example, it could be one broadcast channel and one broadband, or it could be two broadband channels.
MMT has different hybrid service scenarios that are classified into three groups by ISO/IEC 23008-13~\cite{MPEG-H2013Part13}: live and non-live, presentation and decoding, and same/different transport schemes. The first one, live and non-live, refers to the combination of live streaming components or combination of live with pre-stored components. The second group, presentation and decoding, is the combination of the stream components for synchronized presentation or synchronized decoding. The third group, same transport schemes and different transport schemes, supports the combination of just MMT components or MMT components with other format components~(e.g., MPEG-2 TS). An instance of hybrid model comprises of MMT~(as a broadcast channel) and DASH~(as a broadband channel) over heterogeneous networks is also presented in ISO/IEC 23008-13~\cite{MPEG-H2013Part13}. \\

\noindent \textbf{{\color{black}Related technologies.}}~ISO/IEC 23008-1~\cite{MPEG-H2014part1} defined some related MMT technologies. For example, Application Layer Forward Error Correction~(AL-FEC)~\nomenclature{AL-FEC}{Application Layer Forward Error Correction} to repair data, ARQ to retransmit lost data, MMT data model and built-in hypothetical buffer model.

\begin{figure}[!t]
\centering
\includegraphics[width=0.5\textwidth]{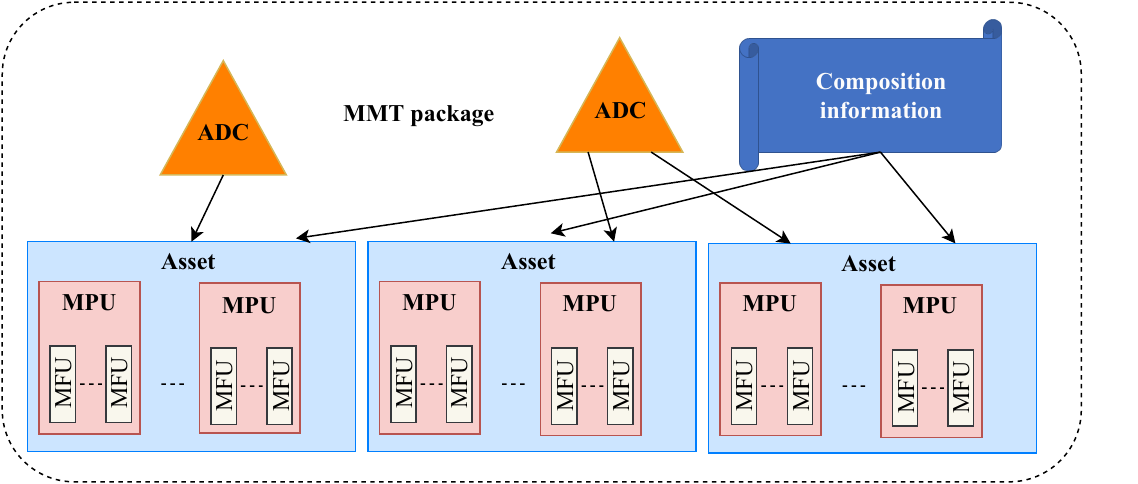}
\caption{{\color{black}Logical MMT package~(source: adapted from~\cite{lim2013mmt}).}}
\label{fig:MMTPackage}
\end{figure}

Regarding the MMT data model~\cite{MPEG-H2014part1}, MMT package is a logical entity, illustrated in Figure~\ref{fig:MMTPackage}, that comprises of one or more assets and required information for video delivery, such as Composition Information~(CI),~\nomenclature{CI}{Composition Information} Presentation Information~(PI)~\nomenclature{PI}{Presentation Information} and Asset Delivery Characteristics~(ADC).~\nomenclature{ADC}{Asset Delivery Characteristics} Asset refers to a logical data entity containing a number of Media Processing Units~(MPUs)\nomenclature{MPU}{Media Processing Unit}. Video, audio, picture, text are some examples of assets. CI provides information on temporal relationships among MPUs written in XML. HTML5 file is referred to PI, which provides initial information on spatial relationships among media elements, and ADC contains QoS information for multiplexing.\\

\vspace{-1.0em}

Built-in hypothetical buffer model~\cite{MPEG-H2014part1} aims to compensate for jitter and multipath delay delivery. In this model, the sending entity runs the {\color{black}Hypothetical Receiver Buffer Model~(HRBM)}~\nomenclature{HRBM}{Hypothetical Receiver Buffer Model} to emulate the receiving entity behavior. In this way, the sending entity determines the required buffering delay and buffer size. Then, sending entity signals this information to the receiving entity. Since at the receiver entity, several buffers exist to reconstruct of MPU from the MMT packets, the received signal is used to define operations of the buffers to ensure that at any time the buffer occupancy is within the buffer size requirement. These buffers are FEC decoding buffer to perform FEC decoding. De-jitter buffer to provide the fixed transmission delay, and MMTP de-capsulation buffer to perform MMT packet processing~(e.g., de-encapsulation, de-fragmentation/de-aggregation).\\



 \noindent {\color{black}\textbf{MMT data transmission.}~Regarding MMT data transmission, each MMTP session consists of one MMTP flow~\cite{jung2015overview}. MMTP flow is defined as all packet flows that are delivered to the same IP and port destination. {\color{black}An} MMT flow may carry multiple assets, which are identified with a unique packet\_id. MMTP packet uses two types of sequence number {\color{black}for} different purposes: packet\_counter and packet\_sequence\_number. packet\_counter represents {\color{black}a} sequence of packets in a delivery session regardless of the value of packet\_id. packet\_counter enables packet loss detection. However, packet\_sequence\_number is the sequence number specific to each packet\_id~(each asset).

Initially, MMT was designed for broadcast networks~(over UDP/IP) with reserved channel capacity. Therefore, congestion control was left to the implementation of the senders. However, MMT inherently supports receiver and sender feedback for stream thinning and bitstream switching. It also may support any Receiver-driven Layered Multicast~(RLM)\nomenclature{RLM}{Receiver-driven Layered Multicast}-based congestion control algorithms~(e.g., WEBRC, TFMCC).}\\

\noindent \textbf{{\color{black}Multipath support.}}~{\color{black}Regarding multipath delivery, Kolan et al.~\cite{kolan2016method} defined a method to establish multipath delivery over MMT, Afzal et al.~\cite{afzal2018novel, afzal2021multipath} proposed a  path-and-content-aware scheduling strategy for packet distribution, and Sohn et al.~\cite{sohn2015synchronization} proposed a synchronization scheme for hierarchical video streams over heterogeneous networks. Next, we  explain each work in more detail.}

kolan et al.~\cite{kolan2016method} defined a method to establish multipath delivery over MMT. In this method, MMT protocol utilizes signaling protocols such as RTSP or HTTP to establish and control multipath sessions between sender and receiver~(transport connection could be either TCP or UDP). For example, in RTSP, the client and the server could be aware of the multipath capability by sending OPTIONS request to each other. This new option tag, called "multipath", could be implemented in the header of the OPTIONS request. The same way, while HTTP is used to set up multipath sessions, the client includes "Multipthid" header to tell the server about its multipath capability. It is also possible to add or drop a network path during the connection. While media is delivering, MMT periodically sends feedbacks to the sender to inform about the path quality information~(e.g., loss, delay and jitter). Therefore, the sender could have a view of different paths' situations and dynamically select better performing paths for packets.

{\color{black}Afzal et al.~\cite{afzal2018novel, afzal2021multipath} proposed a novel path-and-content-aware scheduling strategy for MMT to stream real-time video over heterogeneous wireless networks. The authors claimed that their work is the first {\color{black}attempt} to improve the MMT standard by adding multipath scheduling strategies. The innovation is protected in the patents~\cite{de2021method, de2021methodINPI} and documented as standardization effort in~\cite{MPEG-H2013Part13}. The path-and-content-aware scheduling strategy, implemented at the server side, applies some methods to improve the perceived video quality based on adaptive video traffic split schemes, Markovian-based techniques, in addition to a discard and a content-aware strategy.
Adaptive video traffic split scheme allocates a proper bit rate for each transmission path considering heterogeneous network context with the aim of executing load balancing, relieve congestion, and proper utilizing of each path capacity.  The Markovian-based method estimates path conditions and transition probabilities. Discard strategy reduces congestion by avoiding sending packets that would probably be
lost. Content-aware strategy protects packets with high priority~(I frames and the closest $n$ P frames, named as near-I~(NI) frames in this work) by duplicating or assigning them to the best path. The client constantly monitors the path condition, calculates the path metrics which are sent as feedback information packets to the server through the best path. For this purpose, the feedback signaling mechanisms defined in the MMT standard are leveraged. Finally, the proposed path-and-content-aware scheduling strategy lead to QoE improvements around 12 dB for PSNR and 0.15 for SSIM by significant packet loss rate reductions~($\sim$~90\%).  It  is  important  to  note  that the  approach  does  not  require  any  change  in  the  protocol itself  since  the  scheduler  can  be  implemented  as  part  of  the client/server applications.}

Sohn et al.~\cite{sohn2015synchronization} proposed a synchronization scheme for hierarchical video streams over heterogeneous networks. This scheme {\color{black}combines} MMT~(for broadcasting) and HTTP~(for broadband) video streaming. The work utilizes scalable video streaming. Each layer is segmented in time~(in seconds), and duration value can vary according to the user's definition.
SHVC-encoded stream is used in the experiment with 3-layers: base layer~(HD), first enhanced layer~(Full HD~(FHD):~\nomenclature{FHD}{Full HD} 2K) and second enhanced layer~(UHD: 4K). Base layer and first enhanced layer of video are transferred over the broadcast network~(MMT supports multiplexing on packet level), and the second enhanced layer is transferred over broadband network. If the receiver's display has HD-resolution, it will drop the data of the first enhanced layer among data delivered over the broadcasting channel, and it does not need to have a connection with the server for the second enhanced layer, even if it can connect the networks. PI contains essential information, such as the content resolution, location of content, and MMT eXtension Document~(MXD)\nomenclature{MXD}{MMT eXtension Document}, and can also be transferred on broadcast paths. MXD is inserted in PI and mimics the MPD of DASH-SVC. MXD synchronizes the contents over heterogeneous networks, and organizes content synchronization information. The synchronization scheme is implemented at the receiver side, which requests the segments that can be delivered on time. The expected time to download each segment is computed based on bandwidth calculation and segment size information from MXD. This approach is not aware of video content and there is no scheduling strategy to use the network  paths.

\subsubsection{\textbf{QUIC}}{\label{sec: QUIC}
Quick UDP Internet Connection~(QUIC)~\cite{langley2017quic}\nomenclature{QUIC}{Quick UDP Internet Connection} is an application layer  transport protocol implemented by Google  to reduce the latency of client-server
 communications. QUIC was defined by the IETF working group in 2016~\cite{IETFQUICworkinggroup}.
This protocol has undergone rapid
development~\cite{kakhki2017rigorous} and is supported by all Google services and the Chrome browser.  {\color{black}Chromium report}\footnote{{\color{black}https://blog.chromium.org/2015/04/a-quic-update-on-googles-experimental.html}} points to video services like YouTube achieving 30\% less rebuffering when watching videos over QUIC. Next, we highlight some properties of QUIC, and then, we discuss relevant multipath support efforts.

\noindent \textbf{QUIC properties.}~One of the key characteristics of  QUIC is that it runs on top of UDP. Therefore, QUIC  implementations can be included in the application's libraries, which can be easily updated compared to TCP and other transport layer protocols that need operating system kernel modification. In addition, QUIC is an encrypted end-to-end protocol, covering payload data, and most of the protocol headers. 
QUIC embodies a combination of TCP, TLS and HTTP/2 protocols' features, including zero round trip connection establishment, multiplexed transport with reduced HOL blocking, and improved congestion control. Other features of QUIC are FEC protection and its own retransmission mechanism.
It is important to note that QUIC introduces a new special concept referred {\color{black}to} as "frame"~\cite{michel2018adding}. A QUIC packet is composed of a series of sub-packets called ``frames'' containing either
control information~(acknowledgment, flow control, crypto stuff, etc.) or application data~(STREAM frame).
Details of QUIC design and implementation are provided in~\cite{langley2017quic}.

There are a few efforts enhancing QUIC for video streaming such as~\cite{choi2017streaming} and \cite{li2016mmt}.  The approach in \cite{li2016mmt} proposes a combination of QUIC with MMT for broadband systems to improve QoE.\\

\noindent \textbf{Multipath support.}~The success of MPTCP motivated the investigation of designing  Multipath-enabled QUIC~(MPQUIC)\nomenclature{MPQUIC}{Multipath-enabled QUIC}~\cite{de2017multipath,deconinck-quic-multipath-07} in 2017. Next, we explain some properties of MPQUIC, and then, we highlight some of the main {\color{black}differences between} MPTCP and MPQUIC.

{\color{black}MPQUIC}\footnote{{\color{black}https://multipath-quic.org}} is an extension of QUIC making this protocol capable of using multiple paths over a single connection. MPQUIC introduces path identification~(Path ID) for each path, which uses its own packet number space. Path ID combined with packet number is included in the packet header. This way, MPQUIC exposes paths to the middleboxes to improve reliable data transmissions. Since the Path ID is also added to the ACK QUIC packets, ACKs can be sent over different paths.

MPQUIC has a path manager responsible for adding and removing paths~\cite{de2017multipath}. MPQUIC starts with a secure handshake over the first path and it does not require any handshake over other paths~(in contrast to MPTCP). Since MPQUIC allows both hosts to create paths, paths created by the client have an odd Path ID and paths created by the server have an even Path ID to avoid Path ID clashes.

Default MPQUIC uses the OLIA congestion control scheme~\cite{de2017multipath}. OLIA~\cite{Khalili2013MPTCP} is a window-based coupled congestion-control mechanism.  The heuristic of default MPTCP scheduling strategy is also used for MPQUIC, but with two main differences.  In MPTCP, each packet is sent on the path with the shortest RTT if its congestion window is not full. The first difference is that MPTCP decides a path for a packet or retransmitted packet whereas MPQUIC also selects the path for control frames. The retransmission strategy is also more flexible compared to MPTCP since frames are independent of packets. When a packet is lost, it is {\color{black}unnecessary} to retransmit the contained frames over the same path, as in the case of MPTCP that sends packets in sequence over each path to cope with middleboxes. The second key difference is the way to measure the RTT when a new path is added. In this case, MPQUIC duplicates the traffic through the new path to estimate its RTT. While there is some overhead involved,  this approach results in adding new paths fast without HOL.

Overall, MPQUIC can be considered simpler and cleaner than MPTCP.
Since all frames are encrypted, there is no need to specify a new mechanism to cope with middleboxes.
MPQUIC supports streams, thus, it does not require to specify a new type of sequence number in contrast to MPTCP DSN. MPQUIC is also flexible to add new types of frames to improve the protocol~\cite{michel2018adding}. However, the study in~\cite{przylucki2019simulation} comparing    MPQUIC with MPTCP for adaptive video streaming~(DASH) in a WiFi and LTE testbed environment shows that MPTCP outperforms MPQUIC in terms of available bandwidth and number of video quality switches.
Therefore, MPQUIC may require further work to meet the needs of video streaming. A detailed comparison between MPQUIC and MPTCP can be found in~\cite{de2017multipath}.

Michel et al.~\cite{michel2018adding}  explore Forward Erasure Correction for multipath data transfer and propose a FEC extension to QUIC as an unreliable service delivery named QUIC-FEC.  The proposed design supports XOR, Reed-Solomon, and Random Linear Code schemes. The authors evaluate the performance of QUIC-FEC for single path and multipath communications using different bursty and uniform packet loss conditions.  For multipath data transmission, a new scheduler named HighRB is proposed and can perform path interleaving but only using one path at a time. HighRB selects a path randomly by using the number of remaining bytes computed by the congested window as weights for the random selection. This information is gathered by the loss-based congestion control algorithm of OLIA~\cite{Khalili2013MPTCP}. The reason for using this information as weight is that lossy paths have smaller congestion windows than the other paths which means they have lower number of remaining bytes. Therefore, the probability of selecting such paths becomes lower. Moreover, it leads to the exploration of unused paths in order to check their conditions.
The experimental results show that HighRB yields higher data rates compared to the single path at the cost of increased delivery delay.}

\subsubsection{\textbf{Other Adaptive Streaming Approaches}}

Here we discuss other adaptive streaming approaches that also use HTTP to retrieve data. For example, DAAVI~\cite{johansen2009davvi} has the same core functionality {\color{black}as} DASH, making different bit rate segments on the server, providing MPD for the client, being client logic-based and transferring data over HTTP. However, the MPD structure of DAAVI is different from DASH's MPD. In our surveyed works, the proposed approaches in~\cite{evensen2010quality, evensen2011improving} and~\cite{evensen2012using} are all based on DAAVI. These DAAVI-based approaches are for on-demand and live streaming, and the authors claimed that the solutions could also be implemented in a DASH approach.

All adaptive video streaming approaches have the same challenges explained for DASH-based protocols in Section~\ref{sec: DASH Based Approaches}. One of these challenges is stalling during video playback. The works,~\cite{evensen2010quality, evensen2011improving,evensen2012using}, GreenBag~\cite{bui2013greenbag} and {\color{black}MP-H2~\cite{nikravesh2019MP-H2}} utilized multipath transmission of subsegments to decrease latency, and consequently mitigate the stall issue. As previously explained in Section~\ref{sec: DASH Based Approaches}, fetching subsegments over multiple paths can cause the overhead problem. These works used pipelining techniques~\cite{kaspar2010using} to mitigate the overhead issue.

The works~\cite{evensen2011improving, evensen2012using}, GreenBag~\cite{bui2013greenbag} and {\color{black}MP-H2~\cite{nikravesh2019MP-H2}} also proposed dynamic size subsegment methods to determine the size of each subsegment based on the throughput of each interface. As previously explained in Section~\ref{sec: DASH Based Approaches}, large sized segments increase the out-of-order packet delivery. Instead, small size segments provide smoother video, but impose higher overhead time~\cite{lederer2012dynamic}.
Another problem with using a fixed size subsegment method, instead of a dynamic one, is that a high buffer size is required to compensate for path heterogeneity, which is not desirable. This problem exists in the approach proposed in~\cite{evensen2010quality}.

A feature of GreenBag~\cite{bui2013greenbag} is that it is a middle-ware approach for video streaming over HTTP. Middle-ware approaches are designed to enable multipath interfaces to the current applications without application modifications. Therefore, middle-ware approaches are easy to deploy, but complex to  implement~\cite{habak2015bandwidth}. This middle-ware approach, GreenBag, locates between a local video player and a remote server. The client requests a video file URL normally over HTTP. GreenBag extracts the URL, determines how to download portions of the video~(segments/subsegments), and requests for portions over the decided links. RTT is used to determine when to send the requests for the next segments. Therefore, GreenBag is conventional without requiring any modification in Internet infrastructure or server side.

GreenBag is also an energy-aware bandwidth aggregation approach.
Therefore, when a single path can provide the required QoS, GreenBag stops using multipath and switches to the single path to improve energy efficiency. Besides, the approach has a medium load balancing and a recovery mechanism. Recovery occurs when a subsegment is lagging and it may pass the deadline. Therefore, the rest of the subsegment will be downloaded through both links. Finally, GreenBag leads to mitigate packet reordering problem and decrease latency.

{\color{black} MP-H2~\cite{nikravesh2019MP-H2} uses an HTTP-based multipath scheduling solution where each path is connected to a different CDN server. MP-H2 provides middlebox compatibility, anycast, and load balancing. This solution focuses on minimizing the transfer time of a medium to the large size of a Dropbox file, video chunks, mp3 song, and an image. This way, the authors implement a client-based scheduler on top of HTTP/2, which is responsible for determining when and which chunk should be fetched over which path. MP-H2 uses two main sources of information for its decision making: file size and network condition~(bandwidth and RTT). The scheduler obtains the file size length using a regular HTTP GET request. Regarding the network condition,  the moving average of bandwidth and RTT are computed. Since MP-H2 has not this information from the start of data delivery, firstly, it divides the file into two equal-sized chunks, fetching the data on both paths, LTE and WiFi.
Then, the scheduler uses the transmission of these file chunks to compute bandwidth and RTT, and therefore it calculates sufficient chunk sizes for each path and starts to move some bytes from the slow path to the fast path. Notice that HTTP/2 intra-connection parallelism through multiplexing streams within the same connection avoids any network idle periods between chunks. Experimental results of the video streaming performance over MP-H2 show almost equivalent video quality and less rebuffering ratio when compared to MPTCP using the minRTT scheduling strategy.}

Noteworthy, none of the above adaptive streaming surveyed approaches considers video content features.

\subsection{Transport Layer Approaches}{\label{sec:Transport Layer Approaches}}

Video streaming approaches focusing on transport layer protocols have direct access to the network information. Therefore, they can estimate {\color{black}end-to-end} characteristics of each path, such as capacity and congestion~\cite{raiciu2012hard}, that are useful in multipath scenarios. However, the biggest challenge of these solutions is that they generally require modifications in the standardized multipath transport protocols, which may require changes even in the kernel of operating systems.

There are several works exploiting multipath transmission in transport layer, but MPTCP and SCTP are the two main employed transport protocols with multihomed support. In this subsection, we will discuss surveyed works that are implemented based on UDP, DCCP, TCP, MPTCP and SCTP/CMT. Table~\ref{layered classification} presents each category.

\subsubsection{\textbf{UDP}}

{\color{black}
The User Datagram Protocol~(UDP)\nomenclature{UDP}{User Datagram Protocol}~\cite{postel1980rfc}, standardized by IETF in 1980, is widely used for unidirectional, broadcast, unicast, multicast, and anycast communications.  Next, we provide a brief recap of UDP basics and discuss relevant multipath efforts.

\noindent \textbf{UDP overview.}~UDP was designed to use a single path for data transmission. It is a connectionless protocol, it does not use sequence numbers for data transmission~\cite{habib2016past}, and there is no guarantee for in-ordered and reliable delivery. UDP also has no congestion control for bandwidth adaptation.
These properties make UDP a fast transmission protocol~\cite{fairhurst2017services} upon which  video streaming solutions can be easily implemented. However, the lake of bandwidth adaptation causes UDP to transmit the data with the same bit rate as sent by the application. Therefore, when the network is congested, unless the application holds back,  packets get discarded leading to video distortion and reduced QoE~\cite{hossfeld2014qoe}. Moreover, without congestion control, UDP may occupy a high fraction of the available bandwidth, and consequently, acting unfair to other congestion-avoiding network flows~\cite{huang2012qos}.  \\

\noindent \textbf{Multipath support.}~There are several efforts to add multipath transmission and bandwidth aggregation to UDP for video streaming~\cite{wu2016bandwidth,freris2013distortion,correia2012optimal}.  Note that the approaches proposed in BEMA~\cite{wu2016bandwidth} and~\cite{freris2013distortion} introduced rate balancing methods to avoid network congestion.
}
Wu et al.~\cite{wu2016bandwidth} designed a Bandwidth-Efficient Multipath streAming~(BEMA) protocol and claimed that it was the first work that employed Raptor coding and priority-aware scheduling to stream HD real-time video over heterogeneous wireless networks. This content-aware model sends packets with higher priority on the better-qualified paths and I-frame packets through all available paths. Besides, the approach utilizes {\color{black}FEC} to protect transmission data. {\color{black}BEMA also provides TFRC~\cite{floyd2000equation} in order to guarantee fairness concerning TCP flows.  However, since TFRC may cause unnecessary reduction of transmission rate during wireless losses, BEMA also adds ZigZag scheme~\cite{cen2003end} to distinguish congestion losses from wireless losses}. Only if ZigZag classifies a packet loss as a congestion loss, TFRC will consider it as a lost packet~\cite{cen2003end}. Considering the relevance of the feedback information for the proper scheduling process and its high effect on the performance, it is sent periodically from the client to the server over a reliable TCP connection.

Freris et al.~\cite{freris2013distortion} proposed a distortion-aware scalable video streaming to multiple multihomed clients. The authors claimed that their work is the first that simultaneously considered {\color{black}end-to-end} rate control and scalable stream adaptation for multipath over heterogeneous access networks. In this approach, the requested video stream is divided into substreams on the server side. The authors developed an algorithm to determine the rate of each substream and the packets to be included in each substream considering network information~(e.g., available bandwidth and RTT) and video content features in order to minimize video distortion. Besides, different cost functions are proposed to provide service differentiation and fairness among users.

The authors also developed heuristic algorithms for deterministic packet scheduling. Once it is a scalable streaming approach, each packet is transmitted only if all other related packets in lower layers have been sent before. Substreams integrate into a single scalable video stream at the client. The authors also studied the trade-off between performance and computational complexity and concluded that it works better for a small number of clients because of overhead.

Correia et al.~\cite{correia2012optimal} proposed a video streaming approach for networks with path diversity using MDC as an error resilience technique. The authors proposed a priority classification. A limited number of packets were classified as high priority because they minimize the distortion of the decoded video affected by packet loss. These packets are delivered without {\color{black}loss}. Remaining low priority packets are prone to transmission losses.

\subsubsection{\textbf{TCP}}{\label{sec: Multipath TCP Based Approaches}}

Transmission Control Protocol~(TCP)\nomenclature{TCP}{Transmission Control Protocol}~\cite{postel1981rfc} is a  transport protocol standardized by IETF in 1981. This protocol has been widely adopted for video streaming in {\color{black} Real-Time Communications~(RTC)}\nomenclature{RTC}{Real-Time Communications}\footnote{{\color{black}http://www.webrtc.org/}} and in HTTP-based applications. We previously discussed TCP lack of throughput stability~\cite{wu2016bandwidth} with its negative effect on adaptive bit streaming in Section~\ref{sec: DASH Based Approaches}. Here, we provide more details about TCP and discuss one surveyed work that is based on this protocol.\\

\textbf{{\color{black}TCP overview. }}TCP is designed to use a single path for data transmission.  
Regarding data transmission process, TCP uses sequence numbers to detect losses, guarantee in-order packet delivery, and reconstruct the received data~\cite{habib2016past}. The receiver sends ACKs for the correctly received packets. These ACKs are used to provide reliable communication. Retransmission occurs in two cases. First, when there is no ACK from the receiver, which is detected by using a retransmission timer referred to as Retransmission Time-Out~(RTO)\nomenclature{RTO}{Retransmission Time-Out}. Second, when the sender receives three duplicate ACKs, which means loss occurred. As previously also discussed in Section~\ref{sec: Introduction}, retransmission wastes bandwidth and adds significant delays.
Several protocol improvements have been proposed. For example, Selective Acknowledgements~(SACK)~\cite{floyd2000extension}\nomenclature{SACK}{Selective Acknowledgements}, where the receiver informs the sender all successfully arrived packets, so the sender retransmits only the segments that have actually been lost, and Cumulative ACK, which acknowledge the last successfully received packet to the sender. In addition, Explicit Congestion Notification~(ECN){\color{black}~\cite{black2018rfc}} has been proposed as an optional capability to collect congestion information hop by hop and inform the sender about the congestion levels.

{\color{black} Using congestion control by monitoring packet losses and/or delay variations~\cite{habib2016past}, TCP enables to adapt the data rate to network congestion and leads to minimize packet loss~\cite{hossfeld2014qoe}. In {\color{black}the} case of not enough network bandwidth available, TCP sends video data with a lower bit rate than the required video bit rate. Thus, video transmission takes longer than the video playback, and consequently may cause the playback to stall. While stall has a severe effect on the perceived video quality, in case of VoD delivery, typically,  stall is preferred over video distortions~\cite{hossfeld2014qoe}.
Previously in Section~\ref{sec: Benefits and Challenges}, we explained about HOL issues and liveness strategies used in  TCP-based applications for live or interactive video streaming to cope with stall and delay constraints requirements.
Besides all the explained properties, TCP  {\color{black}also has } the advantage of traversing through firewalls and NATs, a common issue in UDP, altogether turning TCP into a dominant  transport protocol for video services~\cite{sani2017adaptive}.}\\

\textbf{{\color{black}Multipath support. }}
 Sharma et al.~\cite{sharma2008mplot} proposed MultiPath LOss-Tolerant~(MPLOT) protocol based on SACK-based TCP and cumulative ACK. A framework, named Hybrid-ARQ~(HARQ)\nomenclature{HARQ}{Hybrid-ARQ}/FEC, is defined for MPLOT. Based on HARQ/FEC, MPLOT is using adaptive FEC proactively and reactively instead of high retransmissions to recover losses. Proactive FEC~(PFEC)~\nomenclature{PFEC}{Proactive FEC} packets are used to recover losses and when PFEC packets in a block are not enough to recover lost data, then Reactive FEC~(RFEC)~\nomenclature{RFEC}{Reactive FEC} packets need to transmit. This method leads to goodput improvement and decreased recovery latency in high lossy channels~\cite{chow2009ems}. Regarding packet scheduling, paths in MPLOT are categorized into good and bad paths. The channels with ranks higher than a threshold~(median rank) are categorized as good paths. Ranks are calculated based on network parameters, such as congestion window, PLR and RTT. MPLOT provides an uncoupled congestion control which means each path has its own congestion control. ECN is used to find congestion losses~(from faulty/lost channels) and to change the congestion window size. However, MPLOT is deployed for wireless mesh networks and it is not easily expendable on the Internet due to scalability and compatibility issues. The authors assume that a buffer is enough to compensate {\color{black} for out-of-order} delivered packets, which are important in video quality~\cite{chow2009ems,li2014tolerating}. Moreover, the approach is using a CBR coding scheme, which decreases the performance when the path quality decreases sharply~\cite{chow2009ems}.

\subsubsection{\textbf{DCCP}}{\label{sec: Multipath DCCP Based Approaches}}

Datagram Congestion Control Protocol~(DCCP)\nomenclature{DCCP}{Datagram Congestion Control Protocol}~\cite{kohler2006datagram} is a transport protocol standardized by IETF in 2006.  Here, firstly, we provide an overview of DCCP, such as data transmission process, and its  properties. Then, we discuss one surveyed work that is based on this protocol.

\noindent \textbf{{\color{black}DCCP overview.}}~DCCP provides a single path data transmission for  bidirectional and unicast data delivery.
Regarding data transmission process, DCCP uses sequence numbers. Therefore, the client can detect losses and inform them to the sender by ACKs. There is no retransmission method and in-order data delivery. In addition, there is an ability for feature negotiation before or during transmission, such as ECN capability, ACK ratio, and congestion control mechanism.

DCCP has different congestion control mechanisms that are represented by Congestion Control IDentifier~(CCID)\nomenclature{CCID}{Congestion Control IDentifier}, for example, CCID2 and CCID3. CCID2 has a TCP-like Congestion Control. Thus, the sender has a congestion window and sends data until making the window full. Both dropped packets and ECN trigger the congestion algorithm and halve the congestion window. Acknowledgments contain a list of received packets within some window, like SACK-based TCP. Therefore, CCID2~\cite{floyd2006rfcTCP-like} provides quick access to available bandwidth and deals with quick bit rate changing~\cite{huang2012qos}. CCID3~\cite{floyd2006rfcTCP-Friendly} provides TFRC. CCID3 responds to congestion smoothly and maintains steady bit rate~\cite{huang2012qos}.

A comparison among UDP, TCP and DCCP variants~(CCID2 and CCID3) for transferring MPEG4 video, shows that DCCP provides higher throughput and less packet loss compared to UDP while UDP supplies much less delay and jitter. Finally, DCCP comes up with the best QoS compared with TCP and UDP transport protocols {\color{black}over congested network}~\cite{azad2009comparative}. {\color{black}However, since subjective results in the work~\cite{hossfeld2014qoe} shows stalling caused by TCP is preferred over distortion caused by UDP  for VoD streaming,  DCCP without retransmission  may also suffer from video distortion and  may not outperform TCP and UDP  for VoD in terms of QoE. }\\

\textbf{{\color{black}Multipath support. }} In our surveyed works, Huang et al.~\cite{huang2012qos} proposed a Multipath Datagram Congestion Control Protocol~(MP-DCCP) for supplying a multipath transmission to DCCP.  In MP-DCCP, each link has its own DCCP connection, which means that each link can maintain its own congestion control window, sending rate adjustment and CCID. The proposed schedule scheme in MP-DCCP is called QoS-aware Order Prediction Scheduling~(QOPS). QOPS assigns important frames, such as I-frames into paths with less Packet Loss Rate~(PLR). Besides, QOPS predicts the order of packets at the receiver side by estimating the path latency to deal with the out-of-order problem.
Based on the final results, among the {\color{black}{congestion}} control algorithms defined in DCCP standard, conjunction of CCID3 to MP-DCCP is recommended due to its steady transmission.

\subsubsection{\textbf{MPTCP}}{\label{sec: MPTCP Based Approaches}}
{\color{black}
Multipath TCP~(MPTCP)~\cite{ford2009mptcp,Ford2011Architectural}\nomenclature{MPTCP}{Multipath TCP} is a prominent protocol for multipath transmission developed at IETF since 2009. MPTCP has been implemented in the {\color{black}Linux kernel}\footnote{{\color{black}http://www.multipath-tcp.org}}, and also as an experimental kernel patch for {\color{black}FreeBSD-10.x}\footnote{{\color{black}http://caia.swin.edu.au/urp/newtcp/mptcp/}}. Industry has also adopted MPTCP on smartphones~\cite{bonaventure2016multipath} like apple and {\color{black}Android}\footnote{{\color{black}https://multipath-tcp.org/pmwiki.php/Users/Android}}. In the following, we first  provide an overview of MPTCP. Then, we discuss performance problems. Finally, we  survey relevant works based on this protocol.}\\

\begin{figure}[!t]

\centering

\includegraphics[width=0.2\textwidth]{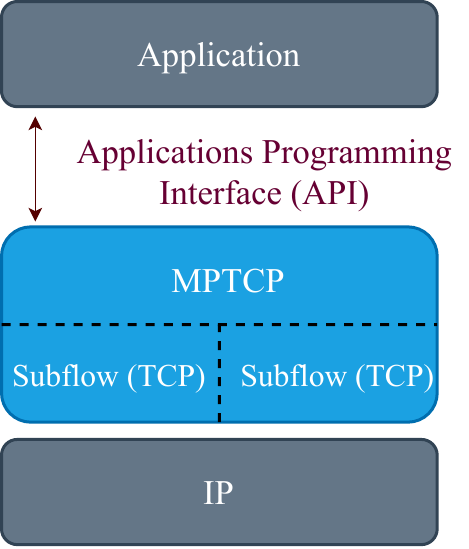}

\caption{{\color{black}MPTCP protocol stack~(source: adapted from~\cite{ford2009mptcp}).}}

\label{fig: MPTCP-protocolStack}

\end{figure}

\noindent \textbf{{\color{black}MPTCP overview.}}~MPTCP was designed to use multiple paths for data transmission. In particular, MPTCP establishes multiple subflows for a single MPTCP session. A subflow is a TCP flow over an individual path and looks similar to a regular TCP connection. 
Besides, there is a MP\_CAPABLE option to identify that the connection is MPTCP rather than TCP. Further, a token is associated to the MPTCP session. This token is used for subflows to add to this particular session. In MPTCP, application layer sees MPTCP connections as unique, as shown in Figure~\ref{fig: MPTCP-protocolStack}. Therefore, sender's transport layer packetizes data to TCP packets and receiver's transport layer reorders and recreates the byte stream without application layer knowing about it. As a result, application layer stays unmodified and a standard socket API is used.\\

\vspace{-1.0em}

Regarding data transmission process, each packet contains two sequence numbers: the Subflow Sequence Number~(SSN) to loss detection~\nomenclature{SSN}{Subflow Sequence Number} and an additional Data Sequence Number~(DSN)~\nomenclature{DSN}{Data Sequence Number} to reconstruct the original data at the receiver.
MPTCP also utilizes ACKs for subflow and connection level. SACK/Cumulative ACKs are used at subflow level and DSN-ACKs are used at connection level~\cite{habib2016past}.
For data transmission protection, MPTCP uses retransmission mechanism as in regular TCP. Besides, in the case of packet loss over a subflow, retransmission could be over another subflow.

{\color{black}

Default MPTCP uses coupled congestion control~(each MPTCP connection has its own congestion control) to avoid an unfair TCP connection. This algorithm provides better congestion balancing than just using TCP congestion control over each subflow~(uncoupled)~\cite{wischik2011design,raiciu2011rfc} because MPTCP over regular TCP connections could behave unfairly.


}

A shared MPTCP receiving buffer is used at {\color{black} the} receiver side to receive and reorder packets of different paths~\cite{corbillon2016cross}. {\color{black}In other words,} there is a single window shared by all subflows at the receiver side.

Because in multipath approaches, packet scheduling strategy has an important role, there are different strategies introduced for MPTCP. Performance comparison of scheduling methods for multipath transfer is analyzed in~\cite{singh2012performance} and different schedulers are implemented and evaluated in~\cite{paasch2014experimental} for MPTCP. Default MPTCP packet scheduling strategy selects the packets in First-In First-Out~(FIFO) order and maps them to the different paths according to RTT-based policy.

MPTCP supports middleboxes and is compatible with the current network infrastructure~\cite{habib2016past}. This is due to this fact that SSN contains a consecutive sequence number for each subflow packet. Therefore, it can pass through middleboxes~\cite{barre2011multipath}. However, in case of conflict, MPTCP handles middleboxes by fallback to the regular TCP~\cite{hesmans2013tcp}. Moreover, MPTCP provides resilience, mobility and load balancing~\cite{fairhurst2017services}.

\noindent \textbf{{\color{black}Performance challenges.}}~Studies in~\cite{deng2014wifi} and~\cite{singh2012performance} show that MPTCP presents performance issues most critically in the case of heterogeneous paths. The reasons of MPTCP performance limitations are discussed below:

\begin{itemize}

\item \textit{Out-of-order {\color{black}packets}:}~MPTCP suffers from out-of-order packet problem. A comparison between Single Path TCP~(SPTCP)\nomenclature{SPTCP}{Single Path TCP} and MPTCP in~\cite{nam2016towards} shows that SPTCP outperforms MPTCP when paths are heavily imbalanced in terms of throughput. MPTCP operates poorly in this case due to a large number of out-of-order delivered packets. Such imbalance throughput could also happen frequently in the case of using 5G network simultaneously with other wireless networks. In our surveyed works, the approach proposed in~\cite{nam2016towards} introduced a dynamic MPTCP path control to remedy out-of-order problem.

\item \textit{HOL blocking due to ARQ mechanism:}~Using ARQ mechanism by MPTCP causes frequently HOL blocking problem, even more than a single TCP connection~\cite{corbillon2016cross}. As previously explained in Section~\ref{sec: Introduction}, HOL incurs large end-to-end delay and low performance.
In our surveyed works, the proposed approaches in ADMIT~\cite{wu2016streaming}, {\color{black}DEAM~\cite{Wu2019Energy},}~\cite{diop2012qos},~\cite{zhang2015multipath} and~\cite{cao2016pr} attempted to solve the retransmission problem to decrease end-to-end delay.

 \item \textit{Frequent throughput fluctuation and  unnecessary fast retransmission:}~MPTCP uses Additive-Increase/Multiplicative-Decrease~(AIMD)~\nomenclature{AIMD}{Additive-Increase/Multiplicative-Decrease} congestion control algorithm to set congestion window sizes. The problem is that AIMD causes frequent throughput fluctuation and significant end-to-end delay~\cite{wu2016streaming, lim2014cross}. For example,  out-of-order packet delivery, which is common in multipath transmission, and losses, which could be wireless loss and not congestion loss, could trigger unnecessary fast retransmission, which impacts undesirable reduction in the size of congestion window and waste useful bandwidth~\cite{habib2016past}. In our surveyed works, ADMIT~\cite{wu2016streaming} and {\color{black}DEAM~\cite{Wu2019Energy}} considered the packet loss differentiation to mitigate this problem.

 \item \textit{Content-agnostic traffic scheduling:}~In MPTCP, availability of multipath connections is unknown to the application. Therefore, MPTCP is unaware of application information and video content features. The approaches proposed in~\cite{corbillon2016cross} and~\cite{han2016mp} introduced {\color{black}cross-layer} solutions to access the video content and deadlines, respectively.

 \item \textit{Fully reliable and ordered service:}~MPTCP is an extension of TCP protocol with inherited fully reliable and ordered services, which are not required by video streaming. In our surveyed works, there are some efforts~\cite{diop2012qos,zhang2015multipath}, PR-MPTCP$^+$\cite{cao2016pr} applying the concept of partial reliability in MPTCP for real-time video delivery. This concept avoids retransmission for acceptable loss rates and provides partial reliable video data transmission to the upper layers~\cite{diop2012qos,zhang2015multipath,cao2016pr}.

 Partial reliability leads to improved network performance parameters~(e.g., delay, bandwidth), and consequently, better QoE~\cite{diop2012qos}.

\end{itemize}

\noindent \textbf{{\color{black}Improved scheduling mechanisms.}}~There are several {\color{black}proposals}  to improve MPTCP regarding the  {\color{black}above-mentioned} problems  through scheduling functions that define the multipath decision. Next, we briefly review them and provide more details. Cross layer works to adapt application/network layer protocols with MPTCP~(e.g.,~\cite{han2016mp},~\cite{corbillon2016cross} and~\cite{nam2016towards}) will be presented {\color{black}later} in Section~\ref{sec: Cross Layer Approaches}.


Wu J. et al. proposed quAlity-Driven MultIpath TCP~(ADMIT) protocol~\cite{wu2016streaming} for streaming high-quality mobile video with multipath TCP in heterogeneous wireless networks.
ADMIT is an extension of MPTCP with {\color{black}inheriting} basic mechanisms from it, including coupled congestion control, the same connection, subflow level {\color{black}acknowledgments}, and retransmission mechanism. The authors claimed that ADMIT is the first MPTCP scheme that incorporates the quality-driven FEC coding and rate allocation to mitigate end-to-end video streaming distortion. The proposed FEC Coding in ADMIT, adaptively chooses FEC redundancy and FEC packet sizes according to the network situations~(e.g., RTT, bandwidth and, packet loss rate) and delay constraint. This adaptive FEC coding leads to remedy the shortcomings of packet retransmission~(e.g., serious delay and performance degradation~\cite{wu2016bandwidth}) by protecting video data. Besides that, the proposed rate allocator algorithm is responsible for load balancing. ZigZag scheme~\cite{cen2003end} is also used in ADMIT. ZigZag has high effect on the FEC coding and rate allocator results due to {\color{black}distinguishing} congestion losses from wireless losses.
Finally, packet scheduling strategy maps FEC packets to the different paths according to the rate allocation vector.
However, there is no mechanism to ACK for reconstructed lost packets in FEC unit. Therefore, the ADMIT protocol keeps sending retransmissions of the lost packets until receiving the ACK. Besides, the packet scheduling strategy is not aware of the frames different priorities. Another problem is that all packets of the Group of Pictures~(GoP) and redundant packets must be received before the GoP frames are processed. Each video unit may consist of several packets and it may also depend on other units.

{\color{black}In other works, Wu J. et al. proposed multipath MPTCP solutions to   achieve the optimal tradeoff between  energy consumption and video quality over heterogeneous wireless networks~\cite{Wu2019Energy,7937943}.} Delay-Energy-quAlity-aware MPTCP~(DEAM) solution~\cite{Wu2019Energy} proposes solutions for the subflow allocation and retransmission control. The subflow allocation scheme is responsible for distributing video packets over the paths considering wireless access network properties~(e.g., bandwidth, RTT, PLR), video distortion, and energy consumption. Notice that, in multipath video streaming, a video in higher quality has less distortion but more energy consumption. Firstly, the subflow allocation scheme should adapt the data rate to ensure the imposed video quality constraint. Secondly, it should allocate subflows in the communication paths to minimize energy consumption. A discard strategy is introduced aiming to drop some selective low priority frames to reduce the transmission rate according to the video quality distortion constraint. Regarding the retransmission control, it protects video packets by retransmitting lost packets through the path with the lowest energy consumption while enforcing the deadline-constrained packet delivery. DEAM does not perform fast retransmission to avoid unnecessary packets. In addition, the ZigZag scheme~\cite{cen2003end} is used by DEAM to differentiate wireless losses from congestion losses. On the analysis of perceived video quality and reduced energy consumption, experimental results show that DEAM achieves significant improvements over the MPTCP and ADMIT default approaches.

{\color{black}Similar to the previous work, an energy distortion-aware MPTCP (EDAM)~\cite{7937943} solution has a flow rate allocation and retransmission controller for
video packet transmission and retransmission, respectively.  The proposed flow rate allocation minimizes the energy consumption under video quality constraints based on utility maximization theory and piece wise linear approximation.  The buffer control algorithm adopts a hybrid NAK (negative acknowledgement) and ACK mechanism to respond to packet losses.}


The works~\cite{diop2012qos,zhang2015multipath}, and PR-MPTCP$^+$\cite{cao2016pr} apply the concept of partial reliability in MPTCP. These works demonstrate that capability of partial reliability for MPTCP outperforms the default MPTCP for real-time video streaming. As a comparison among these works, one can note that the approach in PR-MPTCP$^+$~\cite{cao2016pr} defines that switching between MPTCP and partial reliable capability occurs dynamically based on the network situation. However, in~\cite{zhang2015multipath}, partial reliability is only activated in the initial handshake, and there is no explanation about how switching occurs in~\cite{diop2012qos}. Besides, the works in~\cite{diop2012qos} and PR-MPTCP$^+$~\cite{cao2016pr} used old versions of MPTCP. Finally, these works defined different methods for applying partial reliability, which are explained in more details below.

Diop et al.~\cite{diop2012qos} introduced QoS-ORIENTED MPTCP in order to improve QoS in terms of end-to-end delay. In this work, two QoS-aware mechanisms are implemented with the concept of “partial reliability” in MPTCP for interactive video applications. The first one, MPTCP-SD~(selective discarding), eliminates the least important packets~(B-frames) at the sender side. This could decrease the network traffic and avoid latency and loss of I and P frames. The capability of gathering priority information for MPTCP is implemented by using Implicit Packet Meta Header~(IPMH) interface~\cite{exposito2009building}.

In the second mechanism, a time-aware policy is used. In MPTCP-PR~(time constrained partial reliability), delay of each queued packet on the receiver side is calculated and whenever it gets close to a time limit~(400 ms), packets are sent to the application, and acknowledge would be sent for the missed packets. In addition, delivered packets after a specific time limit are considered as losses, but acknowledgments are sent for them to the sender. The results show that MPTCP-SD provides better video QoS than MPTCP-PR and MPTCP.

Another MPTCP Partial Reliability extension is introduced in~\cite{zhang2015multipath} to provide different required reliability level and recommended for video streaming. There is a threshold for the maximum number of retransmission attempts, or maximum delay of transmission for each packet. In this approach, the sender and receiver negotiate about partial reliability function in the initializing phase. During data transmission whenever a packet exceeds the defined threshold, the sender informs it to the receiver. Therefore, the receiver will not wait anymore to receive that packet. Consequently, the receiver will send a forced acknowledgment and sender eliminates that packet from its buffer similar to the time the packet delivered successfully. The forced acknowledgment also shows losses and congestion in the network and triggers the congestion control algorithm.

Cao et al.~\cite{cao2016pr} proposed Context-aware QoE-oriented MPTCP Partial Reliability extension~(PR-MPTCP$^+$). In this work, sender monitors network congestion and receiver buffer blocking to determine when it should enable partial reliability. To detect network congestion, a function of RTT for each path is proposed and to detect the buffer blocking, advertised receiver window~(rwnd)~\nomenclature{rwnd}{receiver window}is used. In the case of a congested network, only the packets with enough deadline to play would be sent and the packets with the highest priority could be retransmitted. In particular, the concept of context is used to refer to the video content where I-frames have the highest priority. Whenever buffer blocking is detected, a subset of paths  are adaptively selected  based on their quality~(e.g.,  bandwidth). The approach switches to the full MPTCP mode~(standard MPTCP) when there is no buffer blocking. Authors of PR-MPTCP$^+$ demonstrate that this method outperforms the proposed approach in~\cite{diop2012qos} in terms of video performance metric.

\subsubsection{\textbf{SCTP and CMT~(extension of SCTP)}}{\label{sec: SCTP and CMT (extension of SCTP) Based Approaches}}

The first SCTP specification was published in the now obsolete RFC 2960~\cite{stewart2000rfc} in 2000. The current protocol specification is in RFC 4960~\cite{rfc49602007stream} containing updates and standardized by IETF in 2007. SCTP provides multihoming, multistreaming, and there is support for SCTP by different operating systems and platforms~(e.g., FreeBSD, Linux and Android). Here, firstly, we have an overview of SCTP, such as data transmission process, and SCTP properties. Then, we indicate performance limitations. Finally, we discuss the surveyed works that are based on this protocol.

\noindent \textbf{{\color{black}SCTP overview.}}~SCTP is a message-oriented protocol like UDP and supports reliability by using congestion control and retransmission like TCP~\cite{rfc49602007stream}. Default SCTP uses one path as a primary path for transferring data packets, and other paths are used for redundancy transferring~(retransmission and backup packets). Redundant paths are used to have more resilience and reliable data transferring than using only a single path. In particular, SCTP sets up an association with different IP addresses for each end host~\cite{fu2004sctp}. Association, in SCTP, refers to the connection between SCTP end hosts.

{\color{black}

SCTP provides multistreaming capabilities that reduce the HOL blocking problem. 
In SCTP, each stream is a subflow within the overall data flow, where multistreaming refers to the simultaneous transmission of  several independent streams of data in an SCTP association.  SCTP multistreaming works by adding stream sequence numbers to the chunks of each stream. Sequence numbering guarantees the in-order packet delivery inside a stream while unordered delivery can happen across streams. Therefore, arrived data of a stream can be delivered to the application layer even if other streams are blocked because of losses. 
 Default SCTP also uses another sequence space called Transmission Sequence Number~(TSN)~\nomenclature{TSN}{Transmission Sequence Number} for each chunk -- the unit of information within an SCTP packet~\cite{rfc49602007stream}}. TSN is global for all streams with the goal of loss detection and reconstructing the original data at the receiver.  Besides, SACK/Cumulative TSN ACK are leveraged as acknowledgment methods. Cumulative TSN ACK is a  field of SACK to acknowledge the TSN of the last successfully received DATA chunk to the sender.

For data transmission protection, SCTP uses a retransmission mechanism upon two types {\color{black}of} events. First, whenever RTO expires. Second, after four SACK chunks have reported gaps with the same data chunk missing. Besides, SCTP uses uncoupled congestion control, and a shared buffer is used for all paths on the receiver side. \\

\noindent \textbf{{\color{black}SCTP performance limitations.}}~SCTP  presents performance limitations in heterogeneous paths and it is challenging to adopt it for video streaming:

\begin{itemize}

\item \textit{Application modification:}~SCTP requires distinct socket API and applications modifications~\cite{barre2011multipath}.

\item \textit{Lack of middleboxes support:}~SCTP suffers from lack of support in middleboxes~\cite{barre2011multipath}.

\item \textit{Frequent primary path exchange:}~SCTP is slow due to frequent primary path exchanges in case of failure. In SCTP, the process of path primary exchange takes a long time~\cite{da2016preventing} by, for example,
detecting 6 lost packets. In SCTP, a packet is recognized as lost if the sender does not receive ACK at a specific time of RTO. RTO is set to 1 second at the start and after each loss detection, it doubles. Finally, the minimum time to change the path is 63 seconds. Therefore, the process of path primary exchange takes a long time and causes a high delay.
This issue is considered in the works,~\cite{kelly2004delay,okamoto2014performance}, and SRMT~\cite{da2016preventing}.

\item \textit{Lack of load balancing support:}~Default SCTP is not load balancing over multiple paths. Load balancing is an important factor in multipath transmission. Several efforts have been done to add capability of bandwidth aggregation to SCTP, and also adapting this protocol for video streaming. This issue is considered in the surveyed works, CMT-DA~\cite{wu2015distortion}, CMT-CA~\cite{wu2016content} and CMT-QA~\cite{xu2013cmt}.


\item \textit{Unnecessary fast retransmission:}~Out-of-order packet delivery and wireless losses could trigger unnecessary fast retransmissions, decrease goodput sharply, and consequently mitigate transmission efficiency~\cite{xu2013cmt}. This issue is considered in the surveyed works,  CMT-DA~\cite{wu2015distortion}, CMT-CA~\cite{wu2016content} and CMT-QA~\cite{xu2013cmt}.

 \item \textit{ Content-agnostic traffic scheduling:}~While considering video content features in scheduling strategy could improve the QoE and network utilization, default SCTP scheduling treats in a content-agnostic fashion.  This issue is considered in the surveyed work, CMT-CA~\cite{wu2016content}.

\item \textit{Fully  reliable  and  ordered  service:}~SCTP is a fully reliable and in order protocol, which is not required by video streaming. In our surveyed works, PR-SCTP~\cite{sanson2010pr} applied the concept of partial reliability in SCTP for real-time video delivery.



\end{itemize}

\noindent \textbf{{\color{black}Improved  scheduling  mechanisms. }}~To reduce the explained problem of {\color{black}longtime} primary path exchanging in SCTP, Kelly et al.~\cite{kelly2004delay} proposed a delay-centric strategy to set the primary path based on the lowest end-to-end delay and RTT. The solution improves quality, but using this adaptive primary path selection in the lossy wireless environment makes the SCTP slow due to frequent path exchanges. This approach does not use the full ability of all paths and uses the primary path for data transmission and secondary paths as backup.

A more stable solution based on SCTP is in~\cite{okamoto2014performance}. The authors defeated with packet loss by proposing a selective bicasting method. Therefore, instead of sending the same data through two different paths~(bicasting), which would lead to significant congestion and reduce the throughput, the selective bicasting method duplicates only retransmissions.

Da Silva et al.~\cite{da2016preventing} proposed a Selective-Redundancy Multipath Transfer~(SRMT) scheme. In this approach, the primary path is used to transfer data and secondary paths are used to send redundant packets, which have more priority and stronger delay limitation. These redundancies mitigate degradation QoE. There are two key factors for packet selection over secondary paths. The first one is the amount of redundant packets to be transferred, which is calculated based on sRTT of the primary path and the maximum delay tolerated by the application. The second one is the selection of packets, which have to be sent redundantly based on the importance of packets for reconstructing the video~(a content-aware approach). For example, I-frames have the highest priority and among the I-frame packets, the initially ordered ones have more priority than others. P-frames are the next and the lowest priority is for B-frames. Duplicated packets on the receiver side would be discarded. SRMT uses the default SCTP handover scheme to avoid HOL problem.

In order to make reliable SCTP protocol flexible for video streaming, the Partially Reliable SCTP~(PR-SCTP) extension was   specified in~\cite{Stewart2004rfc3758, tuexen2015additionalrfc}. Similar to the explained concept of partial reliability for MPTCP in Section~\ref{sec: MPTCP Based Approaches}, PR-SCTP introduced some {\color{black}policies} for choosing reliability level. PR-SCTP supports choosing the retransmission policy by using either a maximum number or a time for retransmissions, and after that{\color{black},} the packet will not be retransmitted anymore. PR-SCTP shows benefits for time-sensitive applications involving video and audio streaming~\cite{wang2003performance}. In our surveyed works, the proposed approach in~\cite{sanson2010pr} utilized the partial reliability services of PR-SCTP for real-time H.264/AVC video streaming. H.264/AVC has a Network Adaptation Layer~(NAL) feature, which is a layer of abstraction over the actual encoded data. A probabilistic model is developed to find optimum values for the maximum number of retransmissions for different types of frames in order to provide a trade-off between reliability and delay. Retransmissions are over the secondary paths. The result shows that the proposed solution outperforms UDP and TCP.

Another extension solution of SCTP is Concurrent Multipath Transfer~(CMT)~\cite{iyengar2006concurrent}\nomenclature{CMT}{Concurrent Multipath Transfer}.  Most CMT solutions use all the available paths simultaneously for data transferring to increase the throughput and network resiliency {\color{black}\cite{chen2019buffer}}. There are many schemes developed based on CMT, such as CMT-DA~\cite{wu2015distortion}, CMT-CA~\cite{wu2016content} and CMT-QA~\cite{xu2013cmt}. Among these works, CMT does not use any path selection method and uses Round Robin for data distribution.
Using Round Robin for CMT not only increases out-of-order delivery, and HOL blocking at receiver, but also increases SACK overhead and additional unnecessary retransmission. CMT evolved to perform better estimation of the network situation and choosing qualified paths for data transmission in CMT-QA~\cite{xu2013cmt}, CMT-DA~\cite{wu2015distortion} and CMT-CA~\cite{wu2016content}. CMT-CA~\cite{wu2016content} is also fed with video content properties besides the network situation. These works are also different in designing of congestion control and retransmission mechanism. More details will be presented in Section~\ref{sec: Packet Loss Differentiation}.

Xu et al.~\cite{xu2013cmt}  proposed a path and quality-aware adaptive concurrent multipath transfer~(CMT-QA) approach for packet scheduling over network channels. The goal of this scheme is decreasing out-of-order problem by reducing the unnecessary fast retransmissions and reordering delay. To achieve this target, a path quality estimation model~(PQEM)\nomenclature{PQEM}{Path Quality Estimation Model}, an Optimal Retransmission Policy~(ORP)~\nomenclature{ORP}{Optimal Retransmission Policy} and Data Distribution Scheduler~(DDS)~\nomenclature{DDS}{Data Distribution Scheduler} are introduced. PQEM calculates each path quality by estimating the rate of the distributed data, which is a function of sending buffer size and transmission delay.
In PQEM, the shared sender buffer is divided into subbuffers. Each path has its own subbuffer and management independently and the allocation of buffer space size is dynamical. ORP handles packet loss differentiation and retransmits the lost packets over faster paths. DDS predicts the arrival time of data distributed over each path, and determines the amount of data to be transferred based on the congestion control parameters including cwnd, rwnd and sender buffer size. Therefore, DDS distributes data per path in the way that they arrive to the receiver in order.
SACK is used for acknowledgment method. However, the approach does not concern TCP fairness toward other traffic flows~\cite{xu2015cross} and it is not appropriate for video due to the lack of use of video content parameters.

Wu et al.~\cite{wu2015distortion} proposed a distortion-aware concurrent multipath transfer~(CMT-DA) scheme and claimed that this approach was the first work to introduce the video distortion into SCTP for enhancing HD video quality in heterogeneous wireless environments. The goal of this approach is decreasing video distortion by mitigating the effective loss rate for variable bit rate video streaming. To achieve this goal, three main methods are proposed: path status estimation and congestion control, flow rate allocation, and data retransmission control. CMT-DA estimates path situations~(e.g., RTT and available bandwidth) by processing ACK feedbacks, and applies a distortion-aware model at the flow level to schedule the packets. Aggregated feedback packets are sent after each packet delivery. The used SACK/Cumulative ACK feedback packets return to the sender through the most reliable paths to avoid losing or dropping during the network transmission. In addition, the congestion control is designed per path and defined parameters are RTT, cwnd and RTO. ECN detects path congestion and changes the congestion window size. The rate controller is proposed to choose a subset of paths dynamically and assign data transmission rates. The data retransmission control is defined to retransmit the packets which are estimated to arrive at the destination
within the deadline. However, only flow level distortion consideration without analyzing frame priority and decoding dependency of frames is not adequate for video streaming.

In another surveyed work, Wu et al.~\cite{wu2016content} proposed a content-aware CMT~(CMT-CA) scheme and claimed this approach was the first SCTP to incorporate the video content analysis into the scheduling for enhancing HD video quality in heterogeneous wireless environments. The goal of CMT-CA is to accurately estimate the video content parameters and appropriately schedule the video frames to achieve the optimal quality. To achieve this goal, three main methods are proposed: quality evaluation based decision making, congestion control, and data distribution.
Quality evaluation based decision making estimates network situation and frame level distortion. Further, these pieces of information are used for packet scheduling. Similar to what explained for CMT-DA, SACK/Cumulative ACK feedbacks are used for path situation estimation and they are sent after each packet delivery through the most reliable paths. The congestion control for CMT-CA is designed per path, Markov model-based~(MDP), and is TCP-Friendliness. Congestion control parameters are RTT, cwnd, RTO and ssthresh. ZigZag scheme~\cite{cen2003end} detects path congestion and MDP changes the congestion window size. Data distribution is responsible for packet scheduling and different transmission is applied for I and P frames. Therefore, high priority frames can be transmitted first, which helps to decrease video distortion. Besides that, the proposed algorithm drops the video frame if its parent frame cannot be delivered due to  bandwidth restriction. Therefore, this algorithm {\color{black}conserves} network resources. Besides the proposed methods, CMT-CA also utilizes similar data retransmissions methods designed in CMT-DA. For example,  SACK~\cite{floyd2000extension}, which provides a list of correctly/incorrectly received packets to the sender, and cumulative ACK, which informs the last successfully received packet to the sender.

\subsection{Network Layer Approaches }{\label{sec: Network Layer Approaches}}


Video streaming approaches focusing on the network layer have access to the IP level and to useful information in multipath scenarios, such as network, routing and data forwarding information. In addition, network layer multipath approaches take care of data spread over different interfaces without the application awareness about this process. The biggest challenge of these solutions is that they generally require network changes, new infrastructure or modifications in the kernel of operating systems. Our surveyed works are categorized into two groups based on the required network technologies: SDN/OpenFlow-based and Proxy-based approaches. These surveyed works will be discussed in this subsection. Table~\ref{layered classification} presents each category.
\subsubsection{\textbf{SDN/OpenFlow}}

Software-Defined Networking~(SDN)~\nomenclature{SDN}{Software-Defined Networking} is a network architecture based on  a logically centralized control plane~\cite{kreutz2015software} and programmatic abstractions~(e.g., OpenFlow) to define the behaviour of the forwarding devices~(e.g., routers, switches). 
 SDN controllers gather network information including capacity and packet loss rate of links in real-time and dynamically change routing paths based on the network situations and policy definitions. {\color{black}In this survey, we leave out of scope the  topic of how paths are computed. We only cover relevant works on  refactoring and modifying  the networking stack on Android and Linux devices to be able to use multiple network interfaces simultaneously in~\cite{yap2012making}, and we also discuss SDN feedback approach for path decision actions, as proposed by   MARS~\cite{sun2016mars}. }

Yap et al.~\cite{yap2012making} explored how to make use of all the available networks around us. The approach provides seamless HTTP connectivity on heterogeneous networks. In this approach, {\color{black}the application uses one IP source address to transfer data from one application over multiple interfaces.} Then, the networking stack spreads data over multiple interfaces and assigns an IP address for each one. This was implemented using a virtual Ethernet interface to connect the application, with its local IP address, to a special gateway inside the Linux kernel. This gateway combines multiple interfaces together without the application knowledge. To implement the solution, the authors re-factored the networking stack connectivity service of the Android kernel and added a controller Open vSwitch~(OVS)~\nomenclature{OVS}{Open vSwitch} in the kernel of the mobile devices. OVS has an OpenFlow interface and can utilize flow table entries. Hence, the controller and OVS helped to route and re-route the flows and packet controlling.

The goal of Multiple Access Radio Scheduling~(MARS)~\cite{sun2016mars} is solving out-of-order problem and reducing the end-to-end delay. MARS is implemented on separate TCP connections. The authors used SDN for flow aggregation and flow splitting, and also designed a scheduling scheme, named MARS, which is based on relative RTT measurement~(which will be explained in Section~\ref{sec: Which path is the best path to transfer the packet?}). The relative RTT is calculated each fixed period of time to make sure it is always valid. Accordingly, the low-latency paths are chosen for data transmission. In MARS, the controller calculates bandwidth and RTT of each path, and notifies them to the sender. {\color{black}The sender} can also inquiry such information from the controller. This information would be used in scheduler to split video blocks into several paths. These flows combine on edge router close to the client for one-interface receiver, but it can also work for the receiver with two interfaces. However, the approach {\color{black} considers neither packet loss for path quality calculation nor priority of video data units.}

\subsubsection{\textbf{Proxy solutions}}

It is possible to use proxy at one side~(client/server) or at both sides. Using proxy at one side hides multipath transmission from the other side. In the case of using proxy on both sides, each endpoint communicates with the proxy via a normal connection without awareness of the multipath communication. In proxy-based applications, a tunneling IP-in-IP mechanism~(to encapsulate one IP packet as a payload in a new IP packet) is used to redirect data to different paths over routing level.  Consequently, proxy-based approaches are transparent to both transport and application layers and do not require any changes in them~\cite{li2016multipath}.

\begin{figure}[t!]

\centering

\includegraphics[width=0.5\textwidth]{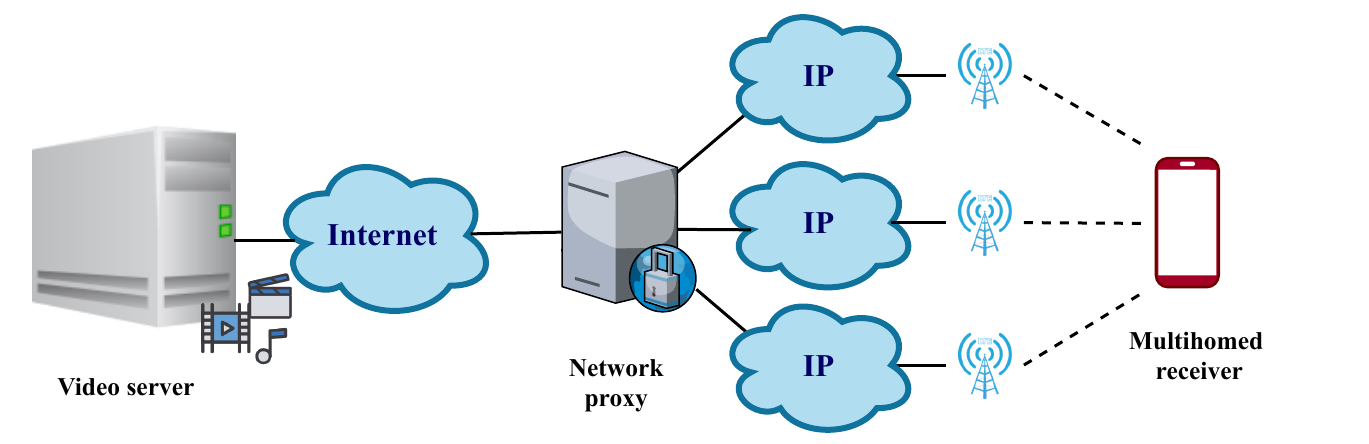}

\caption{{\color{black}BAG~\cite{chebrolu2006bandwidth} system architecture featuring the use of a proxy and IP-in-IP  tunneling between a client and the proxy~(source: adapted from~\cite{chebrolu2006bandwidth}).}}

\label{fig:BAG}

\end{figure}

Chebrolu et al. designed a network layer architecture, Bandwidth Aggregation~(BAG)~\cite{chebrolu2006bandwidth}, to utilize bandwidth aggregation for real-time applications. In BAG, server streams video data to the client by using a UDP socket. In particular, there is a proxy at the client side, which is aware of client interfaces and splits flow over these network interfaces by using IP-in-IP tunneling~(see Figure~\ref{fig:BAG}). The proposed scheduling algorithm, Earliest Delivery Path First~(EDPF)\nomenclature{EDPF}{Earliest Delivery Path First}, estimates the delivery time of each packet over each path and spreads packets over the fastest path in order to avoid packets from missing their deadlines and minimizing packet reordering. Delay and wireless bandwidth between the proxy and the client are used for delivery time estimation. As a result, EDPF is more efficient than Round Robin in avoiding HOL~\cite{li2016multipath}. The advantage of using proxy at the client side is that no change is required at the server side~\cite{li2016multipath}.


\subsection{Cross Layer Approaches}{\label{sec: Cross Layer Approaches}}

Although it is possible to estimate throughput or bandwidth and other network parameters at the application layer, they are not as accurate as the transport or network layer measurements. Different layers have different knowledge levels. For instance, the application layer is aware of video features, player buffer and deadlines. The transport layer is able to calculate the bandwidth and RTT, and it also has a congestion control mechanism. The network layer accesses IP level and routing paths, and the link layer has wireless parameter access.

Therefore, the interaction between different layers has the benefit of utilizing the advantages of different layers by signaling messages among them. This interaction is known as {\color{black}cross-layer} and was epitomized in the Transport Services~(TAPS) working group by IETF~\cite{fairhurst2017services}. Mostly, lower layers gather network information and feed them to higher layers~\cite{li2016multipath}.

In {\color{black}cross-layer} approaches, usually application layer or transport layer becomes the main layer. The main layer could decide a path for data transferring and manage load balancing or apply a method to save energy. The main layer could even change other layers behaviors. For example, application layer could change the TCP window size in order to control throughput, modifies routing tables, disconnect and reconnect the interfaces to manage failure or energy saving~\cite{li2016multipath}.

Therefore, we categorize our surveyed works into two groups: decision by application layer, and decision by transport layer, depending on which layer can be considered the main one, as discussed further in this subsection and summarized in Table~\ref{layered classification}.

\subsubsection{\textbf{Application Layer Decision}} Corbillon et al.~\cite{corbillon2016cross} proposed a {\color{black}cross-layer} approach with interaction between application and transport layer. In this approach, an adaptive mechanism is used to select the segments on application layer and MPTCP is used as transport protocol. The main goal of this approach is to maximize the amount of data that is received on time to destination. Therefore, it utilizes the benefit of being application aware to estimate playback deadline and it only sends the video units that have chance to arrive in time. As there is no {\color{black}cross-layer} feedback available in MPTCP, it is assumed that such a feedback exists and can be used. The feedback should indicate which path should be selected by MPTCP to send the next packet and only after that the {\color{black}cross-layer} scheduler would give MPTCP the data to send on this selected path~(only one packet at a time). Therefore, the scheduler, which is content-aware, can decide if and when a video unit is given to the transport layer.



Ojanpera et al.~\cite{ojanpera2016network} proposed a {\color{black}cross-layer} approach with interaction between application and network layer. The goal of this approach is to improve quality and availability of video streaming. The approach utilizes DASH to provide transparently bit rate adaptation support and MPTCP with default settings~(coupled congestion control and default scheduling strategy) to provide multipath transmission capability. As explained in Section~\ref{sec: DASH Based Approaches}, rate adaptation method available in DASH system could perform more efficiently if it could access accurate network information. Therefore, in this work, a network management system, built upon the Distributed Decision Engine~(DDE) framework, is proposed. DDE provides network information, including QoS, load, and capacity. Consequently, the client is adjusted to support DDE in order to incorporate the gathered network information into the bit rate adaptation decision in order to cope with changes in the network available bandwidth. Then, the MPTCP scheduler on the server side is responsible for mapping data on the different paths. For achieving network load balancing, the operator network management~(of DDE) can dynamically disable the access network for the client by DDE signaling. MPTCP reacts to the event by stopping the usage of the corresponding path and mapping the traffic to other available paths. Finally, the results of the work show that using more network information for client bit rate adaptation decision outperforms standalone throughput-based  by improving the stability of the video.

Wu et al.~\cite{wu2015goodput} developed a model, Goodput-Aware Load distribuTiON~(GALTON), in application-network layer. GALTON optimizes the goodput performance of video streaming over multipath networks. Goodput is an application level throughput, a key parameter for video QoS and refers to the successfully received data at the receiver within the deadline. In GALTON, the receiver monitors network status~(e.g., available bandwidth, RTT, PLR) and informs this information to the sender via feedback. The sender estimates the path quality based on the reported network information and detects congested paths by ZigZag scheme. There is also a proposed flow rate allocator which is responsible for partitioning flows to several subflows and assigning them to the available paths to optimize the aggregated goodput. It is also responsible for performing load balancing. Then, packets scheduled to the same path would be spread out within imposed deadline through the UDP connections. Besides that, scheduler {\color{black}adjusts} probe rate and probing packet sizes dynamically over the congested paths.

Wu et al.~\cite{wu2013joint} proposed a flow rate allocation-based Joint Source and Channel Coding~(FRA-JSCC) approach in an application-physical layer. Joint Source and Channel Coding~(JSCC) is an efficient solution for improving error-resilient in wireless video transmission. Therefore, in this work, JSCC is optimized to a FRA-JSCC for mobile video broadcasting in multipath networks. In FRA-JSCC approach, three main methods are proposed. First, FEC redundancy estimation to protect video data against channel losses. Second, source rate adaptation based on the calculated encoding rate. The encoding rate is concerned because high encoding rate makes more channel distortion and imposes high delay due to heavier load and network congestion. On the other hand, low encoding rate cannot provide the video delay requirements. Third, flow rate allocation dynamically selects the appropriate paths out of all available access networks and assign the transmission rates to them based on Weighted Round Robin~(WRR) scheduling strategy.

 {\color{black}
Another cross-layer optimized approach based on application and physical layers is proposed in Deng et al.~\cite{deng2021cross}. This approach presents a  framework where the client
sequentially requests video segments stored in various CDN servers via DASH technique through different networks~(LTE and 802.11ac). In this framework, the video segment bitrate at the application layer and the Modulation and Coding Scheme~(MCS)~\nomenclature{MCS}{Modulation and Coding Scheme} mode at the physical layer are jointly adapted to improve the video streaming performance. The playback buffer occupancy rate is also considered for bitrate selection and rate allocation between multiple networks.}

\subsubsection{\textbf{Transport Layer Decision}} 
\begin{figure}[!t]

\centering

\includegraphics[width=0.5\textwidth]{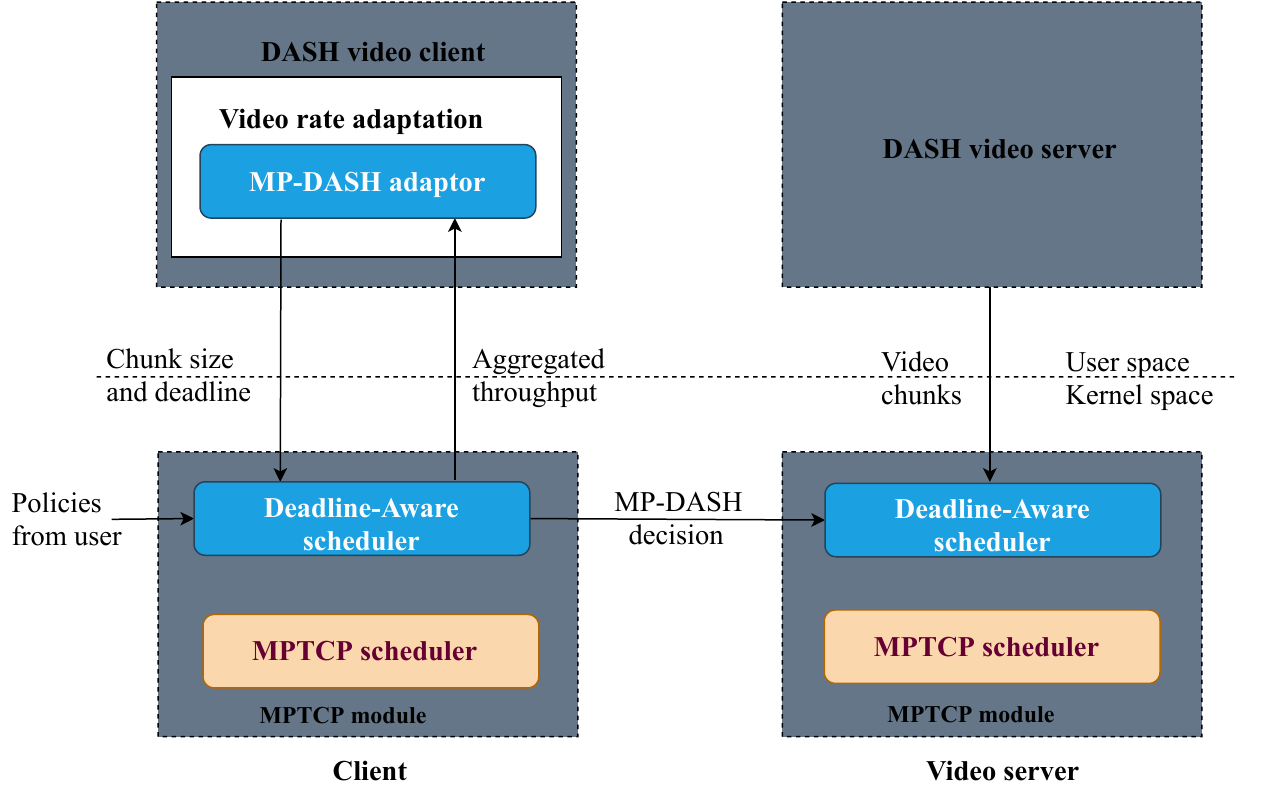}

\caption{{\color{black}MP-DASH system architecture~(source: adapted from~\cite{han2016mp}).}}

\label{fig:MP-DASH}

\end{figure}

Han et al.~\cite{han2016mp} proposed MP-DASH framework, with overall goal of enhancing MPTCP to support adaptive video streaming~(DASH) under user-specified interface preferences. For this goal, MP-DASH is designed as a {\color{black}cross-layer} approach with interaction between application and transport layer. In order to implement MP-DASH two components are designed: MP-DASH scheduler, and MP-DASH video adapter - Figure~\ref{fig:MP-DASH}.

MP-DASH scheduler is implemented with MPTCP scheduler with knowledge of network interface preferences from {\color{black} the} user and aggregated throughput. MP-DASH video adapter component, which is a lightweight add-on, is implemented to integrate the MP-DASH scheduler with DASH rate adaptation. Video adapter exchanges information between video player and MP-DASH scheduler~(segment sizes and deadlines from video player to MP-DASH scheduler, and throughput from MP-DASH scheduler to the video player). This way, DASH algorithms becomes multipath friendly and MP-DASH scheduler becomes aware of delivery deadline. Besides that, MP-DASH splits the MP-DASH scheduling functions into two parts: decision function on the client, and enforcement function on the server. Decision function determines how to manage paths based on information from video player~(e.g., segment sizes and deadlines), and enforcement function operates the decisions. The knowledge of network interface preferences is used to reduce cellular data usage while maintaining video QoE.
Therefore, the approach starts data transferring with WiFi link and checks WiFi throughput dynamically to see if it is sufficient. If WiFi cannot deliver data before deadline time, the cellular network should be enabled. The results of the work show cellular usage reduced up to 99\%, and radio energy consumption reduced up to 85\% compared with the default MPTCP.

The work in~\cite{nam2016towards} proposed a dynamic MPTCP path control using SDN~(which makes {\color{black}cross-layer} approach of transport and network layer). The goal of the approach is to cope with out-of-order delivered packets to speed up download rate and improve video QoE in ABR streaming. In this work, the authors show the feasibility of using SDN platform regarding MPTCP. The SDN controller monitors information and estimates path capacity. Then, the SDN controller communicates periodically with the SDN clients to inform which paths are the best. The SDN platform on the client side removes poor and low capacity links because poor links increase the MPTCP reordering queue size. The removed paths attach again when they return to the proper capacity. Throughput measurement is used to find the available path capacity. It also may consider other multiple factors, such as RTT and delay to compute the best paths depending on the applications~(e.g., video, VoIP or web surfing). Therefore, SDN application dynamically selects the proper paths and adjusts the number of paths in real-time.
The evaluation shows that dynamically switch between MPTCP and SPTCP increases download time. In addition, the results of DASH implementation over the proposed dynamic MPTCP path control shows less switches and rebuffering than without dynamic MPTCP path control.


Cross-layer fairness-driven SCTP-based CMT solution~(CMT-CL/FD) approach~\cite{xu2015cross} is a path quality-aware approach over CMT. In CMT-CL/FD, {\color{black}cross-layer} evaluates path quality by using loss rate information in {\color{black}Effective Signal-to-Noise Ratio~(ESNR)}~\nomenclature{ESNR}{Effective Signal-to-Noise Ratio}~(which is calculated at the link layer), and bandwidth or transmission rate information~(which are estimated at the transport layer). ESNR is an upgrade calculation for signal-to-noise ratio/noise ratio~(SNR)~\nomenclature{SNR}{signal-to-noise ratio} to evaluate wireless communication quality because the default SNR method has some shortcomings. For example, SNR is not accurate in real-time communication, and is not able to capture co-channel interference, frequency-selective fading and signal multipath effects~\cite{wu2013csi}. Then, CMT-CL/FD distributes data intelligently over different paths depending {\color{black}on} their estimated quality. A loss-cause dependent retransmission~(RTX) policy is also introduced to distinguish wireless loss from congestion loss. Consequently, in case of congested network, cwnd is changed and retransmission occurs~(as explained in Section~\ref{sec: SCTP and CMT (extension of SCTP) Based Approaches}). Finally, this proposed approach mitigates reordering, losses, and consequently decreases HOL problem.

{\color{black} Xing et al.~\cite{xing2021low} designed a packet scheduling scheme for Multipath TCP~(MPTCP), called OverLapped Scheduler~(OLS), to reduce out-of-order issues and transmission latency in asymmetric networks with unpredictable jitter. OLS schedules packets based on their arrival time with a controlled number of redundant packets. OLS is implemented with two scheduling algorithms~\textit{(i)} DElay-and-Jitter-Aware~(DEJA), which schedules packets based on their arrival time, estimated based on the size of the current send buffer, cwnd, RTT, and jitter. Additionally, DEJA also sends the same packet through the other channels with the same arrival-time interval.  Then, even though some packets may be delayed on some paths due to network jitter, other redundant packets will fill the holes, and all packets arrive in their preserved order. \textit{(ii)} Throughput Assurance~(TPA) algorithm limits the number of redundant packets and ensures a sufficient number of packets carrying video data. The idea is that throughput higher than the target is unnecessary, considering live streaming does not support content caching in advance. To do so, TPA needs cross-layer information on the target throughput from the application layer.  The results show that redundant packets effectively compensate for the impact of inaccurate arrival-time estimates.} However, none of these works use video content features for the scheduling strategy.

\section{Scheduling, Resilience, and Path Selection}{\label{sec: Scheduler, Resilience, and Routing Functions in Multipath Video}}
{\color{black}A key characteristic of video data is that packets may have unequal importance based on the en/decoding technology (e.g., I-frames vs. P-frames). Different error protection levels can be applied, considering the importance of each packet.} In addition, packets can be sent over different network paths based on paths quality to meet real-time deadlines, increase reliability, minimize out-of-order packet delivery, circumventing path heterogeneity issues~\cite{li2016multipath}, as discussed in Section~\ref{fig:Challenges causes and effects}.
Therefore, {\color{black} wireless} multi-path video scheduling strategies need to consider, at least, three  main functional aspects; packet selection, packet protection and path selection.

We now revisit the works surveyed in Section~\ref{sec: MultiPath Mobile Video Streaming Approaches} through the  new classification presented in Tables~\ref{table:Scheduler Functions-application layer},~\ref{table:Scheduler Functions-transport layer},~\ref{table:Scheduler Functions-network layer}, and~\ref{table:Scheduler Functions-cross layer} based on the following questions:

\begin{itemize}
\item Which packet should be sent next?
\item How to protect the packet?
\item Which is the best path to send the packet?
\end{itemize}

\definecolor{ForestGreen}{RGB}{34,139,34}

\begin{table*}[!htb]

\caption{{\color{black}APPLICATION LAYER APPROACHES TO MULTIPATH WIRELESS SCHEDULING FUNCTIONS}}
\label{table:Scheduler Functions-application layer}
\scalebox{0.81}{
\begin{tabular}{|l|c|c|c|c|c|c|c|c|c|c|c|} \hline

\cellcolor[HTML]{C0C0C0}                                                                     & \multicolumn{2}{c|}{\cellcolor[HTML]{C0C0C0}\textbf{Which packet?}}                                                                                                                                                                & \multicolumn{4}{c|}{\cellcolor[HTML]{C0C0C0}\textbf{How to protect the packet?}}                                                                                                                                                                        & \multicolumn{5}{c|}{\cellcolor[HTML]{C0C0C0}\textbf{Which path?}}                                                                                                                                                                                                                                                                                                                                                        \\ \cline{2-12}

\cellcolor[HTML]{C0C0C0}                                                                     & \cellcolor[HTML]{C0C0C0}                                                                                         & \cellcolor[HTML]{C0C0C0}                                                                                        & \multicolumn{2}{l|}{\cellcolor[HTML]{C0C0C0}\textbf{\begin{tabular}[c]{@{}l@{}}JSCC/\\ Channel Level\end{tabular}}} & \multicolumn{2}{l|}{\cellcolor[HTML]{C0C0C0}\textbf{\begin{tabular}[c]{@{}l@{}}Error \\ Resilience/\\ Source Level\end{tabular}}} & \cellcolor[HTML]{C0C0C0}                               & \cellcolor[HTML]{C0C0C0}                               & \cellcolor[HTML]{C0C0C0} \textbf{Bandwidth/}                                                                                          & \cellcolor[HTML]{C0C0C0}                                                                                      & \cellcolor[HTML]{C0C0C0}\textbf{Video} {\color[HTML]{656565} }                   \\ \cline{4-7}

\multirow{-3}{*}{\cellcolor[HTML]{C0C0C0}\textbf{Works}}                                     & \multirow{-2}{*}{\cellcolor[HTML]{C0C0C0}\textbf{\begin{tabular}[c]{@{}l@{}}Content\\ Awareness\end{tabular}}} & \multirow{-2}{*}{\cellcolor[HTML]{C0C0C0}\textbf{\begin{tabular}[c]{@{}l@{}} Video Distortion\\~(Frame Level)\end{tabular}}} & \cellcolor[HTML]{C0C0C0}\textbf{ARQ}                                              & \cellcolor[HTML]{C0C0C0}\textbf{FEC}                                            & \cellcolor[HTML]{C0C0C0}\textbf{Scalability}                                            &\cellcolor[HTML]{C0C0C0} \textbf{MDC}                                                    & \multirow{-2}{*}{\cellcolor[HTML]{C0C0C0}\textbf{\begin{tabular}[c]{@{}l@{}}RTT/\\ Delay\end{tabular}}} & \multirow{-2}{*}{\cellcolor[HTML]{C0C0C0}\textbf{PLR}} & \multirow{-2}{*}{\cellcolor[HTML]{C0C0C0}\textbf{\begin{tabular}[c]{@{}l@{}}Throughput/\\  Goodput\end{tabular}}} & \multirow{-2}{*}{\cellcolor[HTML]{C0C0C0}\textbf{\begin{tabular}[c]{@{}l@{}}Delay\\ Constraint\end{tabular}}} & \multirow{-2}{*}{\cellcolor[HTML]{C0C0C0}{   \textbf{\begin{tabular}[c]{@{}l@{}}Distortion \\(Flow Level) \end{tabular}}}} \\
\rowcolor[HTML]{EFEFEF} { \begin{tabular}[c]{@{}l@{}} MRTP\\ ~\cite{mao2006mrtp}\end{tabular}}         & {\color{red}\ding{55}}                                                                                                                & {\color{red}\ding{55}}                                                                                                               & \color{ForestGreen}\ding{51}                                             & \color{ForestGreen}\ding{51}                                            & \color{ForestGreen}\ding{51}                                                        & \color{ForestGreen}\ding{51}                                               & \color{ForestGreen}\ding{51}                                                      & \color{ForestGreen}\ding{51}                                                      & \color{ForestGreen}\ding{51}                                                                                                                  &\color{red}\ding{55} &\color{red}\ding{55}                                                                                                                                   \\
{\color[HTML]{333333} {}\begin{tabular}[c]{@{}l@{}}MPRTP\\ ~\cite{singh2013mprtp}\end{tabular}}     & \color{ForestGreen}\ding{51}                                                                                                                & {\color{red}\ding{55}}                                                                                                               & \color{ForestGreen}\ding{51}                                             & {\color{red}\ding{55}}                                            & {\color{red}\ding{55}}                                                        & {\color{red}\ding{55}}                                               & \color{ForestGreen}\ding{51}                                                      & \color{ForestGreen}\ding{51}                                                      & \color{ForestGreen}\ding{51}                                                                                                                  &\color{red}\ding{55}                                                                                                                                  &\color{red}\ding{55}\\

\rowcolor[HTML]{EFEFEF} {\begin{tabular}[c]{@{}l@{}}Xing et al.\\ ~\cite{xing2012rate}\end{tabular}} & {\color{red}\ding{55}}                                                                                                                & {\color{red}\ding{55}}                                                                                                               & \color{ForestGreen}\ding{51}                                             & {\color{red}\ding{55}}                                            & {\color{red}\ding{55}}                                                        & {\color{red}\ding{55}}                                               & {\color{red}\ding{55}}                                                      & {\color{red}\ding{55}}                                                      & \color{ForestGreen}\ding{51}                                                                                                                  &\color{red}\ding{55}                                                                                                                                   &\color{red}\ding{55} \\
{\color[HTML]{333333}\begin{tabular}[c]{@{}l@{}}RTRA\\~\cite{xing2014real}\end{tabular}}        & {\color{red}\ding{55}}                                                    & {\color{red}\ding{55}}                                                                                                               & \color{ForestGreen}\ding{51}                                             & {\color{red}\ding{55}}                                            & \color{ForestGreen}\ding{51}                                                        & {\color{red}\ding{55}}                                               & {\color{red}\ding{55}}                                                      & {\color{red}\ding{55}}                                                      & \color{ForestGreen}\ding{51}                                                                                                                  &\color{red}\ding{55}                                                                                                                                    &\color{red}\ding{55}\\
\rowcolor[HTML]{EFEFEF} {\begin{tabular}[c]{@{}l@{}}Houz{\'e} et al.\\~\cite{houze2016applicative}\end{tabular}}    & {\color{red}\ding{55}}                                                                                                                & {\color{red}\ding{55}}                                                                                                               & \color{ForestGreen}\ding{51}                                             & {\color{red}\ding{55}}                                            & {\color{red}\ding{55}}                                                        & {\color{red}\ding{55}}                                               & \color{ForestGreen}\ding{51}                                                      & {\color{red}\ding{55}}                                                      & {\color{red}\ding{55}}                                                                                                                  &\color{ForestGreen}\ding{51}                                                                                                                                    &\color{red}\ding{55}\\
{\begin{tabular}[c]{@{}l@{}}{\color{black}Go et al.}\\~\cite{go2019hybrid}\end{tabular}}    & {\color{red}\ding{55}}                                                                                                                & {\color{red}\ding{55}}                                                                                                               & \color{ForestGreen}\ding{51}                                             & \color{ForestGreen}\ding{51}                                            & {\color{red}\ding{55}}                                                        & {\color{red}\ding{55}}                                               & \color{ForestGreen}\ding{51}                                                      & \color{ForestGreen}\ding{51}                                                      & \color{ForestGreen}\ding{51}                                                                                                                  &\color{red}\ding{55}                                                                                                                                    &\color{red}\ding{55}\\
\rowcolor[HTML]{EFEFEF} {\begin{tabular}[c]{@{}l@{}}{\color{black}Afzal et al.}\\{\color{black}~\cite{afzal2018novel,afzal2021multipath}}\end{tabular}}& \color{ForestGreen}\ding{51}                                                                                                                & {\color{red}\ding{55}}                                                                                                               & {\color{red}\ding{55}}                                             & {\color{red}\ding{55}}                                            & {\color{red}\ding{55}}                                                        & {\color{red}\ding{55}}                                               &   \color{ForestGreen}\ding{51} &                         \color{ForestGreen}\ding{51} &                         \color{ForestGreen}\ding{51} &                         {\color{red}\ding{55}} &             \color{red}\ding{55}                                                  \\

{\color[HTML]{333333}   \begin{tabular}[c]{@{}l@{}}Sohn et al.\\~\cite{sohn2015synchronization}\end{tabular}}& {\color{red}\ding{55}}                                                                                                                & {\color{red}\ding{55}}                                                                                                               & \color{ForestGreen}\ding{51}                                             & {\color{red}\ding{55}}                                            & \color{ForestGreen}\ding{51}                                                        & {\color{red}\ding{55}}                                               & \multicolumn{5}{l|}{{\color[HTML]{333333} Paths pre-selected}}                                                                                                                                                                                                                                                                                                                                                                                   \\
\rowcolor[HTML]{EFEFEF} {\begin{tabular}[c]{@{}l@{}}{\color{black}QUIC-FEC}\\~\cite{michel2018adding}\end{tabular}}    & {\color{red}\ding{55}}                                                                                                                & {\color{red}\ding{55}}                                                                                                               & {\color{red}\ding{55}}                                             & \color{ForestGreen}\ding{51}                                            & {\color{red}\ding{55}}                                                        & {\color{red}\ding{55}}                                               & {\color{red}\ding{55}}                                                      & {\color{red}\ding{55}}                                                      & {\color{red}\ding{55}}                                                                                                                  & {  \color{red}\ding{55}} &\color{red}\ding{55}                                                                                                            \\
{\color[HTML]{333333}\begin{tabular}[c]{@{}l@{}}Evensen et al.\\~\cite{evensen2010quality}\end{tabular}} & {\color{red}\ding{55}}                                                                                                                & {\color{red}\ding{55}}                                                                                                               & \color{ForestGreen}\ding{51}                                             & {\color{red}\ding{55}}                                            & {\color{red}\ding{55}}                                                        & {\color{red}\ding{55}}                                               & \color{ForestGreen}\ding{51}                                                      & {\color{red}\ding{55}}                                                      & \color{ForestGreen}\ding{51}                                                                                                                  &\color{red}\ding{55}                                                                                                                                   &\color{red}\ding{55}\\
\rowcolor[HTML]{EFEFEF} { \begin{tabular}[c]{@{}l@{}}Evensen et al.\\~\cite{evensen2011improving}\end{tabular}} & {\color{red}\ding{55}}                                                                                                                & {\color{red}\ding{55}}                                                                                                               & \color{ForestGreen}\ding{51}                                             & {\color{red}\ding{55}}                                            & {\color{red}\ding{55}}                                                        & {\color{red}\ding{55}}                                               & \color{ForestGreen}\ding{51}                                                      & {\color{red}\ding{55}}                                                      & \color{ForestGreen}\ding{51}                                                                                                                  &\color{ForestGreen}\ding{51}                                                                                                                                    &\color{red}\ding{55}\\
{\begin{tabular}[c]{@{}l@{}}Evensen et al.\\~\cite{evensen2012using}\end{tabular}} & {\color{red}\ding{55}}                                                                                                                & {\color{red}\ding{55}}                                                                                                               & \color{ForestGreen}\ding{51}                                             & {\color{red}\ding{55}}                                            & {\color{red}\ding{55}}                                                        & {\color{red}\ding{55}}                                               & \color{ForestGreen}\ding{51}                                                      & {\color{red}\ding{55}}                                                      & \color{ForestGreen}\ding{51}                                                                                                                  &\color{ForestGreen}\ding{51}                                                                                                                                    &\color{red}\ding{55}\\
\rowcolor[HTML]{EFEFEF} {\begin{tabular}[c]{@{}l@{}}Greenbag\\ ~\cite{bui2013greenbag}\end{tabular}} & {\color{red}\ding{55}}                                                                                                                & {\color{red}\ding{55}}                                                                                                               & \color{ForestGreen}\ding{51}                                             & {\color{red}\ding{55}}                                            & {\color{red}\ding{55}}                                                        & {\color{red}\ding{55}}                                               & \color{ForestGreen}\ding{51}                                                      & {\color{red}\ding{55}}                                                      & \color{ForestGreen}\ding{51}                                                                                                                  &\color{ForestGreen}\ding{51}                                                                                                                                    &\color{red}\ding{55}\\
{\color[HTML]{333333}\begin{tabular}[c]{@{}l@{}}{\color{black}MP-H2}\\ ~\cite{nikravesh2019MP-H2}\end{tabular}} & {\color{red}\ding{55}}                                                                                                                & {\color{red}\ding{55}}                                                                                                               & \color{ForestGreen}\ding{51}                                             & {\color{red}\ding{55}}                                            & {\color{red}\ding{55}}                                                        & {\color{red}\ding{55}}                                               & \color{ForestGreen}\ding{51}                                                      & {\color{red}\ding{55}}                                                      & \color{ForestGreen}\ding{51}                                                                                                                  &\color{ForestGreen}\ding{51}                                                                                                                                    &\color{red}\ding{55}\\ \hline

\end{tabular}}
\end{table*}

\begin{table*}[!htb]

\caption{{\color{black}TRANSPORT LAYER APPROACHES TO MULTIPATH WIRELESS SCHEDULING FUNCTIONS}}
\label{table:Scheduler Functions-transport layer}
\scalebox{0.81}{
\begin{tabular}{|l|c|c|c|c|c|c|c|c|c|c|c|}
\hline

\cellcolor[HTML]{C0C0C0}                                                                     & \multicolumn{2}{c|}{\cellcolor[HTML]{C0C0C0}\textbf{Which packet?}}                                                                                                                                                                & \multicolumn{4}{c|}{\cellcolor[HTML]{C0C0C0}\textbf{How to protect the packet?}}                                                                                                                                                                        & \multicolumn{5}{c|}{\cellcolor[HTML]{C0C0C0}\textbf{Which path?}}                                                                                                                                                                                                                                                                                                                                                        \\ \cline{2-12}

\cellcolor[HTML]{C0C0C0}                                                                     & \cellcolor[HTML]{C0C0C0}                                                                                         & \cellcolor[HTML]{C0C0C0}                                                                                        & \multicolumn{2}{l|}{\cellcolor[HTML]{C0C0C0}\textbf{\begin{tabular}[c]{@{}l@{}}JSCC/\\ Channel Level\end{tabular}}} & \multicolumn{2}{l|}{\cellcolor[HTML]{C0C0C0}\textbf{\begin{tabular}[c]{@{}l@{}}Error \\ Resilience/\\ Source Level\end{tabular}}} & \cellcolor[HTML]{C0C0C0}                               & \cellcolor[HTML]{C0C0C0}                               & \cellcolor[HTML]{C0C0C0} \textbf{Bandwidth/}                                                                                          & \cellcolor[HTML]{C0C0C0}                                                                                      & \cellcolor[HTML]{C0C0C0}\textbf{Video} {\color[HTML]{656565} }                   \\ \cline{4-7}

\multirow{-3}{*}{\cellcolor[HTML]{C0C0C0}\textbf{Works}}                                     & \multirow{-2}{*}{\cellcolor[HTML]{C0C0C0}\textbf{\begin{tabular}[c]{@{}l@{}}Content\\ Awareness\end{tabular}}} & \multirow{-2}{*}{\cellcolor[HTML]{C0C0C0}\textbf{\begin{tabular}[c]{@{}l@{}} Video Distortion\\ (Frame Level)\end{tabular}}} & \cellcolor[HTML]{C0C0C0}  \textbf{ARQ}                                              & \cellcolor[HTML]{C0C0C0}  \textbf{FEC}                                            &\cellcolor[HTML]{C0C0C0}   \textbf{Scalability}                                            & \cellcolor[HTML]{C0C0C0}  \textbf{MDC}                                                    & \multirow{-2}{*}{\cellcolor[HTML]{C0C0C0}\textbf{\begin{tabular}[c]{@{}l@{}}RTT/\\ Delay\end{tabular}}} & \multirow{-2}{*}{\cellcolor[HTML]{C0C0C0}\textbf{PLR}} & \multirow{-2}{*}{\cellcolor[HTML]{C0C0C0}\textbf{\begin{tabular}[c]{@{}l@{}}Throughput/\\  Goodput\end{tabular}}} & \multirow{-2}{*}{\cellcolor[HTML]{C0C0C0}\textbf{\begin{tabular}[c]{@{}l@{}}Delay\\ Constraint\end{tabular}}} & \multirow{-2}{*}{\cellcolor[HTML]{C0C0C0}{   \textbf{\begin{tabular}[c]{@{}l@{}}Distortion \\(Flow Level) \end{tabular}}}} \\ \hline

\rowcolor[HTML]{EFEFEF}{\begin{tabular}[c]{@{}l@{}}\textbf{BEMA}\\\textbf{ ~\cite{wu2016bandwidth}}\end{tabular}}     & \color{ForestGreen}\ding{51}                                                                                                                & \color{ForestGreen}\ding{51}                                                                                                               & {\color{red}\ding{55}}                                             & \color{ForestGreen}\ding{51}                                            & {\color{red}\ding{55}}                                                        & {\color{red}\ding{55}}                                               & \color{ForestGreen}\ding{51}                                                      & \color{ForestGreen}\ding{51}                                                      & \color{ForestGreen}\ding{51}                                                                                                                  &\color{ForestGreen}\ding{51}    &\color{red}\ding{55}                                                                                                                                \\
{\color[HTML]{333333}         \begin{tabular}[c]{@{}l@{}}\textbf{Freris at al.}\\\textbf{~\cite{freris2013distortion}}\end{tabular}}& \color{ForestGreen}\ding{51}                                                                                                                & \color{ForestGreen}\ding{51}                                                                                                               & {\color{red}\ding{55}}                                             & {\color{red}\ding{55}}                                            & \color{ForestGreen}\ding{51}                                                        & {\color{red}\ding{55}}                                               & \color{ForestGreen}\ding{51}                                                      & \color{ForestGreen}\ding{51}                                                      & \color{ForestGreen}\ding{51}                                                                                                                  &\color{red}\ding{55}  &\color{red}\ding{55}                                                                                                                                  \\
\rowcolor[HTML]{EFEFEF}{\begin{tabular}[c]{@{}l@{}}\textbf{Correia at al.}\\\textbf{~\cite{correia2012optimal}}\end{tabular}} & \color{ForestGreen}\ding{51}                                                                                                                & {\color{red}\ding{55}}                                                                                                               & {\color{red}\ding{55}}                                             & {\color{red}\ding{55}}                                            & {\color{red}\ding{55}}                                                        & \color{ForestGreen}\ding{51}                                               & \multicolumn{5}{l|}{{\color[HTML]{333333} Paths pre-selected}}                                                                                                                                                                                                                                                                                                                                                                                   \\

{\color[HTML]{333333} \begin{tabular}[c]{@{}l@{}}\textbf{MPLOT }\\ \textbf{~\cite{sharma2008mplot}}\end{tabular}}   & {\color{red}\ding{55}}                                                                                                                & {\color{red}\ding{55}}                                                                                                               & \color{ForestGreen}\ding{51}                                             & \color{ForestGreen}\ding{51}                                            & {\color{red}\ding{55}}                                                        & {\color{red}\ding{55}}                                               & \color{ForestGreen}\ding{51}                                                      & \color{ForestGreen}\ding{51}                                                      & {\color{red}\ding{55}}                                                                                                                  &\color{red}\ding{55}                                                                                                                                   &\color{red}\ding{55} \\
\rowcolor[HTML]{EFEFEF}{\begin{tabular}[c]{@{}l@{}}\textbf{MP-DCCP} \\\textbf{ \cite{huang2012qos}}\end{tabular}}   & \color{ForestGreen}\ding{51}                                                                                                                & {\color{red}\ding{55}}                                                                                                               & {\color{red}\ding{55}}                                             & {\color{red}\ding{55}}                                            & {\color{red}\ding{55}}                                                        & {\color{red}\ding{55}}                                               & \color{ForestGreen}\ding{51}                                                      & \color{ForestGreen}\ding{51}                                                      & {\color{red}\ding{55}}                                                                                                                  &\color{red}\ding{55}
                    &\color{red}\ding{55}\\
{\color[HTML]{333333} \begin{tabular}[c]{@{}l@{}}\textbf{ADMIT}\\\textbf{ ~\cite{wu2016streaming}}\end{tabular}}    & {\color{red}\ding{55}}                                                                                                                & {\color{red}\ding{55}}                                                                                                               & \color{ForestGreen}\ding{51}                                             & \color{ForestGreen}\ding{51}                                            & {\color{red}\ding{55}}                                                        & {\color{red}\ding{55}}                                               & \color{ForestGreen}\ding{51}                                                      & \color{ForestGreen}\ding{51}                                                      & \color{ForestGreen}\ding{51}                                                                                                                  & {  \color{ForestGreen}\ding{51}}&\color{ForestGreen}\ding{51}                                                                                                             \\
\rowcolor[HTML]{EFEFEF}{\begin{tabular}[c]{@{}l@{}}{\color{black}\textbf{DEAM}}\\\textbf{ ~\cite{Wu2019Energy}}\end{tabular}}    & \color{ForestGreen}\ding{51}                                                                                                                & {\color[HTML]{333333} {\color{red}\ding{55}}}                                                                                                               & \color{ForestGreen}\ding{51}                                             & {\color{red}\ding{55}}                                            & {\color{red}\ding{55}}                                                        & {\color{red}\ding{55}}                                               & \color{ForestGreen}\ding{51}                                                      & \color{ForestGreen}\ding{51}                                                      & \color{ForestGreen}\ding{51}                                                                                                                  & {  \color{ForestGreen}\ding{51}}&\color{ForestGreen}\ding{51}                                                                                                             \\

{\begin{tabular}[c]{@{}l@{}}{\color{black}EDAM}\\ \textbf{ {\color{black}~\cite{7937943}}}\end{tabular}}  & \color{ForestGreen}\ding{51}                                                                                                                & {\color{red}\ding{55}}                                                                                                               & \color{ForestGreen}\ding{51}                                             & {\color{red}\ding{55}}                                            & {\color{red}\ding{55}}                                                        & {\color{red}\ding{55}}                                               & \color{ForestGreen}\ding{51}                                                      & {\color{ForestGreen}\ding{51}}                                                      & \color{ForestGreen}\ding{51}                                                                                                                  & {  \color{ForestGreen}\ding{51}}                                                                                                             &\color{red}\ding{55}\\
\rowcolor[HTML]{EFEFEF}{\begin{tabular}[c]{@{}l@{}}\textbf{MPTCP-SD} \\ \textbf{~\cite{diop2012qos}}\end{tabular}}    & \color{ForestGreen}\ding{51}                                                                                                                & {\color{red}\ding{55}}                                                                                                               & \color{ForestGreen}\ding{51}                                             & {\color{red}\ding{55}}                                            & {\color{red}\ding{55}}                                                        & {\color{red}\ding{55}}                                               &  \color{ForestGreen}\ding{51}                                                      & {\color{red}\ding{55}}                                                                                                                  & {  \color{red}\ding{55}}    &                                                   {  \color{red}\ding{55}}                                                              &\color{red}\ding{55}\\
{\color[HTML]{333333} {}\begin{tabular}[c]{@{}l@{}}\textbf{MPTCP-PR} \\ \textbf{ ~\cite{diop2012qos}}\end{tabular}}   & {\color{red}\ding{55}}                                                                                                                & {\color{red}\ding{55}}                                                                                                               & \color{ForestGreen}\ding{51}                                             & {\color{red}\ding{55}}                                            & {\color{red}\ding{55}}                                                        & {\color{red}\ding{55}}                                                   &  \color{ForestGreen}\ding{51}                                                      & {\color{red}\ding{55}}                                                                                                                  & {  \color{red}\ding{55}}    &                                                   {  \color{red}\ding{55}}                                                                             &\color{red}\ding{55}\\
\rowcolor[HTML]{EFEFEF}{\begin{tabular}[c]{@{}l@{}}\textbf{PR-MPTCP$^+$}\\ \textbf{ ~\cite{cao2016pr}}\end{tabular}}  & \color{ForestGreen}\ding{51}                                                                                                                & {\color{red}\ding{55}}                                                                                                               & \color{ForestGreen}\ding{51}                                             & {\color{red}\ding{55}}                                            & {\color{red}\ding{55}}                                                        & {\color{red}\ding{55}}                                               & \color{ForestGreen}\ding{51}                                                      & {\color{red}\ding{55}}                                                      & \color{ForestGreen}\ding{51}                                                                                                                  & {  \color{ForestGreen}\ding{51}}                                                                                                             &\color{red}\ding{55}\\

{\color[HTML]{333333} \begin{tabular}[c]{@{}l@{}}\textbf{SRMT}\\\textbf{ ~\cite{da2016preventing}}\end{tabular}}    & \color{ForestGreen}\ding{51}                                                                                                                & {\color{red}\ding{55}}                                                                                                               & \color{ForestGreen}\ding{51}                                             & {\color{red}\ding{55}}                                            & {\color{red}\ding{55}}                                                        & {\color{red}\ding{55}}                                               & \multicolumn{5}{l|}{{Paths pre-selected}}                                                                                                                                                                                                                                                                                                                                                           \\

\rowcolor[HTML]{EFEFEF}{\begin{tabular}[c]{@{}l@{}}\textbf{CMT-QA}\\ \textbf{~\cite{xu2013cmt}}\end{tabular}}         & {  \color{red}\ding{55}}                                                                                                                & {  \color{red}\ding{55}}                                                                                                               & {  \color{ForestGreen}\ding{51}}                                             & {  \color{red}\ding{55}}                                            & {  \color{red}\ding{55}}                                                        & {  \color{red}\ding{55}}                                               & {  \color{ForestGreen}\ding{51}}                                                      & {  \color{red}\ding{55}}                                                      & {  \color{red}\ding{55}}                                                                                                                  & {  \color{red}\ding{55}}                                                                                                             &\color{red}\ding{55}\\
{\color[HTML]{333333} \begin{tabular}[c]{@{}l@{}}\textbf{CMT-DA}\\\textbf{ ~\cite{wu2015distortion}}\end{tabular}}  & {\color{red}\ding{55}}                                                                                                                & {\color{red}\ding{55}}                                                                                                               & \color{ForestGreen}\ding{51}                                             & {\color{red}\ding{55}}                                            & \color{ForestGreen}\ding{51}                                                        & {\color{red}\ding{55}}                                               & \color{ForestGreen}\ding{51}                                                      & \color{ForestGreen}\ding{51}                                                      & \color{ForestGreen}\ding{51}                                                                                                                  & {  \color{red}\ding{55}*}&\ding{51}                                                                                                             \\
\rowcolor[HTML]{EFEFEF}{\begin{tabular}[c]{@{}l@{}}\textbf{CMT-CA}\\ \textbf{ ~\cite{wu2016content}}\end{tabular}}    & \color{ForestGreen}\ding{51}                                                                                                                & \color{ForestGreen}\ding{51}                                                                                                               & \color{ForestGreen}\ding{51}                                             & {\color{red}\ding{55}}                                            & {\color{red}\ding{55}}                                                        & {\color{red}\ding{55}}                                               & \color{ForestGreen}\ding{51}                                                      & \color{ForestGreen}\ding{51}                                                      & \color{ForestGreen}\ding{51}                                                                                                                  & {  \color{ForestGreen}\ding{51}} &\color{red}\ding{55}                                                                                                            \\ \hline

\end{tabular}}
\end{table*}

\begin{table*}[!htb]

\caption{{\color{black}NETWORK LAYER APPROACHES TO MULTIPATH WIRELESS SCHEDULING FUNCTIONS}}
\label{table:Scheduler Functions-network layer}
\scalebox{0.83}{
\begin{tabular}{|l|c|c|c|c|c|c|c|c|c|c|c|}
\hline

\cellcolor[HTML]{C0C0C0}                                                                     & \multicolumn{2}{c|}{\cellcolor[HTML]{C0C0C0}\textbf{Which packet?}}                                                                                                                                                                & \multicolumn{4}{c|}{\cellcolor[HTML]{C0C0C0}\textbf{How to protect the packet?}}                                                                                                                                                                        & \multicolumn{5}{c|}{\cellcolor[HTML]{C0C0C0}\textbf{Which path?}}                                                                                                                                                                                                                                                                                                                                                        \\ \cline{2-12}
\rowcolor[HTML]{C0C0C0}
\cellcolor[HTML]{C0C0C0}                                                                     & \cellcolor[HTML]{C0C0C0}                                                                                         & \cellcolor[HTML]{C0C0C0}                                                                                        & \multicolumn{2}{l|}{\cellcolor[HTML]{C0C0C0}\textbf{\begin{tabular}[c]{@{}l@{}}JSCC/\\ Channel Level\end{tabular}}} & \multicolumn{2}{l|}{\cellcolor[HTML]{C0C0C0}\textbf{\begin{tabular}[c]{@{}l@{}}Error \\ Resilience/\\ Source Level\end{tabular}}} & \cellcolor[HTML]{C0C0C0}                               & \cellcolor[HTML]{C0C0C0}                               & \cellcolor[HTML]{C0C0C0} \textbf{Bandwidth/}                                                                                          & \cellcolor[HTML]{C0C0C0}                                                                                      & \cellcolor[HTML]{C0C0C0}\textbf{Video} {\color[HTML]{656565} }                   \\ \cline{4-7}
\rowcolor[HTML]{C0C0C0}
\multirow{-3}{*}{\cellcolor[HTML]{C0C0C0}\textbf{Works}}                                     & \multirow{-2}{*}{\cellcolor[HTML]{C0C0C0}\textbf{\begin{tabular}[c]{@{}l@{}}Content\\ Awareness\end{tabular}}} & \multirow{-2}{*}{\cellcolor[HTML]{C0C0C0}\textbf{\begin{tabular}[c]{@{}l@{}} Video Distortion\\ (Frame Level)\end{tabular}}} & \textbf{ARQ}                                              & \textbf{FEC}                                            & \textbf{Scalability}                                            & \textbf{MDC}                                                    & \multirow{-2}{*}{\cellcolor[HTML]{C0C0C0}\textbf{\begin{tabular}[c]{@{}l@{}}RTT/\\ Delay\end{tabular}}} & \multirow{-2}{*}{\cellcolor[HTML]{C0C0C0}\textbf{PLR}} & \multirow{-2}{*}{\cellcolor[HTML]{C0C0C0}\textbf{\begin{tabular}[c]{@{}l@{}}Throughput/\\  Goodput\end{tabular}}} & \multirow{-2}{*}{\cellcolor[HTML]{C0C0C0}\textbf{\begin{tabular}[c]{@{}l@{}}Delay\\ Constraint\end{tabular}}} & \multirow{-2}{*}{\cellcolor[HTML]{C0C0C0}{   \textbf{\begin{tabular}[c]{@{}l@{}}Distortion \\(Flow Level) \end{tabular}}}} \\ \hline

\rowcolor[HTML]{EFEFEF}{\begin{tabular}[c]{@{}l@{}}\textbf{Yap at al.}\\\textbf{~\cite{yap2012making}}\end{tabular}} & {  \color{red}\ding{55}}                                                                                                                & {  \color{red}\ding{55}}                                                                                                               & {  \color{ForestGreen}\ding{51}}                                             & {  \color{red}\ding{55}}                                            & {  \color{red}\ding{55}}                                                        & {  \color{red}\ding{55}}                                               & \multicolumn{5}{l|}{{   Paths pre-selected}}                                                       \\
{\begin{tabular}[c]{@{}l@{}}\textbf{MARS}\\ \textbf{~\cite{sun2016mars}}\end{tabular}}         & {  \color{red}\ding{55}}                                                                                                                & {  \color{red}\ding{55}}                                                                                                               & {  \color{ForestGreen}\ding{51}}                                             & {  \color{red}\ding{55}}                                            & {  \color{red}\ding{55}}                                                        & {  \color{red}\ding{55}}                                               & {  \color{ForestGreen}\ding{51}}                                                      & {  \color{red}\ding{55}}                                                      & {  \color{ForestGreen}\ding{51}}                                                                                                                  & {  \color{red}\ding{55}}                                                                                                             &\color{red}\ding{55}\\
\rowcolor[HTML]{EFEFEF}{   \begin{tabular}[c]{@{}l@{}}\textbf{BAG}\\\textbf{~\cite{chebrolu2006bandwidth}}\end{tabular}}         & {  \color{red}\ding{55}}                                                                                                                & {  \color{red}\ding{55}}                                                                                                               & {  \color{red}\ding{55}}                                             & {  \color{red}\ding{55}}                                            & {  \color{red}\ding{55}}                                                        & {  \color{red}\ding{55}}                                               & {  \color{ForestGreen}\ding{51}}                                                      & {  \color{red}\ding{55}}                                                      & {  \color{ForestGreen}\ding{51}}                                                                                                                  & {  \color{red}\ding{55}}                                                                                                             &\color{red}\ding{55}\\ \hline
\end{tabular}}
\end{table*}

\begin{table*}

\caption{{\color{black}CROSS LAYER APPROACHES TO MULTIPATH WIRELESS SCHEDULING FUNCTIONS}}
\label{table:Scheduler Functions-cross layer}
\scalebox{0.81}{
\begin{tabular}{|l|c|c|c|c|c|c|c|c|c|c|c|}
\hline
\rowcolor[HTML]{C0C0C0}
\cellcolor[HTML]{C0C0C0}                                                                     & \multicolumn{2}{c|}{\cellcolor[HTML]{C0C0C0}\textbf{Which packet?}}                                                                                                                                                                & \multicolumn{4}{c|}{\cellcolor[HTML]{C0C0C0}\textbf{How to protect the packet?}}                                                                                                                                                                        & \multicolumn{5}{c|}{\cellcolor[HTML]{C0C0C0}\textbf{Which path?}}                                                                                                                                                                                                                                                                                                                                                        \\ \cline{2-12}
\rowcolor[HTML]{C0C0C0}
\cellcolor[HTML]{C0C0C0}                                                                     & \cellcolor[HTML]{C0C0C0}                                                                                         & \cellcolor[HTML]{C0C0C0}                                                                                        & \multicolumn{2}{l|}{\cellcolor[HTML]{C0C0C0}\textbf{\begin{tabular}[c]{@{}l@{}}JSCC/\\ Channel Level\end{tabular}}} & \multicolumn{2}{l|}{\cellcolor[HTML]{C0C0C0}\textbf{\begin{tabular}[c]{@{}l@{}}Error \\ Resilience/\\ Source Level\end{tabular}}} & \cellcolor[HTML]{C0C0C0}                               & \cellcolor[HTML]{C0C0C0}                               & \cellcolor[HTML]{C0C0C0} \textbf{Bandwidth/}                                                                                          & \cellcolor[HTML]{C0C0C0}                                                                                      & \cellcolor[HTML]{C0C0C0}\textbf{Video} {\color[HTML]{656565} }                   \\ \cline{4-7}
\rowcolor[HTML]{C0C0C0}
\multirow{-3}{*}{\cellcolor[HTML]{C0C0C0}\textbf{Works}}                                     & \multirow{-2}{*}{\cellcolor[HTML]{C0C0C0}\textbf{\begin{tabular}[c]{@{}l@{}}Content\\ Awareness\end{tabular}}} & \multirow{-2}{*}{\cellcolor[HTML]{C0C0C0}\textbf{\begin{tabular}[c]{@{}l@{}} Video Distortion\\ (Frame Level)\end{tabular}}} & \textbf{ARQ}                                              & \textbf{FEC}                                            & \textbf{Scalability}                                            & \textbf{MDC}                                                    & \multirow{-2}{*}{\cellcolor[HTML]{C0C0C0}\textbf{\begin{tabular}[c]{@{}l@{}}RTT/\\ Delay\end{tabular}}} & \multirow{-2}{*}{\cellcolor[HTML]{C0C0C0}\textbf{PLR}} & \multirow{-2}{*}{\cellcolor[HTML]{C0C0C0}\textbf{\begin{tabular}[c]{@{}l@{}}Throughput/\\  Goodput\end{tabular}}} & \multirow{-2}{*}{\cellcolor[HTML]{C0C0C0}\textbf{\begin{tabular}[c]{@{}l@{}}Delay\\ Constraint\end{tabular}}} & \multirow{-2}{*}{\cellcolor[HTML]{C0C0C0}{   \textbf{\begin{tabular}[c]{@{}l@{}}Distortion \\(Flow Level) \end{tabular}}}} \\ \hline

\rowcolor[HTML]{EFEFEF}{\begin{tabular}[c]{@{}l@{}}\textbf{Corbillon et al.}\\\textbf{~\cite{corbillon2016cross} }\end{tabular}} & {  \color{ForestGreen}\ding{51}}                                                                                                                & {  \color{red}\ding{55}}                                                                                                               & {  \color{ForestGreen}\ding{51}}                                             & {  \color{red}\ding{55}}                                            & {  \color{red}\ding{55}}                                                        & {  \color{red}\ding{55}}                                               & {  \color{ForestGreen}\ding{51}}                                                      & {  \color{ForestGreen}\ding{51}}                                                      & {  \color{ForestGreen}\ding{51}}                                                                                                                  & {  \color{ForestGreen}\ding{51}}                                                                                                             &\color{red}\ding{55}\\

{                                                 \begin{tabular}[c]{@{}l@{}}\textbf{Ojanper{\"a} et al.}\\\textbf{~\cite{ojanpera2016network}}\end{tabular}}  & {  \color{red}\ding{55}}                                                                                                                & {  \color{red}\ding{55}}                                                                                                               & {  \color{ForestGreen}\ding{51}}                                             & {  \color{red}\ding{55}}                                            & {  \color{red}\ding{55}}                                                        & {  \color{red}\ding{55}}                                               & {  \color{ForestGreen}\ding{51}}                                                      & {  \color{red}\ding{55}}                                                      & {  \color{red}\ding{55}}                                                                                                                  & {  \color{red}\ding{55}}                                                                                                             &\color{red}\ding{55}\\
\rowcolor[HTML]{EFEFEF}{\begin{tabular}[c]{@{}l@{}}\textbf{GALTON}\\\textbf{ ~\cite{wu2015goodput}}\end{tabular}}     & {  \color{red}\ding{55}}                                                                                                                & {  \color{red}\ding{55}}                                                                                                               & {  \color{red}\ding{55}}                                             & {  \color{ForestGreen}\ding{51}}                                            & {  \color{ForestGreen}\ding{51}}                                                        & {  \color{red}\ding{55}}                                               & {  \color{ForestGreen}\ding{51}}                                                      & {  \color{ForestGreen}\ding{51}}                                                      & {  \color{ForestGreen}\ding{51}}                                                                                                                  & {  \color{ForestGreen}\ding{51}}                                                                                                             &\color{red}\ding{55}\\
{   \begin{tabular}[c]{@{}l@{}}\textbf{FRA-JSCC}\\ \textbf{~\cite{wu2013joint}}\end{tabular}}     & {  \color{red}\ding{55}}                                                                                                                & {  \color{red}\ding{55}}                                                                                                               & {  \color{red}\ding{55}}                                             & {  \color{ForestGreen}\ding{51}}                                            & {  \color{ForestGreen}\ding{51}}                                                        & {  \color{red}\ding{55}}                                               & {  \color{ForestGreen}\ding{51}}                                                      & {  \color{ForestGreen}\ding{51}}                                                      & {  \color{ForestGreen}\ding{51}}                                                                                                                  & {  \color{ForestGreen}\ding{51}}                                                                                                             &\ding{51}\\

\rowcolor[HTML]{EFEFEF}{\begin{tabular}[c]{@{}l@{}} {\color{black}\textbf{Deng et al.}}\\{\color{black}\textbf{~\cite{deng2021cross}}}\end{tabular}}     & {  \color{red}\ding{55}}                                                                                                                & {  \color{red}\ding{55}}                                                                                                               & {  \color{ForestGreen}\ding{51}}                                             & {  \color{red}\ding{55}}                                            & {  \color{red}\ding{55}}                                                        & {  \color{red}\ding{55}}                                               & {  \color{red}\ding{55}}                                                      & {  \color{red}\ding{55}}                                                      & {  \color{ForestGreen}\ding{51}}                                                                                                                  & {  \color{red}\ding{55}}                                                                                                             &\color{red}\ding{55}\\
{   \begin{tabular}[c]{@{}l@{}}\textbf{MP-DASH}\\\textbf{~\cite{han2016mp}}\end{tabular}}                                                   & {  \color{red}\ding{55}}                                                                                                                & {  \color{red}\ding{55}}                                                                                                               & {  \color{ForestGreen}\ding{51}}                                             & {  \color{red}\ding{55}}                                            & {  \color{red}\ding{55}}                                                        & {  \color{red}\ding{55}}                                               & {  \color{red}\ding{55}}                                                      & {  \color{red}\ding{55}}                                                      & {  \color{ForestGreen}\ding{51}}                                                                                                                  & {  \color{ForestGreen}\ding{51}}                                                                                                             &\color{red}\ding{55}\\
\rowcolor[HTML]{EFEFEF}{\begin{tabular}[c]{@{}l@{}}\textbf{Nam et al.}\\\textbf{~\cite{nam2016towards}}\end{tabular}}& {  \color{red}\ding{55}}                                                                                                                & {  \color{red}\ding{55}}                                                                                                               & {  \color{ForestGreen}\ding{51}}                                             & {  \color{red}\ding{55}}                                            & {  \color{red}\ding{55}}                                                        & {  \color{red}\ding{55}}                                               & {  \color{red}\ding{55}}                                                      & {  \color{red}\ding{55}}                                                      & {  \color{ForestGreen}\ding{51}}                                                                                                                  & {  \color{red}\ding{55}}                                                                                                             &\color{red}\ding{55}\\
{                                            \begin{tabular}[c]{@{}l@{}}\textbf{CMT-CL/FD}\\\textbf{~\cite{xu2015cross}}\end{tabular}} & {\color{red}\ding{55}}                                                                                                                & {\color{red}\ding{55}}                                                                                                               & \color{ForestGreen}\ding{51}                                             & {\color{red}\ding{55}}                                            & {\color{red}\ding{55}}                                                        & {\color{red}\ding{55}}                                               & \color{ForestGreen}\ding{51}                                                      & \color{ForestGreen}\ding{51}                                                      & \color{ForestGreen}\ding{51}                                                                                                                  & {  \color{red}\ding{55}}                                                                                                             &\color{red}\ding{55}\\
\rowcolor[HTML]{EFEFEF}{\begin{tabular}[c]{@{}l@{}}{\color{black} \textbf{OLS~\cite{xing2021low}} }\end{tabular}} & {\color{red}\ding{55}}                                                                                                                & {\color{red}\ding{55}}                                                                                                               & \color{ForestGreen}\ding{51}                                             & {\color{red}\ding{55}}                                            & {\color{red}\ding{55}}                                                        & {\color{red}\ding{55}}                                               & \color{ForestGreen}\ding{51}                                                      & {\color{red}\ding{55}}                                                      & \color{ForestGreen}\ding{51}                                                                                                                  & {  \color{red}\ding{55}}                                                                                                             &\color{red}\ding{55}\\ \hline
\end{tabular}}
\end{table*}

\subsection{Which packet should be sent next? }
One important scheduling task is selecting the next packet to be sent. Content awareness and video distortion at frame level are key features to select the proper packets. These features will be discussed in this subsection. {\color{black}Tables~\ref{table:Scheduler Functions-application layer},~\ref{table:Scheduler Functions-transport layer},~\ref{table:Scheduler Functions-network layer}, and~\ref{table:Scheduler Functions-cross layer} present each category related to the protocol layer. }

Note that, generally, ABR approaches rely on HTTP and separate TCP connections do not consider each one packet for data transmission and proper path  for a DASH segment/subsegment needs to be determined instead of packet~(e.g.,~\cite{xing2012rate,xing2014real,houze2016applicative,evensen2010quality,evensen2011improving,evensen2012using,bui2013greenbag}).  However, when using MPTCP for HTTP-based ABR video, the MPTCP scheduler performs  its own transport-level scheduling for the received DASH data stream.

\subsubsection{\textbf{Content Awareness}}
Considering video content features in the scheduling strategy helps to define the priority of each packet, and subsequently {\color{black}to} choose the frame packets with higher priority to send {\color{black}them} first or via more qualified paths.
In video streaming, some frames have higher effect on video quality, and large frame inter-dependency. For example, I-frames have highest priority among other frames. These strategies are generally referred to as content-aware scheduling strategies.
In addition, a content-aware scheduling strategy could use {\color{black}more robust} packet protection for higher priority packets than the less priority packets, for example, by applying adaptive FEC, which will be explained in {\color{black}the} next subsection.
On the other hand, if the scheduler is unaware of the video content features, the sending buffer would transmit data packets in the same order as they arrived in the buffer~(FIFO\nomenclature{FIFO}{First-In First-Out}) without considering the priority of packets ~(e.g., MPTCP scheduler).

Video content features are considered  as inputs to the scheduling strategy in  the  following  works:
MPRTP~\cite{singh2013mprtp},
{\color{black}~\cite{afzal2018novel},}
BEMA~\cite{wu2016bandwidth},
~\cite{freris2013distortion},
~\cite{correia2012optimal},
MP-DCCP~\cite{huang2012qos},
DEAM~\cite{Wu2019Energy}, {\color{black}EDAM~\cite{7937943}},
MPTCP-SD~\cite{diop2012qos},
PR-MPTCP$^+$~\cite{cao2016pr},
CMA-CA~\cite{wu2016content},
~\cite{corbillon2016cross}.
 In SRMT~\cite{da2016preventing}, the primary path is used for all data
while the secondary paths are used to send redundant packets, which are, in turn, chosen based on their priority~(e.g., I-frame packets have highest priority).

\subsubsection{\textbf{Video  Distortion (Frame Level)}}
{\label{Video  Distortion (Frame Level)}}
Video distortion impacts perceived video quality.
Generally, video distortion is considered at both frame level and flow level. In this section, we study the frame level video distortion because it assesses inter-frame dependencies and analyzes each specific video frame, including the frame priority and decoding dependency~\cite{freris2013distortion}. We will discuss flow level video distortion in Section~\ref{Video  Distortion (Flow Level)}. In particular, frame level distortion refers to the quality degradation of
each frame of GoP after data transmission and video decoding process~\cite{wu2016bandwidth}. This way, the frame level distortion is calculated as a total of truncation and drifting distortion.
{\color{black}The truncation distortion refers to the video quality degradation caused by packet drops during transferring data. The drifting distortion refers to the video quality distortion that occurred by imperfect reconstruction of parent frames used for inter-frame prediction.}  In the surveyed works, frame level distortion is used by BEMA~\cite{wu2016bandwidth} for calculating FEC coding parameters~(e.g., code rate and symbol size), and also it is used by~\cite{freris2013distortion} to assign higher priority values to the pictures which minimize the distortion of the decoded video affected by packet loss. Such information could also be used for path selection in CMT-CA~\cite{wu2016content}.

\subsection{How to protect the packet?}{\label{sec: How to protect the packet?}}
Providing packet protection techniques to the scheduler leads to data loss rate decreases, and consequently,  better video streaming throughput and QoE. In fact, inter-dependency among video frames causes a compressed video to be very sensitive to data loss. By this idea~\cite{le1991mpeg}, individual frames of pictures are grouped together, which is called GoP\nomenclature{GoP}{Group of Picture}. Each GoP consists of one initial Intra~(I)-frame, several Predicted~(P)-frames and possibly Bidirectional~(B)-frames. While an I-frame is encoded without reference to any other video frames, a P-frame is encoded with reference to previous I or P-frames, and a B-frame is encoded with reference to both immediate previous and forward I or P-frames. Therefore, in the decoding process, loss of some frames may preclude proper decoding, especially in the miss of I-frames. Thus, it is important to protect frames~(especially I-frames) in lossy wireless channels.
For this purpose, some JSCC/Channel  Level and Error Resilience/Source Level techniques have been implemented. These techniques will be discussed in this subsection. {\color{black}Tables~\ref{table:Scheduler Functions-application layer},~\ref{table:Scheduler Functions-transport layer},~\ref{table:Scheduler Functions-network layer}, and~\ref{table:Scheduler Functions-cross layer} present each category of such techniques divided by the protocol layer.}\\

\subsubsection{ \textbf{Joint Source and Channel Coding~(JSCC)/Channel  Level techniques}\nomenclature{JSCC}{ Joint Source and Channel Coding}}{\label{sec: Joint Source and Channel Coding (JSCC)}}

The channel level techniques for JSCC are ARQ and FEC. Automatic Repeat reQuest~(ARQ)\nomenclature{ARQ}{Automatic Repeat reQuest} retransmits requests to provide reliable data transmission. The retransmission occurs in case of packets lost or received with bit error. Inherently, all protocols atop or extensions of TCP~(e.g., HTTP, DASH, MPTCP) use ARQ. However, the retransmission wastes bandwidth, causing network congestion, and consequently, increasing end-to-end delay. For example, in efforts to mitigate these problems, {\color{black}DEAM~\cite{Wu2019Energy} {\color{black}and EDAM~\cite{7937943}},  consider retransmission based on energy-consumption and delay}, CMT-QA~\cite{xu2013cmt} retransmits packets over the path with minimum transfer delay; CMT-DA~\cite{wu2015distortion} and CMT-CA~\cite{wu2016content} retransmit only the estimated packets to arrive at the destination within the deadline; and CMT-CL/FD~\cite{xu2015cross} selects the path with the largest cwnd for the retransmission, which sends the lost packet before all the other packets that exist in the path buffer.
In addition, considering the existence of many clients in multicast communications, responding to the retransmission requests of all clients might be difficult for the server.

Other surveyed works, which utilize ARQ as JSCC technique are
MRTP~\cite{mao2006mrtp}, MPRTP~\cite{singh2013mprtp},~\cite{xing2012rate}, RTRA~\cite{xing2014real}, ~\cite{houze2016applicative}, {\color{black}Go et al.~\cite{go2019hybrid}},~\cite{sohn2015synchronization},~\cite{evensen2010quality},~\cite{evensen2011improving},~\cite{evensen2012using}, Greenbag~\cite{bui2013greenbag},
MP-H2~\cite{nikravesh2019MP-H2},
MPLOT~\cite{sharma2008mplot},
ADMIT~\cite{wu2016streaming},
MPTCP-SD~\cite{diop2012qos},
MPTCP-PR~\cite{diop2012qos},
PR-MPTCP$^+$~\cite{cao2016pr},
SRMT~\cite{da2016preventing},~\cite{yap2012making},
MARS~\cite{sun2016mars},~\cite{corbillon2016cross},~\cite{ojanpera2016network}, {\color{black}~\cite{deng2021cross}}, MP-DASH~\cite{han2016mp}, ~\cite{nam2016towards} and {\color{black} OLS~\cite{xing2021low} }.

Forward Error Correction~(FEC) \nomenclature{FEC}{Forward Error Correction} appeared to remedy the shortcoming of packet retransmission and delay constraints, especially for live video streaming. FEC can be applied to circumvent packet erasures/loss by cross-packets FEC in the application or transport layer~(inter-packet FEC), and/or to handle bit errors in the physical layer~\cite{zhai2004rate}~(intra-packet FEC).  Wireless networks can have packet loss and packet truncation due to congestion. Therefore, either the packets are dropped by the network routers or the receiver due to excessive delay. There {\color{black} also exists} bit errors due to noisy channels. Next, more details about inter- and intra-packet FEC techniques are provided.

In inter-packet FEC, redundant/parity packets are commonly generated in addition to source packets to perform cross-packet FEC, which is usually achieved by erasure codes.  These allow the receiver to detect error packets and correct data without retransmission. The capability of FEC to recover the lost data depends on the added redundant symbols.
Among the many existent erasure codes, the most commonly studied ones are Reed-Solomon~(RS)\nomenclature{RS}{Reed-Solomon}~\cite{frossard2001fec}, Low-Density Generator Matrix~(LDGM)~\cite{nakachi2013next}\nomenclature{LDGM}{Low-Density Generator Matrix}, Raptor codes~\cite{shokrollahi2006raptor}, {\color{black}XOR and Random Linear CODE~(RLC)\nomenclature{RLC}{Random Linear CODE}~\cite{RLC-16}}.
In our surveyed works, ADMIT~\cite{wu2016streaming}, GALTON~\cite{wu2015goodput} and FRA-JSCC~\cite{wu2013joint} utilize RS due to stringent delay constraint. {\color{black}QUIC-FEC~\cite{michel2018adding} applies RS, XOR and RLC FEC schemes.}
MPEG-H part 10 defines several MMT AL-FEC algorithms, including RS codes and LDGM.
Raptor coding is used in BEMA~\cite{wu2016bandwidth} {\color{black} and~\cite{go2019hybrid}} due to low processing time and high error correction capability. Such erasure codes could be applied at frame level, GoP level, or subGoP level for video protection~\cite{wu2013joint}.

In frame level~\cite{kuo2014modeling}, the frames in each GoP are classified in terms of their type and their distance from the leading I-frame. Then, FEC is applied on the frames according to their priority. In GoP level,
each GoP packetizes in  source packets. Then, FEC encoding maps source packets to some encoded packets. In SubGoP level~\cite{wu2016bandwidth}, each GoP consists of several subgroups, each mapped to a source block.
In our surveyed works, GoP level is used in ADMIT~\cite{wu2016streaming}, GALTON~\cite{wu2015goodput}, FRA-JSCC~\cite{wu2013joint}, and SubGoP level is used in BEMA~\cite{wu2016bandwidth}.

A trade-off between bandwidth/end-to-end delay and FEC redundancy is required. In particular, a smaller FEC packet size indicates a larger FEC block size due to the larger number of redundant packets~\cite{wu2016streaming}. While higher redundancy leads to better recoverability, it also increases overhead rate and bandwidth consumption. Consequently, congestion, packet reordering, FEC decoding delay and end-to-end delay have their probability increased, especially in the presence of burst losses. Therefore, an adaptive FEC is required to minimize these problems, and maximize the recoverability by adaptively changing FEC parameters~(e.g., adequate FEC packet size and FEC redundancy) according to the network channel status, application delay characteristics, or based on the importance of content data.
For example, a stronger FEC would be used in a more lossy channel while not required in a more stable channel with less loss rate percentage, or more robust FEC could also be used only for I-frames rather than B or P-frames.

Adaptive FEC is used in several of our surveyed works, like FRA-JSCC~\cite{wu2013joint} and GALTON~\cite{wu2015goodput} to find FEC redundancy, ADMIT~\cite{wu2016streaming} to adjust FEC redundancy and code rate, BEMA~\cite{wu2016bandwidth} {\color{black} and~\cite{go2019hybrid}} to set code rate and symbol size. Moreover, MPLOT~\cite{sharma2008mplot} also adaptively chooses block sizes, considering the usage of large block sizes in order to reduce bursty loss for delay-tolerant applications. We also identified
FEC usage in MRTP~\cite{mao2006mrtp}.

Besides using FEC method, an adequate technique is also requested to distinguish losses due to traffic congestion with the ones caused by wireless channel disturbances and impairments.
It is based on the fact that FEC redundancy in wireless lossy networks leads to better packet recovery; however, adding more FEC redundancy in a congested network worsens network situation since it pushes higher congestion and more losses~\cite{wallace2012review} due to bit stuffing operations. More technical details on packet loss differentiation are provided in Section~\ref{sec: Packet Loss Differentiation}.

In intra-packet FEC, channel coding is applied to correct bit errors in the physical layer. Turbo Codes~(parallel Concatenated Constitutional coding) and Low-Density Parity-Check~(LDPC)\nomenclature{LDPC}{Low-Density Parity-Check} codes are generally used. Error detection is performed at the link layer, based on Cyclic Redundancy Check~(CRC)\nomenclature{CRC}{Cyclic Redundancy Check}. 
Due to this approach, only packets passing CRC stage are visible on the network/Internet layer.

 Moreover, a joint-ARQ and FEC usage approach can enhance efficiency,  depending on the adopted strategy to couple both techniques. For example, in Go et al.~\cite{go2019hybrid}, the recovery of lost packets using the FEC scheme occurs only by retransmitting packets that cannot be recovered. ADMIT~\cite{wu2016streaming} also utilizes FEC for data reconstruction. However, there is no additional help to mitigate the number of retransmissions {\color{black}in this work, and bandwidth} consumption increases drawbacks since there is no ACK message sending to inform the server that the data is successfully reconstructed. Therefore, the MPTCP protocol on the server keeps sending retransmission of each lost packet until it receives the ACK from the receiver. This scenario outlines a motivation for a proper ARQ-and-FEC joint approach, using  FEC for data protection while retransmitting events only occurs when there is no way to perform data reconstruction.

\subsubsection{\textbf{Error Resilience/Source Level}}
Besides employing JSCC techniques to recover from packet loss and bit errors, increasing the error resilience of the video sequence itself is also an important task. To provide this functionality, error resilience techniques embrace, among others, the usage of Scalable Video Coding~(SVC)\nomenclature{SVC}{Scalable Video Coding} and Multiple Description Coding~(MDC)\nomenclature{MDC}{Multiple Description Coding} methods.

In SVC~\cite{kazemi2014review}, source video is encoded in one base layer and several enhancement layers. These layers are hierarchically dependent to each other. This means that, at the receiver, each layer can be decoded only when its lower layers have been correctly received. Therefore, video quality is improved based on the number of received enhancement layers. {\color{black}To improve the efficiency of SVC, the base layer packets are often protected by FEC or transmitted through more reliable paths due to their importance.} In the proposed approach~\cite{freris2013distortion}, each packet is transmitted to the network only if all other related packets in lower layers have been sent before.
Other surveyed works, which utilize SVC as Error Resilience are
MRTP~\cite{mao2006mrtp},
RTRA~\cite{xing2014real},~\cite{sohn2015synchronization},
CMT-DA~\cite{wu2015distortion},
GALTON~\cite{wu2015goodput} and FRA-JSCC~\cite{wu2013joint}.

In MDC~\cite{kazemi2014review}, source video is encoded into several independent compressed streams which are called descriptions. Each description can be decoded independently and shall provide acceptable quality. When one or more descriptions arrive at the receiver, a video with a certain quality level would be made by the decoder. MDC is a good alternative to retransmission to remedy the delay constraint in real-time video streaming.

According to a reviewed work about MDC techniques for video streaming~\cite{kazemi2014review}, MDC is {\color{black} more valuable} than FEC in the case of high lossy networks since FEC uses long code block sizes, increasing bandwidth consumption as well. MDC also outperforms SVC in high lossy networks, but SVC is more proper than MDC in low loss rate networks due to overhead reduction.
MDC is also recommended for multicast with heterogeneous receivers~\cite{kobayashi2009robust}. Accordingly, works like \cite{correia2012optimal} and MRTP~\cite{mao2006mrtp} utilize MDC as error resilience technique.

\subsection{Which is the best path to send the packet?}{\label{sec: Which path is the best path to transfer the packet?}}
{\color{black}Before discussing how to select the proper path to transfer the packet, it is worth mentioning that using many paths for data transmission does not always lead to better QoE since many paths for video delivery make large overheads due to parallel connections~\cite{habib2016past}.} According to~\cite{mitzenmacher2001power}, it is possible to achieve maximum multipath benefits {\color{black} by} just two paths when   using a proper scheduling strategy.

The simplest scheduling strategy is Round Robin~\cite{li2016multipath}. This strategy sorts paths and sends data to the next available path in circular order without taking into account the heterogeneous paths' characteristics. In Round Robin strategy, slow channels would be overloaded while fast channels remain underutilized~(e.g., CMT~\cite{iyengar2006concurrent}).

Obviously, scheduling strategies that are aware of path characteristics~(e.g., RTT, packet loss rate) generate wiser scheduling decisions. {\color{black}These strategies are generally referred to as path-aware scheduling strategies.}
For example, Weighted Round Robin~(WRR)\nomenclature{WRR}{Weighted Round Robin} is a scheduling strategy {\color{black} that} assigns weight to each path. Weight shows path capability regarding available bandwidth/delay/packet loss rate. This way, data distribution is proportional to the path transmission capability~(e.g., MPTCP and FRA-JSCC~\cite {wu2013joint}). Earliest Delivery Path First~(EDPF) is another scheduling strategy that estimates the delivery time of each packet over each path. Then, the packets are transmitted over the fastest path to prevent missing their deadlines and minimizing packet-reordering~(e.g., BAG~\cite{chebrolu2006bandwidth} and MPLOT~\cite{sharma2008mplot}).

Finding end-to-end path capability of real-time video traffic communication leads to estimate path quality or path reliability{\color{black}~\cite{afzal2018novel},}~\cite{wu2016bandwidth,wu2016streaming,xu2013cmt}.
Therefore, scheduling strategy could map higher priority packets to the more reliable or qualified paths~(assume that it is a combination of content-aware and path-aware scheduling strategy).

It is important to note here that mapping many packets to {\color{black}the} most qualified or reliable paths pushes congestion over that path, and consequently decreases video quality, which is called load imbalance problem~\cite{wu2015distortion}. Therefore, using a method to balance the data over the available paths is required.
In our surveyed works,  BEMA~\cite{wu2016bandwidth},~\cite{freris2013distortion},
ADMIT~\cite{wu2016streaming},
{\color{black}DEAM~\cite{Wu2019Energy},}  {\color{black}EDAM~\cite{7937943}}, CMT-CA~\cite{wu2016content}, CMT-DA~\cite{wu2015distortion} and GALTON~\cite{wu2015goodput} use load balancing mechanism to avoid imbalance problem.

Most network characteristics that are used to find the quality or reliability probability of network channels are RTT/Delay, PLR, Available bandwidth/Throughput/Goodput. There are also some other metrics that lead to better path selection and scheduling decision, such as delay constraint and video distortion at flow level. These network characteristics and metrics will be discussed in this subsection. {\color{black}Tables~\ref{table:Scheduler Functions-application layer},~\ref{table:Scheduler Functions-transport layer},~\ref{table:Scheduler Functions-network layer}, and~\ref{table:Scheduler Functions-cross layer} present each category per protocol layer.}

\subsubsection{\textbf{RTT/Delay}}
Round Trip Time~(RTT)\nomenclature{RTT}{Round-Trip Time} is the time required for a packet to be sent plus the time it takes to receive an ACK of that packet~\cite{wu2015goodput,wu2016streaming}. Therefore, RTT consists of the packet transmission time and path propagation delay~\cite{wu2016streaming}. To avoid sudden variations of RTT, some approaches~(e.g., MPTCP and SCTP) apply a smoothing factor to the RTT~(sRTT). In the approaches without ACK method, for example, UDP-based approaches, one-way delay could be considered instead of RTT.

Considering RTT/Delay for path scheduling decreases the probability of expired arrival packets, stall or out-of-order packet delivery.  In our surveyed works, {\color{black}
MP-H2~\cite{nikravesh2019MP-H2} utilizes HTTP/2 PING, as defined by the HTTP/2 specification~\cite{rfc7540}, to calculate application layer RTT. Generally, HTTP/2 PING is used to check whether an idle connection is still alive. In specific, MP-H2 utilizes HTTP/2 PING to estimate  the application layer RTT, including the server buffer delay. This would be the same delay as the HTTP response delay.}

Since RTCP protocol, which is generally used by RTP to transfer monitored information, is possible to calculate RTT by using sender and receiver reports,  the multipath transmission approaches over RTP, such as MRTP~\cite{mao2006mrtp} and MPRTP~\cite{singh2013mprtp} extended RTCP  in order to calculate RTT in multipath transmission solutions.

MARS~\cite{sun2016mars}, which is implemented over separate TCP connections, utilized a relative RTT measurement method based on OpenFlow protocol. In this approach, duplicated packets~(probes) are sent through different interfaces.
The probes would return to the sender through the common reverse path from the edge switch close to the client side. The transfer process can be implemented with the tables of OpenFlow at the edge switch. The approach measures the relative delay of forward paths instead of their absolute delay because, in the case of absolute forward path delays, the tight clock synchronization between sender and receiver is required. More information and comparison details between relative and absolute delay can be found at~\cite{ribeiro2006minimum}.

In SCTP protocol, the sent packet acknowledgment~(SACK)  can be transmitted over different paths.  Mostly the acknowledgment  packet returns through the most reliable path to mitigate the probability of dropped or overdue feedback packets. Since paths have different delay characteristics, the estimated RTT is incorrect and using this estimated RTT to find the path quality leads to the wrong result. For this reason, CMT-QA~\cite{xu2013cmt} does not use RTT directly. Instead, it uses transmission delay. Transmission delay refers to the time difference between the time of the first chunk entering each path sender buffer from a group of distributed data chunks and the time of the last chunk leaving the path sender buffer.
CMT-CL/FD~\cite{xu2015cross} utilizes the SCTP heartbeat mechanism to calculate RTT. In this mechanism, the HEARTBEAT-ACKs have to return through the same path used to send the HEARTBEAT messages.

 FRA-JSCC~\cite{wu2013joint} and BAG~\cite{chebrolu2006bandwidth}, which are the approaches that use UDP as transport protocol, utilize propagation delay.  FRA-JSCC~\cite{wu2013joint} calculates propagation delay network characteristic by using the existing time stamp in each packet header.

RTT/Delay is also used for packet loss differentiation decision in CMT-QA~\cite{xu2013cmt}, CMT-CL/FD~\cite{xu2015cross}, BEMA~\cite{wu2016bandwidth}, ADMIT~\cite{wu2016streaming}, GALTON~\cite{wu2015goodput}, and CMT-CA~\cite{wu2016content}. More technical details on packet loss differentiation are provided in Section~\ref{sec: Packet Loss Differentiation}. Besides, RTT/Delay can also be used for other tasks. For example, MRTP~\cite{mao2006mrtp} sets retransmission timeout value by RTT, and Greenbag~\cite{bui2013greenbag} utilizes RTT to determine when to send requests for the next segments.

Other surveyed works considered RTT/Delay network characteristic for their scheduling decision are~\cite{houze2016applicative},  {\color{black}Go et al.~\cite{go2019hybrid},}
{\color{black}~\cite{afzal2018novel},}
~\cite{evensen2010quality},
~\cite{evensen2011improving},
~\cite{evensen2012using},
~\cite{freris2013distortion},
MPLOT~\cite{sharma2008mplot},
MP-DCCP~\cite{huang2012qos},
DEAM~\cite{Wu2019Energy}, {\color{black}EDAM~\cite{7937943}},
MPTCP-SD~\cite {diop2012qos},
MPTCP-PR~\cite{diop2012qos},
PR-MPTCP$^+$~\cite{cao2016pr},
CMT-DA~\cite{wu2015distortion},
~\cite{corbillon2016cross},
~\cite{ojanpera2016network}, and {\color{black} OLS~\cite{xing2021low} }.

\subsubsection{\textbf{PLR}}
Packet Loss Rate~(PLR)\nomenclature{PLR}{Packet Loss Rate} comprises of network transmission lost packets, which are lost/error arrived packets during the communication paths,
and the expired arrival packets~(overdue)~\cite{wu2015goodput}.
Three basic reasons cause packet losses~\cite{xu2013cmt}; 1) congestion due to limited bandwidth or buffer size, 2) noise or interference in the wireless networks, 3) path failure or handover.
Therefore, sending highest priority frame packets on the paths with less PLR leads to better QoE. Besides,
PLR network characteristic and distinguishing packet loss differentiation are key factors for adaptively FEC protection~(Section~\ref{sec: Joint Source and Channel Coding (JSCC)}), avoiding unnecessary fast retransmission~(Section~\ref{sec:Transport Layer Approaches}), and video distortion estimation~(Section~\ref{Video  Distortion
(Frame Level)}, and Section~\ref{Video  Distortion (Flow Level)}).

PLR is considered for scheduling decision in the following works:
MRTP~\cite{mao2006mrtp},
MPRTP~\cite{singh2013mprtp}, {\color{black}Go et al.~\cite{go2019hybrid},}{\color{black}~\cite{afzal2018novel},}
BEMA~\cite{wu2016bandwidth},~\cite{freris2013distortion},   MPLOT~\cite{sharma2008mplot},
MP-DCCP~\cite{huang2012qos},
ADMIT~\cite{wu2016streaming},
DEAM~\cite{Wu2019Energy}, {\color{black}EDAM~\cite{7937943}},
CMT-DA~\cite{wu2015distortion}, CMT-CA~\cite{wu2016content},~\cite{corbillon2016cross},
GALTON~\cite{wu2015goodput},
FRA-JSCC~\cite{wu2013joint}, and
CMT-CL/FD~\cite{xu2015cross}. \\

\subsubsection{\textbf{Available bandwidth/Throughput/Goodput}}

Available bandwidth is defined as the maximum video rate that can be transmitted over end-to-end path~\cite{wu2016streaming}.
Different methods are introduced to estimate available bandwidth in the literature{\color{black}~\cite{bentaleb2019bandwidth}},~\cite{paul2016enhanced,zhou2008new}. Some approaches {\color{black}utilize} throughput or goodput for this purpose.
The amount of data that could traverse through a path is known as throughput. Throughput refers to all useful and not useful data, including data retransmissions and overhead data~(e.g., headers). If the scheduler considers only throughput among all network characteristics, it may distribute packets over high loss rate channels, and consequently, serious degrade of goodput performance and video quality occurs~\cite{wu2016delay}.
Goodput refers to the amount of useful data~(exclusive protocol overhead or retransmission) delivered successfully to the destination within the imposed specific deadline\cite{wu2015goodput}. Goodput is also known as application level throughput.
Regarding~\cite{wu2016streaming}, the approaches over HTTP/TCP could estimate the available bandwidth by using the observed TCP throughput. In our surveyed works,~\cite{freris2013distortion} measures bandwidth by using {\color{black}Abing}\footnote{{\color{black}http://iphome.hhi.de/suehring/tml/download/}}.
GALTON~\cite{wu2015goodput} and FRA-JSCC~\cite{wu2013joint} implement pathChirp algorithm~\cite{ribeiro2003pathchirp} for this purpose. CMT-CL/FD~\cite{xu2015cross}  computes available bandwidth as the ratio between the average packet length and average inter-packet sending time. CMT-CA~\cite{wu2016content} and CMT-DA~\cite{wu2015distortion} believe that cwnd has effect on bandwidth, therefore, these works calculate it as $(cwnd/RTT)$. In RTRA~\cite{xing2014real}, once a segment has been successfully downloaded, the transmission bandwidth would be calculated as division of the total size of transmitted data over the transmission time, and then, a Markov channel model is used to estimate future available bandwidth.

Other surveyed works, which consider Available bandwidth/Throughput/Goodput network characteristic for their scheduling decision are
MRTP~\cite{mao2006mrtp}, MPRTP~\cite{singh2013mprtp},~\cite{xing2012rate}, {\color{black}Go et al.~\cite{go2019hybrid},}{\color{black}~\cite{afzal2018novel},}~\cite{evensen2010quality},~\cite{evensen2011improving}
,~\cite{evensen2012using}
,
Greenbag~\cite{bui2013greenbag}
,
{\color{black}MP-H2~\cite{nikravesh2019MP-H2},}
DEAM~\cite{Wu2019Energy}, {\color{black}EDAM~\cite{7937943}},     ADMIT~\cite{wu2016streaming},
{\color{black}DEAM~\cite{Wu2019Energy},}
PR-MPTCP$^+$~\cite{cao2016pr},
MARS~\cite{sun2016mars},
BAG~\cite{chebrolu2006bandwidth},~\cite{corbillon2016cross},
 {\color{black}~\cite{deng2021cross}}, MP-DASH~\cite{han2016mp},~\cite{nam2016towards}, and {\color{black} OLS~\cite{xing2021low}}. \\

\subsubsection{\textbf{Delay Constraint}}
A real-time video application imposes a decoding deadline. In this manner, the overdue packets cannot handle at the decoder, even if they arrive successfully. Therefore, the end-to-end delay has to be less than delay constraint~\cite{wu2016streaming}. Besides that, considering delay constraint in scheduling strategy could also avoid playback buffer starvation~\cite{wu2016bandwidth}.

In our surveyed works, the delay constraint of GALTON~\cite{wu2015goodput}, ADMIT~\cite{wu2016streaming},
DEAM~\cite{Wu2019Energy}, {\color{black}EDAM~\cite{7937943}},
FRA-JSCC~\cite{wu2013joint} and CMT-CA~\cite{wu2016content} are set with values 300, 500, {\color{black}500,} 250 and 100 ms for each video frame respectively. This value in BEMA~\cite{wu2016bandwidth} is set equal to its playback duration, so the delay constraint should be 40 ms if the video is encoded at 25 frames per second. GALTON~\cite{wu2015goodput} uses delay constraint to compute transmission intervals in order to mitigate consecutive losses. ADMIT~\cite{wu2016streaming} calculates the rate allocation vector and FEC coding parameters {\color{black}with respect to} delay constraint. FRA-JSCC~\cite{wu2013joint} finds source rate adaption under delay constraint. CMT-CA~\cite{wu2016content} finds the optimal congestion window sizes and frame scheduling vector to mitigate video distortion. While CMT-DA~\cite{wu2015distortion} is not appropriate for the video streaming with stringent delay constraint but the retransmission method is based on the delay constraint. In the work~\cite{freris2013distortion}, the same deadline time is assumed for all users, which is determined as a system parameter by the service provider. Then, this is used to find packet loss probability.







Proposed approaches in~\cite{houze2016applicative},~\cite{evensen2011improving, evensen2012using}, GreenBag~\cite{bui2013greenbag} {\color{black}and MP-H2~\cite{nikravesh2019MP-H2}} are application-aware, therefore, they are aware of buffer level at the receiver in order to calculate the delay constraint. These approaches utilize adaptive streaming over multiple separate TCP connections, and mostly path selection is integrated with the adaptation logic.
In works~\cite{evensen2011improving} and~\cite{ evensen2012using}, the delay constraint is calculated by the client to select the suited bit rate. The client calculates the amount of already received content to playout in the buffer~(transfer-deadline) and estimates how long it takes to receive the already requested data~(pipeline-deadline). The difference between pipeline-deadline and transfer-deadline shows the amount of time that the client can wait to receive the next segment without interruption. Then, this estimation is compared with the estimation of the times it takes to receive the desired segment in the different bit rates, and the most proper bit rate is selected. After that, the segment is divided into subsegments. The size of each subsegment is decided based on the measured throughput of each interface that it will be requested through. The approach in~\cite{houze2016applicative} finds suited segment bit rate {\color{black}by} checking the size of the first frame in each segment representation. It chooses the representation with the highest bit rate and high probability of {\color{black}getting the} frame on time. Then, it finds the best size of byte range per path dynamically based on paths' RTT. GreenBag~\cite{bui2013greenbag} utilizes paths' delay and available bandwidth to determine per path subsegment size. If one path received its subsegment within a segment, but the other path is significantly lagging, so, the former path takes over some portion of the problematic path to recover. {\color{black}Similarly, when estimating byte ranges MP-H2~\cite{nikravesh2019MP-H2} also considers delay and bandwidth to realize data transmission over multiple paths simultaneously.} The {\color{black}above-mentioned} approaches could achieve zero or close to zero interruption during playback time.

Two more other application-aware approaches concerning delay constraint are~\cite{corbillon2016cross} and MP-DASH~\cite{han2016mp}. These approaches utilize adaptive streaming over MPTCP paths. MP-DASH~\cite{han2016mp} feed the modified MPTCP with the deadline of each video data unit in order to further use and path selection.
The approach in~\cite{corbillon2016cross}  understands the display time of each video unit with access to the Picture Order Counts~(POC)\nomenclature{POC}{Picture Order Counts} and the coding identifier of each frame~(because it is content awareness). Therefore, the approach estimates the deadlines and ignores transmission of packets {\color{black}that} will miss their playback deadline and instead, assigns more priority to the packets which their deadline time is close. The high priority packets can be spread through less RTT paths. This helps to use bandwidth more efficiently and experience less video distortion.
In PR-MPTCP$^+$~\cite{cao2016pr}, when {\color{black}the} network is detected as congested, only the packets with enough deadline time to play would be sent.

\subsubsection{\textbf{Video  Distortion~(Flow Level)}}{\label{Video  Distortion (Flow Level)}}
We previously discussed frame level video distortion in Section~\ref{Video  Distortion
(Frame Level)}.
Here, we study flow level video distortion. {\color{black}End-to-end} video distortion at flow level~(intra-coding) is calculated as total of source and channel distortion~\cite{wu2015distortion}.
Source distortion is determined by the video source rate and video sequence parameters because of their impact on the efficiency of video codec. For example, in case of the same video encoding rate, a more complex video sequence has higher distortion. As another example, increasing the video encoding rate causes decreasing distortion. Channel distortion refers to the packet losses during the network transmission
and expired arrivals.
Some other features including the frame structure and GoP size also have an impact on both the source and the channel distortion.
Flow level video distortion is considered for scheduling strategy in the following surveyed works: ADMIT~\cite{wu2016streaming},
DEAM~\cite{Wu2019Energy}, {\color{black}EDAM~\cite{7937943}},
CMT-DA~\cite{wu2015distortion} and
FRA-JSCC~\cite{wu2013joint}.

Although most important network characteristics and metrics for path selection were discussed, but there are some other parameters that are used directly or indirectly~(to calculate RTT, PLR or other metrics) by different approaches.
For example, cwnd is used in MPLOT~\cite{sharma2008mplot}, MP-DCCP~\cite{huang2012qos}~(CCID2),
{\color{black}DEAM~\cite{Wu2019Energy},}
CMT-CA~\cite{wu2016content} and CMT-DA~\cite{wu2015distortion}, sending rate is used in MP-DCCP~\cite{huang2012qos}~(CCID3) and CMT-CL/FD~\cite{xu2015cross}, cost function is utilized in MP-DASH~\cite{han2016mp} and GreenBag~\cite{bui2013greenbag}. In MP-DASH, cost can be data usage, energy consumption or both, and in GreenBag~\cite{bui2013greenbag}, cost refers to energy consumption. Other useful factors can be buffer size, packet size, packet count and etc.

\section{Analysis and Comparison of Methods and Techniques}{\label{sec: Analysis and Comparison of Candidate Methods and Techniques}}
In the previous two sections, we analyzed different multipath {\color{black}wireless} video streaming works based on layer dependency and scheduling functions. In this section, we study other effected features and related methods that are used in these works. Table~\ref{compression} re-classified the candidate previously explained surveyed works based on the features or methods the authors used.

\begin{table*}
\centering
\caption{{\color{black}COMPARISON REGARDING KEY FEATURES AND EXPERIMENTAL EVALUATION. (N/D STANDS FOR NOT DEFINED)}}

\label{compression}
\resizebox{1.8\columnwidth}{!}{%
\begin{tabular}{|l|c|c|l|l|l|l|l|l}
\hline
\cellcolor[HTML]{C0C0C0}\textbf{Works}                                                    & \cellcolor[HTML]{C0C0C0}\textbf{\begin{tabular}[c]{@{}l@{}}Packet loss\\ Differentiation\\ Method\end{tabular}} &\cellcolor[HTML]{C0C0C0}\textbf{\begin{tabular}[c]{@{}l@{}}{\color{black}Fairness}\end{tabular}} & \cellcolor[HTML]{C0C0C0}\textbf{\begin{tabular}[c]{@{}l@{}}Video\\ Compression\end{tabular}} & \cellcolor[HTML]{C0C0C0}\textbf{\begin{tabular}[c]{@{}l@{}}Error\\ Concealment\end{tabular}} &  \cellcolor[HTML]{C0C0C0}\textbf{\begin{tabular}[c]{@{}l@{}}{\color{black}Experimental}\\ {\color{black}Environment}\end{tabular}} & \cellcolor[HTML]{C0C0C0}\textbf{\begin{tabular}[c]{@{}l@{}}Performance\\ Metrics\end{tabular}}                                                                            & \cellcolor[HTML]{C0C0C0}\textbf{\begin{tabular}[c]{@{}l@{}}{\color{black}Video}\\ {\color{black}Services}\end{tabular}} \\ \hline
\rowcolor[HTML]{EFEFEF}{\begin{tabular}[c]{@{}l@{}}\textbf{MRTP}\\ \textbf{~\cite{mao2006mrtp}}\end{tabular}}                                                                & {  \color{red}\ding{55}}       &\color{red}\ding{55}                                                                          & {   N/D}                                                           & {   N/D}                                                            & {   OPNET}                                                                 & {   \begin{tabular}[c]{@{}l@{}}PSNR,\\
Bandwidth utilization,\\Buffer overflow \\ probability,\\ Playout buffer size\end{tabular}}                                                &  Real-time \\ \hline
{   \begin{tabular}[c]{@{}l@{}}\textbf{MPRTP}\\ \textbf{~\cite{singh2013mprtp}}\end{tabular}}                                                               & {   }  \color{ForestGreen}\ding{51}  &\color{ForestGreen}\ding{51}                                                                             & {   H.264/AVC}                                                             & {   x264}                                                                   & {   \begin{tabular}[c]{@{}l@{}}Realistic testbed,\\ NetEm,\\Disjoiont paths,\\Client interfaces:\\ WiFi and 3G \\ or multiple 3G\end{tabular}}    & {   \begin{tabular}[c]{@{}l@{}}PSNR,\\ Loss rate,\\ Bandwidth utilization, \\ Connection setup time\end{tabular}}                                                          & \begin{tabular}[c]{@{}l@{}}Live,\\ Real-time\end{tabular}  \\

 \rowcolor[HTML]{EFEFEF}{\begin{tabular}[c]{@{}l@{}}\textbf{Xing et al.}\\\textbf{~\cite{xing2012rate}} \end{tabular}}
 &\color{red}\ding{55}            &\color{red}\ding{55}                                                                                             & H.264/AVC                                                                                    & \begin{tabular}[c]{@{}l@{}}x264 \\ encoder,\\ FFmpeg \\ decoder\end{tabular}                  & \begin{tabular}[c]{@{}l@{}}Realistic  testbed,\\ Android framework,\\Disjoint paths,\\
Client interfaces:\\ WiFi and 3G\end{tabular}              & \begin{tabular}[c]{@{}l@{}} Playback fluency average,\\ Playback quality,\\ Quality switch,\\ Average 3G traffic,\\ Playback traces,\\ Buffer occupancy\end{tabular}       & N/D \\
\begin{tabular}[c]{@{}l@{}}\textbf{RTRA}\\ \textbf{~\cite{xing2014real}}\end{tabular}                                                                                                        &\color{red}\ding{55}  &\color{red}\ding{55}                                                                                                       & H.264/SVC                                                                                    & JSVM                                                                                         & \begin{tabular}[c]{@{}l@{}}Realistic testbed,\\ Android framework,\\ Client interfaces:\\ WiFi and blacktooth
\end{tabular}               & \begin{tabular}[c]{@{}l@{}}PSNR,\\ Startup delay,\\ Playback fluency average,\\ Playback quality,\\ Quality switch,\\ Bandwidth utilization,\\
Playback traces,\\Buffer occupancy\end{tabular} &  Real-time\\
\rowcolor[HTML]{EFEFEF}{\begin{tabular}[c]{@{}l@{}}\textbf{Houz{\'e} et al.}\\\textbf{~\cite{houze2016applicative}} \end{tabular}}                                                                                                                                      &\color{red}\ding{55}     &\color{red}\ding{55}                                                                                                     & HEVC                                                                                         & HM                                                                                           & \begin{tabular}[c]{@{}l@{}}NS3,\\Client interfaces:\\ five homogeneous\\ xDSL links \\ \end{tabular}                                                                                     & \begin{tabular}[c]{@{}l@{}}Cumulative Distribution\\ Function\\ (CDF)\nomenclature{CDF}{Cumulative Distribution Function} of frame sizes,\\ QoE (SAMVIQ method)\end{tabular}                                                    & Live \\

{   \begin{tabular}[c]{@{}l@{}}{\color{black}\textbf{Go et al.}}\\\textbf{~\cite{go2019hybrid} }\end{tabular}}                                                                                                                                      & Spike    &\color{ForestGreen}\ding{51}                                                                                                     & H.264/AVC                                                                                         & FFMPEG                                                                                           & \begin{tabular}[c]{@{}l@{}}Dummynet,\\Android framework,\\Client interfaces:\\ WiFi and LTE \\ \end{tabular}                                                                                     & \begin{tabular}[c]{@{}l@{}}Raptor decoding\\ energy consumption,\\Raptor decoding delay,\\Wireless network interface\\ energy consumption,\\Requested segment data,\\Symbol size,\\Requested redundant data,\\Quality switch\end{tabular}                                                    & N/D \\

\rowcolor[HTML]{EFEFEF}{\begin{tabular}[c]{@{}l@{}}\textbf{{\color{black}Afzal et al.}}\\\textbf{{\color{black}~\cite{afzal2018novel,afzal2021multipath}}} \end{tabular}}                                                                                                                                      &\color{red}\ding{55}     &\color{ForestGreen}\ding{51}                                                                                                     & H.264                                                                                         & \begin{tabular}[c]{@{}l@{}}FFMPEG\end{tabular}                                                                                           & \begin{tabular}[c]{@{}l@{}}NS3-DCE,\\Client interfaces:\\ WiFi (802.11n) and LTE \\ \end{tabular}                                                                                     & \begin{tabular}[c]{@{}l@{}}PSNR, SSIM, Goodput,\\ Loss rate,\\ I and NI \\frame packet loss rate,\\Delay \end{tabular}                                                    & Real-time \\
\begin{tabular}[c]{@{}l@{}}\textbf{Sohn et al.}\\\textbf{ ~\cite{sohn2015synchronization}}\end{tabular}  &\color{red}\ding{55}     &\color{red}\ding{55}                                                                                                    & SHVC                                                                                         & JSVM                                                                                          & \begin{tabular}[c]{@{}l@{}}Own visual studio\\ implementation,\\ Client interfaces: \\WiFi and Ethernet\end{tabular}                   & \begin{tabular}[c]{@{}l@{}}Throughput,\\Play time for base layer,\\ Quality switch\end{tabular}                                                                                                      & \begin{tabular}[c]{@{}l@{}}Live,\\ VoD\end{tabular}  \\
\rowcolor[HTML]{EFEFEF}\begin{tabular}[c]{@{}l@{}}\textbf{Evensen et al.}\\\textbf{~\cite{evensen2010quality}} \end{tabular}                                                                                                                                                  &\color{red}\ding{55}                     &\color{red}\ding{55}                                                                                   & N/D                                                                                  & N/D                                                                                    & \begin{tabular}[c]{@{}l@{}}Realistic testbed,\\ Ubuntu framework,\\ NetEm,\\Client interfaces: \\WiFi (IEEE 802.11b) and\\ Cellular (HSDPA)\end{tabular}       & \begin{tabular}[c]{@{}l@{}}Quality distribution,\\ Missed deadlines,\\ Throughput\end{tabular}                                                                           & Live\\
\begin{tabular}[c]{@{}l@{}}\textbf{Evensen et al.}\\\textbf{~\cite{evensen2011improving} }\end{tabular}                                                                                                                                                     &\color{red}\ding{55}                          &\color{red}\ding{55}                                                                                & N/D                                                                                  & N/D                                                                                 & \begin{tabular}[c]{@{}l@{}}Realistic testbed,\\  Ubuntu framework,\\  NetEm,\\Client interfaces: \\WiFi (IEEE 802.11b) and \\Cellular (HSDPA)\end{tabular}     & \begin{tabular}[c]{@{}l@{}}Quality distribution,\\ Missed deadlines, \\ Throughput\end{tabular}                                                                          &  Live\\

\rowcolor[HTML]{EFEFEF}\begin{tabular}[c]{@{}l@{}}\textbf{Evensen et al.}\\\textbf{~\cite{evensen2012using}} \end{tabular} &\color{red}\ding{55}     &\color{red}\ding{55}                                                                                                    & N/D                                                                                  & N/D                                                                                   & \begin{tabular}[c]{@{}l@{}}Realistic testbed,\\  Ubuntu framework,\\  NetEm,\\Client interfaces: \\WiFi (IEEE 802.11b) and \\Cellular (HSDPA)\end{tabular}      & \begin{tabular}[c]{@{}l@{}}Quality distribution,\\ Missed deadlines, \\ Throughput\end{tabular}                                                                          & \begin{tabular}[c]{@{}l@{}}Live,\\ VoD\end{tabular}  \\
\begin{tabular}[c]{@{}l@{}}\textbf{GreenBag}\\ \textbf{~\cite{bui2013greenbag}}\end{tabular}                                                                                                       &\color{red}\ding{55}        &\color{red}\ding{55}                                                                                                   & N/D                                                                                  & N/D                                                                                 & \begin{tabular}[c]{@{}l@{}}Realistic testbed,\\Own C and\\ JAVA implementation,\\ Android framework,\\ NetEm,\\Client interfaces:\\ WiFi and LTE\end{tabular}       & \begin{tabular}[c]{@{}l@{}}Playback time, \\Interruption time, \\ Energy consumption,\\ Buffer size,\\ In-order data\end{tabular}                                                          & Real-time \\ \hline

\end{tabular}}
\end{table*}

\begin{table*}
\centering
\resizebox{1.8\columnwidth}{!}{%
\begin{tabular}{|l|c|c|l|l|l|l|l|l}
\hline

\cellcolor[HTML]{C0C0C0}\textbf{Works}                                                      & \cellcolor[HTML]{C0C0C0}\textbf{\begin{tabular}[c]{@{}l@{}}Packet loss\\ Differentiation\\ Method\end{tabular}} & \cellcolor[HTML]{C0C0C0}\textbf{\begin{tabular}[c]{@{}l@{}}{\color{black}Fairness}\end{tabular}} & \cellcolor[HTML]{C0C0C0}\textbf{\begin{tabular}[c]{@{}l@{}}Video\\ Compression\end{tabular}} & \cellcolor[HTML]{C0C0C0}\textbf{\begin{tabular}[c]{@{}l@{}}Error\\ Concealment\end{tabular}} & \cellcolor[HTML]{C0C0C0}\textbf{\begin{tabular}[c]{@{}l@{}}{\color{black}Experimental}\\ {\color{black}Environment}\end{tabular}}                                                                                           & \cellcolor[HTML]{C0C0C0}\textbf{\begin{tabular}[c]{@{}l@{}}Performance\\ Metrics\end{tabular}}                                                              & \cellcolor[HTML]{C0C0C0}\textbf{\begin{tabular}[c]{@{}l@{}}{\color{black}Video}\\{\color{black} Services}\end{tabular}}  \\ \hline

\rowcolor[HTML]{EFEFEF}\begin{tabular}[c]{@{}l@{}}{\color{black}\textbf{MP-H2}}\\\textbf{ ~\cite{nikravesh2019MP-H2}}\end{tabular}                                                                                                       &\color{red}\ding{55}        & {\color{black}\ding{51}}                                                                                                   & N/D                                                                                  & N/D                                                                                 & \begin{tabular}[c]{@{}l@{}}Realistic testbed,\\Own JAVA\\ implementation,\\ Android framework,\\Client interfaces:\\ WiFi and LTE\end{tabular}       & \begin{tabular}[c]{@{}l@{}}Quality distribution,\\Rebuffering\end{tabular}                                                          & N/D \\

\begin{tabular}[c]{@{}l@{}}\textbf{BEMA}\\\textbf{ ~\cite{wu2016bandwidth}}\end{tabular}                                                                                                 & ZigZag       &\color{ForestGreen}\ding{51}                                                                                             & H.264/AVC                                                                                     & JM                                                                                            & \begin{tabular}[c]{@{}l@{}}Exata,\\ Client interfaces:\\Cellular,\\ WiFi (802.11a/g) and\\ WiMAX (802.16)\end{tabular}                                                                                       & \begin{tabular}[c]{@{}l@{}}PSNR,\\ end-to-end delay,\\ Goodput,\\ Streaming rate,\\ Number of frames lost, \\ Inter-packet delay, \\Bandwidth utilization, \\ Loss rate \end{tabular}                                                                      & {\color{black}Real-time}  \\

\rowcolor[HTML]{EFEFEF}\begin{tabular}[c]{@{}l@{}}\textbf{Freris at al.}\\~\cite{freris2013distortion} \end{tabular}&\color{red}\ding{55}        & \begin{tabular}[c]{@{}l@{}}Users' \\ fairness\end{tabular}                                                                                                   & H.264/SVC                                                                                    & N/D                                                                                  & \begin{tabular}[c]{@{}l@{}}NS2,\\ Matlab for subroutins,\\Client interfaces:\\
Ethernet, WiFi (802.11b) and\\ WiFi (802.11g)\end{tabular}                     & \begin{tabular}[c]{@{}l@{}}PSNR,\\ Streaming rate,\\ Packet delivery delay,\\ Delivery ratio,\\ Run time, \\ Cost functions  evaluation \\ (service differentiation)\end{tabular}                                                  &  VoD\\
\begin{tabular}[c]{@{}l@{}}\textbf{Correia at al.}\\\textbf{~\cite{correia2012optimal}} \end{tabular}                                                                                                                                                 &\color{red}\ding{55}                                   &\color{red}\ding{55}                                                                          & H.264 /AVC                                                                                   & \begin{tabular}[c]{@{}l@{}}Picture-Copy\\ Method\end{tabular}                                 & N/D                                                                                  & \begin{tabular}[c]{@{}l@{}}PSNR\end{tabular}                                                                                                          &  N/D\\

\rowcolor[HTML]{EFEFEF}\begin{tabular}[c]{@{}l@{}}\textbf{MPLOT }\\ \textbf{~\cite{sharma2008mplot}}\end{tabular}                                                                                             & ECN         &\color{ForestGreen}\ding{51}                                                                                                      & N/D                                                                                  & N/D                                                                                  &\begin{tabular}[c]{@{}l@{}}NS2\end{tabular}                                                                                           & \begin{tabular}[c]{@{}l@{}}Bandwidth utilization,\\ Congestion window size\\ (fairness test), \\ Goodput,\\ Effect of loss correlations\end{tabular}                                                                                                                                                                     &  N/D \\

\begin{tabular}[c]{@{}l@{}}\textbf{MP-DCCP}\\\textbf{~\cite{huang2012qos}}\end{tabular}                                                                                                    & ECN      &\begin{tabular}[c]{@{}l@{}}\color{red}\ding{55} \end{tabular}                                                                                                        & H.264 /AVC                                                                                & N/D                                                                                  & \begin{tabular}[c]{@{}l@{}}NS2,\\Disjoint Paths,\\Client interfaces:\\WiFi, 3G and Ethernet\end{tabular}                                                                                          & \begin{tabular}[c]{@{}l@{}}Decodable ratio of\\ transmitted frames  \end{tabular}                                                                                                                                                                 &  Live\\

\rowcolor[HTML]{EFEFEF}\begin{tabular}[c]{@{}l@{}}\textbf{ADMIT}\\\textbf{ ~\cite{wu2016streaming}}\end{tabular}                                                                                                      & ZigZag     &\color{ForestGreen}\ding{51}                                                                                                & H.264/AVC                                                                                    & JM                                                                                           & \begin{tabular}[c]{@{}l@{}}EXata,\\Client Interfaces:\\WiFi, Cellular and WiMAX\end{tabular}                                                                                           & \begin{tabular}[c]{@{}l@{}}PSNR, end-to-end delay,\\ Goodput, \\ Congestion window size\\ (fairness test), \\ Inter-packet delay, \\ FEC redundancy, \\ Out-of-order packets \end{tabular}                                                                                               & {\color{black}Real-time} \\

\begin{tabular}[c]{@{}l@{}}{\color{black}\textbf{DEAM}}\\\textbf{ ~\cite{Wu2019Energy}}\end{tabular}                                                                                                      & ZigZag     &\color{ForestGreen}\ding{51}                                                                                                & H.264/AVC                                                                                    & JM                                                                                           & \begin{tabular}[c]{@{}l@{}}EXata,\\Client Interfaces:\\WiFi, Cellular and WiMAX\end{tabular}                                                                                           & \begin{tabular}[c]{@{}l@{}}PSNR,\\ Energy consumption,\\ Power dissipation,\\ Allocated rate\\ for each path, \\ Total retransmission,\\ Buffering rate, \\ Out-of-order packets \end{tabular}                                                                                               & Real-time \\

\rowcolor[HTML]{EFEFEF}\begin{tabular}[c]{@{}l@{}}{\color{black}\textbf{EDAM}}\\\textbf{{\color{black} ~\cite{7937943}}}\end{tabular}                                                                                                      & ZigZag     &\color{ForestGreen}\ding{51}                                                                                                & H.264/AVC                                                                                    & JM                                                                                           & \begin{tabular}[c]{@{}l@{}}EXata and realistic testbed,\\Client Interfaces:\\WiFi and LTE\end{tabular}                                                                                           & \begin{tabular}[c]{@{}l@{}}PSNR,\\ Energy and power consumption,\\ Allocated rate\\ for each path, \\ PSNR, SSIM, \\ Retransmission,\\ Goodput,\\  Buffering rate, \\ Out-of-order packets \end{tabular}                                                                                               & Real-time \\

\begin{tabular}[c]{@{}l@{}}\textbf{MPTCP-SD}  \\\textbf{ ~\cite{diop2012qos}}\end{tabular}                                                                                                         &\color{red}\ding{55}     &\color{ForestGreen}\ding{51}                                                                                                    & H.264/AVC                                                                                    & N/D                                                                                  & \begin{tabular}[c]{@{}l@{}}NS2,\\Disjoint paths,\\Client Interfaces:\\ 3G and 3G \end{tabular}                                                                                        & PSNR                                                                                                                                                                      & \begin{tabular}[c]{@{}l@{}}Real-time\\ (interactive)\end{tabular} \\

\rowcolor[HTML]{EFEFEF} {  \begin{tabular}[c]{@{}l@{}}\textbf{MPTCP-PR} \\ \textbf{~\cite{diop2012qos}}\end{tabular} }                                                                                                            & {  \color{red}\ding{55}}             &\color{ForestGreen}\ding{51}                                                                     & {   H.264/AVC}                                                             & {   N/D}                                                           & {   \begin{tabular}[c]{@{}l@{}}NS2,\\Disjoint paths,\\Client Interfaces:\\ 3G and 3G \end{tabular}} &
{   \begin{tabular}[c]{@{}l@{}}PSNR\end{tabular}}                                        &  \begin{tabular}[c]{@{}l@{}}Real-time\\ (interactive)\end{tabular}\\
{   {}\begin{tabular}[c]{@{}l@{}}\textbf{PR-MPTCP$^+$} \\\textbf{~\cite{cao2016pr}}\end{tabular} }                                                                                                            & {  \color{red}\ding{55}}             &\color{ForestGreen}\ding{51}                                                                     & {   N/D}                                                             & {   N/D}                                                           & {   \begin{tabular}[c]{@{}l@{}}NS3,\\Disjoint paths,\\Client interfaces: \\WiFi and LTE\end{tabular}} &
{   \begin{tabular}[c]{@{}l@{}}PSNR, VQM,  SSIM,\\ Number of frames\\ received or dropped\end{tabular}}                                        &  Real-time\\
\rowcolor[HTML]{EFEFEF}{\begin{tabular}[c]{@{}l@{}}\textbf{SRMT}\\\textbf{ ~\cite{da2016preventing}}\end{tabular}}                                                               & {  \color{red}\ding{55}}         &\color{red}\ding{55}                                                                         & {   H.264/AVC}                                                            & {   \begin{tabular}[c]{@{}l@{}}N/D \end{tabular}}                                                           & {   \begin{tabular}[c]{@{}l@{}}Simulator N/D,\\ Client interfaces:\\ WiFi (802.11g), 3G\\ Or WiFi, ADSL\end{tabular}}                                                                                                                                                     & {   \begin{tabular}[c]{@{}l@{}}PSNR, SSIM,  Goodput, \\ Delay distribution\end{tabular}}                                                                                 & \begin{tabular}[c]{@{}l@{}}Live,\\ VoD\end{tabular} \\
{
\begin{tabular}[c]{@{}l@{}}\textbf{PR-SCTP}\\\textbf{~\cite{sanson2010pr}} \end{tabular}}                                                                                                                & {  \color{red}\ding{55}}                &\color{red}\ding{55}                                                                  & {   H.264/AVC}                                                             & {   N/D}                                                           & {   \begin{tabular}[c]{@{}l@{}}Realistic testbed,\\ FreeBSD framework,\\Netem\end{tabular}}                                                                                  &  {   \begin{tabular}[c]{@{}l@{}}Successful frame\\ transmission ratio, \\ Frame late index\end{tabular}}                                                                                                                     & Real-time \\
\rowcolor[HTML]{EFEFEF}{ \begin{tabular}[c]{@{}l@{}}\textbf{CMT-QA}\\ \textbf{~\cite{xu2013cmt}}\end{tabular}}                                                                    & {   ORP}               &\color{red}\ding{55}                                                                        & {   H.264/AVC}                                                             & {   N/D}                                                           & {  \begin{tabular}[c]{@{}l@{}}NS2,\\ Disjoint paths,\\ Client interfaces: \\3G, WiMAX (802.16)\\ and WiFi (802.11)\end{tabular}}                                                                                                                                                              & {   \begin{tabular}[c]{@{}l@{}}PSNR,  VQM,  SSIM,\\ Number of frames lost,\\ Out-of-order packets, \\ Average retransmission,\\ Average throughput\end{tabular}} &  Real-time\\ \hline

\end{tabular}}
\end{table*}

\begin{table*}
\centering
\resizebox{1.8\columnwidth}{!}{%
\begin{tabular}{|l|c|c|l|l|l|l|l|l}

\hline
\cellcolor[HTML]{C0C0C0}\textbf{Works}                                                      & \cellcolor[HTML]{C0C0C0}\textbf{\begin{tabular}[c]{@{}l@{}}Packet loss\\ Differentiation\\ Method\end{tabular}} & \cellcolor[HTML]{C0C0C0}\textbf{\begin{tabular}[c]{@{}l@{}}{\color{black}Fairness}\end{tabular}} & \cellcolor[HTML]{C0C0C0}\textbf{\begin{tabular}[c]{@{}l@{}}Video\\ Compression\end{tabular}} & \cellcolor[HTML]{C0C0C0}\textbf{\begin{tabular}[c]{@{}l@{}}Error\\ Concealment\end{tabular}} & \cellcolor[HTML]{C0C0C0}\textbf{\begin{tabular}[c]{@{}l@{}}{\color{black}Experimental}\\ {\color{black}Environment}\end{tabular}}                                                                                           & \cellcolor[HTML]{C0C0C0}\textbf{\begin{tabular}[c]{@{}l@{}}Performance\\ Metrics\end{tabular}}                                                              & \cellcolor[HTML]{C0C0C0}\textbf{\begin{tabular}[c]{@{}l@{}}{\color{black}Video}\\ {\color{black}Services}\end{tabular}}  \\ \hline

\rowcolor[HTML]{EFEFEF}{\begin{tabular}[c]{@{}l@{}}\textbf{CMT-DA}\\\textbf{~\cite{wu2015distortion}}\end{tabular}}                                                                                                    & {   ECN}     &\color{red}\ding{55}                                                                                  & {   H.264/SVC}                                                             & {   JSVM}                                                                  & {   \begin{tabular}[c]{@{}l@{}}EXata,\\Client interfaces: \\Cellular, WiFi and WiMAX \end{tabular}}                                                                                                                                                           & {   \begin{tabular}[c]{@{}l@{}}PSNR,\\ Inter-packet delay,\\ Goodput,\\ Loss rate, \\Out-of-order packets\end{tabular}}                                             & Real-time \\

{   \begin{tabular}[c]{@{}l@{}}\textbf{CMT-CA} \\ \textbf{~\cite{wu2016content}}\end{tabular}}                                                            & {   ZigZag }                                                                  &\color{ForestGreen}\ding{51}           & {   H.264/AVC}                                                             & {   FFmpeg}                                                                & {  \begin{tabular}[c]{@{}l@{}}EXata,\\Client interfaces: \\Cellular, WiFi and WiMAX \end{tabular}}                                                                                                                                                           & {   \begin{tabular}[c]{@{}l@{}}PSNR,\\ end-to-end delay,\\ CDF of\\ inter-packet delay,\\ Out-of-order packets, \\ Goodput,\\Number of\\ frames (I,P) lost\end{tabular}}                                                          &\begin{tabular}[c]{@{}l@{}}Real-time, \\ Live\end{tabular}  \\

\rowcolor[HTML]{EFEFEF}{\begin{tabular}[c]{@{}l@{}}{\color{black}\textbf{QUIC-FEC}}\\\textbf{ ~\cite{michel2018adding}}\end{tabular}}                                                            & {  \color{red}\ding{55} }                                                                  & {\color{black}\ding{51}}           & {  \begin{tabular}[c]{@{}l@{}}Not\\ compressed\end{tabular}}                                                             & {  N/D }                                                                & {  \begin{tabular}[c]{@{}l@{}}Mininet,\\NetEm,\\Two client interfaces \end{tabular}}                                                                                                                                                           & {   \begin{tabular}[c]{@{}l@{}}CDF of\\ data received, \\CDF of\\ total rebuffering time,\\one-way delay \end{tabular}}                                                          &\begin{tabular}[c]{@{}l@{}}Live\end{tabular}  \\

{   \begin{tabular}[c]{@{}l@{}}\textbf{Yap at al.}\\\textbf{~\cite{yap2012making}}\end{tabular} }                                                                                                             & {  \color{red}\ding{55}}      &\color{red}\ding{55}                                                                            & {   N/D}                                                           & {   N/D}                                                           & {   \begin{tabular}[c]{@{}l@{}}Realistic testbed, \\ Android and Ubuntu\\ framework,\\ Real access networks,\\Up to 10 client interfaces\\ composed of:\\3G (HSPA, CDMA),\\ WiMAX and\\ WiFi (802.11a/g)\end{tabular}}                                               & {   \begin{tabular}[c]{@{}l@{}}Throughput,\\ Goodput,\\ CPU load,\\ Power consumption,\\ RTT\end{tabular}}                                &  N/D\\

\rowcolor[HTML]{EFEFEF}{\begin{tabular}[c]{@{}l@{}}\textbf{MARS}\\\textbf{ ~\cite{sun2016mars}}\end{tabular}}                                                            & {  \color{red}\ding{55}}                        &\color{red}\ding{55}                                                          & {   N/D}                                                           & {   N/D}                                                           & {   \begin{tabular}[c]{@{}l@{}}Own JAVA socket\\  implementation,\\ Four client interfaces\\ composed of:\\ WiFi and LTE\end{tabular}}                                                                                       & {   \begin{tabular}[c]{@{}l@{}}Out-of-order packets, \\ Reordering delay,\\ end-to-end delay,\\ Throughput\end{tabular}}                                           &  Real-time\\
{   \begin{tabular}[c]{@{}l@{}}\textbf{BAG}\\ \textbf{~\cite{chebrolu2006bandwidth}}\end{tabular}}                                                            & {  \color{red}\ding{55}}             &\color{red}\ding{55}                                                                     & {   H.263}                                                           & {   N/D}                                                           & {   \begin{tabular}[c]{@{}l@{}}Realistic testbed,\\ Up to five client interfaces\\ composed of 3G\end{tabular}}                                                                                       & {   \begin{tabular}[c]{@{}l@{}} Delay distribution,\\ Lost frame ratio,\\ Required Bandwidth,\\ Video disruption\\ (glitch statistics)\end{tabular}}                                           &
{   \begin{tabular}[c]{@{}l@{}}Real-time\\ (interactive)\end{tabular}}\\
\rowcolor[HTML]{EFEFEF}{\begin{tabular}[c]{@{}l@{}}\textbf{Corbillon et al.}\\\textbf{~\cite{corbillon2016cross}}\end{tabular} }                                                                                                        & {  \color{red}\ding{55}}      &\color{ForestGreen}\ding{51}                                                                            & {   HEVC}                                                                  & {   FFmpeg}                                                                & {   \begin{tabular}[c]{@{}l@{}}Own C++\\ implementation,\\ Disjoint paths,\\ Client interfaces:\\ 3G and WiFi\end{tabular}}                                                                                                & {   \begin{tabular}[c]{@{}l@{}}PSNR,\\ MS-SSIM,\\ Received frame ratio,\\ Received tile ratio\end{tabular}}                               &  \begin{tabular}[c]{@{}l@{}} Live, \\ VoD\end{tabular}\\

\begin{tabular}[c]{@{}l@{}}\textbf{Ojanper{\"a} et al.}\\\textbf{~\cite{ojanpera2016network}}\end{tabular}                                                                                                                                                  &\color{red}\ding{55}       &\color{ForestGreen}\ding{51}                                                                                                  & H.264/AVC                                                                                    & FFmpeg                                                                                       & \begin{tabular}[c]{@{}l@{}}Realistic testbed,\\ Ubuntu framework,\\ Client interfaces:\\ WiFi (802.11g) and\\ WiFi (802.11a)\end{tabular}                & \begin{tabular}[c]{@{}l@{}}Throughput,\\ Quality switch,\end{tabular}                                                                                                     &  N/D\\

\rowcolor[HTML]{EFEFEF}{\begin{tabular}[c]{@{}l@{}}\textbf{GALTON}\\ \textbf{~\cite{wu2015goodput}}\end{tabular}}                                                               & {   ZigZag }                &\color{red}\ding{55}                                                             & {   H.264/SVC}                                                             & {   JSVM}                                                                  & {   \begin{tabular}[c]{@{}l@{}}EXata,\\  Client interfaces:\\ WiFi, WiMAX, \\Cellular (HSDPA)\\ or multiple wired interfaces\end{tabular} }                                                                                                                                                           & {   \begin{tabular}[c]{@{}l@{}}PSNR,\\ Goodput,\\ end-to-end delay,\\ Loss rate\end{tabular}}                                             &  Real-time\\
\begin{tabular}[c]{@{}l@{}}\textbf{FRA-JSCC}\\\textbf{ ~\cite{wu2013joint}}\end{tabular}                                                                                                         &\color{red}\ding{55}    &\color{red}\ding{55}                                                                                                    & H.264/SVC                                                                                    & JSVM                                                                                         & \begin{tabular}[c]{@{}l@{}}EXata,\\  Client interfaces:\\ WiFi (802.11b), WiMAX and\\ Cellular\end{tabular}                                                                                                                                                                                  & \begin{tabular}[c]{@{}l@{}}PSNR,\\ end-to-end delay,\\ Loss rate, \\ Available bandwidth\end{tabular}                                                                               & Real-time \\

\rowcolor[HTML]{EFEFEF} \begin{tabular}[c]{@{}l@{}} {\color{black}\textbf{Deng et al.}}\\{\color{black}\textbf{~\cite{deng2021cross}}}\end{tabular}                                                                                                         &\color{red}\ding{55}    &\color{red}\ding{55}                                                                                                    & H.264/AVC                                                                                    & JM                                                                                         & \begin{tabular}[c]{@{}l@{}}Matlab,\\  Client interfaces:\\ LTE and WiFi (802.11ac)\end{tabular}                                                                                                                                                                                  & \begin{tabular}[c]{@{}l@{}} Selected MCS and segment bitrates, \\ PSNR, \\end-to-end delay,\\ Playback bitrate and rebuffering\end{tabular}                                                                               & N/D \\
{   \begin{tabular}[c]{@{}l@{}}\textbf{MP-DASH}\\\textbf{ ~\cite{han2016mp}}\end{tabular}}                                                               & {  \color{red}\ding{55}}                &\color{red}\ding{55}                                                                  & {   H.264/AVC}                                                             & {   N/D}                                                           & {   \begin{tabular}[c]{@{}l@{}}Realistic testbed,\\ Ubuntu framework,\\ Real access networks,\\ Client interfaces: \\ WiFi and Cellular\end{tabular}}                                                            & {   \begin{tabular}[c]{@{}l@{}}Throughput,\\ Energy consumption,\\ Download time,\\ Average 3G traffic\end{tabular}}                     &  N/D\\
\rowcolor[HTML]{EFEFEF}{ \begin{tabular}[c]{@{}l@{}}\textbf{Nam et al.}\\\textbf{~\cite{nam2016towards}} \end{tabular}}                                                                                                              & {  \color{red}\ding{55}}     &\color{ForestGreen}\ding{51}                                                                             & {   H.264/AVC}                                                             & {   N/D}                                                           & {   \begin{tabular}[c]{@{}l@{}}Realistic testbed, \\ Ubuntu framework,\\ Real MPEG-DASH platform, \\ Mininet over WiFi for SDN, \\ Real access networks,\\Client interfaces:\\ WiFi (802.11g)\\ and WiFi (802.11a) \end{tabular}} &
{   \begin{tabular}[c]{@{}l@{}}Played bit rate,\\ Rebuffering,\\ Out-of-order packets\end{tabular}}                                        & Real-time \\

{   \begin{tabular}[c]{@{}l@{}}\textbf{CMT-CL/FD}\\ \textbf{~\cite{xu2015cross}}\end{tabular}}                                                              & {   RTX}  &\color{ForestGreen}\ding{51}                                                                                   & {   N/D}                                                           & {   N/D}                                                           & {   \begin{tabular}[c]{@{}l@{}}NS2, \\Disjoint paths,\\ Server and client interfaces:\\ 3G (WCDMA), WiMAX  and \\WiFi (802.11) \end{tabular}}                                                                                                                                                             & {   \begin{tabular}[c]{@{}l@{}}PSNR,\\ Video buffer underflow,\\ Throughput, \\Fairness test \end{tabular}}                                         & Real-time \\
\rowcolor[HTML]{EFEFEF}{  \begin{tabular}[c]{@{}l@{}}{\color{black} \textbf{OLS~\cite{xing2021low}} }\end{tabular}}                                                              &\color{red}\ding{55}    &\color{red}\ding{55}                                                                                   & {   N/D}                                                           & {   N/D}                                                           & {  \begin{tabular}[c]{@{}l@{}}Realistic testbed, \\Disjoint paths,\\ Server and client interfaces:\\ 4G and WiFi \end{tabular}}                                                                                                                                                             & {   \begin{tabular}[c]{@{}l@{}}RTT,\\ Out-of-order queue size,\\ Throughput, \\Bytes received \end{tabular}}                                         & Live \\ \hline

\end{tabular}}
\end{table*}

\subsection{Packet Loss Differentiation}{\label{sec: Packet Loss Differentiation}}

A packet loss differentiation method can distinguish congestion losses from wireless losses. In heterogeneous wireless networks, packet losses due to lost channels, handover, noise or interface in the wireless network occur more than losses due to congestion~\cite{xu2013cmt}. Identifying {\color{black}the} reason for losses is essential.
For example, if losses occur because of congestion in the network, then retransmission or adding more FEC redundancy pushes worse congestion and more losses~\cite{wallace2012review}~(Section~\ref{sec: How to protect the packet?}). But, decreasing cwnd mitigates congestion.
On the other hand, if losses occur because of
wireless lossy network, then decreasing cwnd drops goodput sharply~(Section~\ref{sec:Transport Layer Approaches}). But, adding more FEC redundancy leads to better recovery. Therefore, with an accurate loss differentiation method could react properly to the network situation.

In our surveyed works, MPRTP~\cite{singh2013mprtp} categorizes a path as a lossy one if feedback reports show only transmission losses and no discards~(overdue packets) over that path. A path is categorized as a mildly congested one if feedback reports show both transmission losses and discards either in a single or consecutive reports. If this behavior occurs in more than three consecutive reports, it means that the path is congested.
CMT-QA~\cite{xu2013cmt} handles the packet loss differentiation by proposing optimal retransmission policy~(ORP). In ORP, when a loss occurs, $(RTT/cwnd)$ is calculated, and the result would be compared with a threshold. This threshold is defined as path quality. Therefore, if $(RTT/cwnd)$ is more than the threshold, the loss is due to wireless loss. Otherwise, it is a congestion loss. If losses occur more than once and consecutively, then congestion is the reason.
CMT-CL/FD~\cite{xu2015cross} proposed loss-cause
dependent retransmission~(RTX) policy. In RTX, two cases are considered; 1) When the loss is detected by fast retransmission. Thus, the residual capacity of the path is calculated. If it is a positive value, it means that the path is underused and wireless loss {\color{black}has occurred}. Otherwise, if the residual path value is negative, congestion is the reason.  2) When the loss is detected by expiring RTO. In this case, the path is failed or severe congestion {\color{black}has occurred}.
CMT-DA~\cite{wu2015distortion}, MPLOT~\cite{sharma2008mplot} and~\cite{correia2012optimal} utilize Explicit Congestion Notification~(ECN)\nomenclature{ECN}{Explicit Congestion Notification} to distinguish loss differentiation. ECN is defined by IETF~\cite{ramakrishnan2001rfc} in 2001. ECN-aware routers informs congestion by setting a mark in the IP header, without dropping any packet.
{\color{black}The work proposed in \cite{go2019hybrid} differentiates losses utilizing the Spike scheme~\cite{cen2003end}. In Spike, the Relative One-way Trip Time~(ROTT)\nomenclature{ROTT}{Relative One-way Trip Time} is the time a packet takes to transmit from the sender to the receiver, so-called relative due to the clock skew between them. It is used to identify the actual connection state as: when the connection stands in the \textit{spike state}, losses are caused by congestion; otherwise, they are caused by wireless losses. The spike state derives its name from the fact that plots of ROTT versus time present spikes during the periods of congestion.}
BEMA~\cite{wu2016bandwidth}, ADMIT~\cite{wu2016streaming},
{\color{black}DEAM~\cite{Wu2019Energy}}, {\color{black}EDAM~\cite{7937943}},  GALTON~\cite{wu2015goodput} and CMT-CA~\cite{wu2016content} use ZigZag scheme introduced in~\cite{cen2003end}. ZigZag classifies losses as wireless
based on the number of losses and on the difference between relative one-way trip times and the mean of relative one-way trip times.
For further information about the effect of different types of losses like random loss or bursty loss on video streaming quality refer to~\cite{apostolopoulos2000reliable}.

\subsection{Fairness}
Table~\ref{compression} summarizes the surveyed works that consider fairness, which were previously introduced in Section~\ref{sec: MultiPath Mobile Video Streaming Approaches}. These works address fairness in terms of consumed resources by the proposed {\color{black}congestion control mechanisms~(e.g., \cite{afzal2018novel}, MP-H2~\cite{nikravesh2019MP-H2},} MPRTP~\cite{singh2013mprtp}, {\color{black}\cite{afzal2021multipath}}, MPLOT~\cite{sharma2008mplot}, CMT-CL/FD~\cite{xu2015cross},) as adopted by TFRC~(e.g., {\color{black}\cite{go2019hybrid},} BEMA~\cite{wu2016bandwidth}, CMT-CA~\cite{wu2016content}) or in terms of  MPTCP coupled congestion control~(e.g., ADMIT~\cite{wu2016streaming},
{\color{black}DEAM~\cite{Wu2019Energy},} {\color{black}EDAM~\cite{7937943}},
MPTCP-SD/PR~\cite{diop2012qos}, PR-MPTCP$^+$~\cite{cao2016pr},  {\color{black}QUIC-FEC~\cite{michel2018adding},} Corbillon et al.~\cite{corbillon2016cross}, Ojanper{\"a} et al.~\cite{ojanpera2016network}, Nam et al.~\cite{nam2016towards}). Besides, in our surveyed works, Freris et al.~\cite{freris2013distortion} considers user fairness of network resources.

\subsection{Video Compression and Error Concealment}

Several video codecs were used in the surveyed works cited in Table~\ref{compression}, such as H.263~\cite{rijkse1996h}, H.264/AVC~\cite{h264avc}, H.264/SVC~\cite{h264svc}, HEVC\nomenclature{HEVC}{High Efficiency Video Coding}~\cite{hevc}, and SHVC~\cite{shvc}. After video transmission, if protection methods are not able to recover the lost packets, the decoder itself can employ error concealment. 
This way, decoder exploits correlations in the previously received video sequence to conceal the lost information. JM, for instance, performs frame copy while FFmpeg performs temporal interpolation.
According to~\cite{chang2012network}, in case of whole-frame losses, when isolated B-frames were lost and concealed by either JM or FFmpeg, about 40\% of the losses were not even noticed by observers.
{\color{black}Our surveyed works used JM\footnote{{\color{black}http://iphome.hhi.de/suehring/tml/download/}}, x264\footnote{{\color{black}http://www.videolan.org/developers/x264.html}}, JSVM\footnote{{\color{black}https://github.com/floriandejonckheere/jsvm}}, FFmpeg\footnote{{\color{black}https://ffmpeg.org}}, HM\footnote{{\color{black}https://hevc.hhi.fraunhofer.de/svn/svn\_HEVCSoftware/}} for error concealment.}

\subsection{Experimental environment}
Table~\ref{compression} shows that experimental evaluation is mostly dominated by  network simulators, such as {\color{black}OPNET}\footnote{{\color{black}https://opnetprojects.com/opnet-network-simulator/}}, {\color{black}Dummynet}\footnote{{\color{black}http://info.iet.unipi.it/~luigi/dummynet/}}, {\color{black}NS2}\footnote{{\color{black}http://www.isi.edu/nsnam/ns/}},  {\color{black}NS3}\footnote{{\color{black}https://www.nsnam.org/}}, {\color{black}EXata}\footnote{{\color{black}https://www.scalable-networks.com/products/exata-network-emulator-software/}},  {\color{black}NetEm}\footnote{{\color{black}https://openwrt.org/docs/guide-user/network/traffic-shaping/sch\_netem}}, {\color{black}MATLAB}\footnote{{\color{black}https://www.mathworks.com/products/matlab.html}}. Only few works, mainly due to costs, scale, and scope, carried their evaluation on real testbeds. Wireless-enabled network emulators like Mininet-WiFi~\cite{mn-wifi} are also another category of experimental environments. {\color{black} We also cover some additional implementation details. For example, which type of network interfaces are used in experiments, or if the simulation uses disjoint paths~(no common link or node). Using disjoint paths improves bandwidth aggregation and has the benefit of additional fault-tolerance compared with non-disjoint paths~\cite{singh2015survey}, altogether contributing to the users video experience.}

\subsection{ Performance Metrics}
Several performance metrics were used in the surveyed works cited in Table~\ref{compression}. Most of them are explained in Section~\ref{sec:Transport Layer Approaches}. We have added some additional video quality metrics, such as Peak Signal-to-Noise Ratio~(PSNR)\nomenclature{PSNR}{Peak Signal-to-Noise Ratio}, Video  Quality
Metric~(VQM)\nomenclature{VQM}{Video  Quality
Metric}, Structural SIMilarity~(SSIM)~\cite{wang2004image}\nomenclature{SSIM}{Structural SIMilarity}, MultiScale Structural SIMilarity~(MS-SSIM)~\cite{wang2003multiscale}\nomenclature{MS-SSIM}{MultiScale Structural SIMilarity}, and Subjective Assessment Methodology for VIdeo Quality~(SAMVIQ)~\cite{blin2006new}\nomenclature{SAMVIQ}{Subjective Assessment Methodology for VIdeo Quality}.

\subsection{Video Services}
The last column of {\color{black}Table~\ref{compression} presents for each of surveyed works  which type of video service was considered by the authors, such as VoD, live, and {\color{black}real-time as an upper set including interactive video streaming applications}.  As discussed in Sections~\ref{sec: Benefits and Challenges} and~\ref{sec: MultiPath Mobile Video Streaming Approaches}, each type of video {\color{black}service} has different QoS requirements, such as delay sensitivity.}

\section{Open Research Issues}{\label{sec: Open Research Issues}}
Many research avenues around {\color{black} multipath wireless} video streaming are open. In the following, we present some relevant evolving aspects along {\color{black}with} potential future work opportunities. \\
\noindent \textbf{Standardization developments.}~{\color{black}MMT is a recent standard protocol with potential abilities discussed in the survey. 
Future work could evaluate the performance of  MMT over MPTCP or {\color{black}MPQUIC} utilizing multipath scheduling methods defined in these protocols for video streaming over heterogeneous networks.}
 HTTP/2 provides noticeable features such as the ability to push content in advance and frame multiplexing. 
Therefore, further attention on multipath delivery over HTTP/2 shall be pursuit~\cite{frommgen2017programming}. Another standards related topic would be the use of HEVC, especially SHVC, which do not seem to be widespread in the  networking literature despite being  widespread in the video coding community.\\

\noindent \textbf{Network Softwarization.}~Attempts to integrate SDN with multipath video streaming~(Section~\ref{sec: Network Layer Approaches}) promise effectiveness for path-aware strategies due to its ability to programmatically define the {\color{black} end-to-end} network behaviour. While OpenFlow is considered the mostly accepted interface between control and data planes~\cite{kreutz2015software}, alternative means for southbound interaction of controllers and datapath devices~(e.g., P4 programmable data planes),  {\color{black}including SDN protocol extensions relevant for wireless communications~(e.g.,~\cite{7437151, Lee:2014:MME:2609908.2609948}) deserve further research efforts.
  SDN and NFV as enabling technologies of multi-domain network service orchestration~\cite{SARAIVADESOUSA201969} will certainly keep attracting research attention and will play a critical role in the realization of multipath strategies for video streaming and other types of services.  } \\

\noindent \textbf{5G.}~As Fifth generation~(5G)\nomenclature{5G}{Fifth Generation} cellular wireless are rolling-out, services that require extreme bandwidth and ultra-low latency are expected to benefit from multi-homing~\cite{dandachi2017multihoming} and eventual multi-path communications~\cite{ibnalfakih2016multi,karimi2017evaluating}. Furthermore, there are several studies and efforts for MPTCP operation in 5G~\cite{I-D.purkayastha-mptcp-considerations-for-nextgen,karimi2017evaluating}. In addition, studies show that emerging technologies such as SDN to MPTCP in 5G networks could improve the transmission performance due to SDN capability to control the subflows by monitoring network condition~\cite{lei2018ndn,barakabitze2018novel}.  {\color{black}  Therefore, multi-path video streaming over 5G networks en route to 6G is an important research area where innovative solutions may leverage  diverse multihoming solutions along {\color{black}with} SDN/NFV-based technologies.} \\

\noindent \textbf{WiFi Evolution.}~In the near future~\cite{ghasempour2017ieee,oughton2021revisiting,khorov2020current}, significant enhancements in WiFi communication focus on increasing the performance of wireless networks, such as the proposals for 802.11ad and 802.11ay on 60 GHz, 802.11ax concurrently on 2.4 GHz and 5 GHz,{\color{black}~or 802.11be\footnote{{\color{black}https://www.ieee802.org/11/Reports/ehtsg\_update.htm}} in the 2.4, 5 and 6 GHz frequency bands}. Such technologies aim to achieve high throughput and ultra-low latency. Moving forward, more open questions stand in exploring the impact of these WiFi technologies on multipath video streaming.\\

{\color{black}
\noindent \textbf{IoT networks.}~Many Internet of Things~(IoT)~\cite{sultana2019choice} applications require multimedia communications dealing with high bandwidth and low latency requirements, for example, real-time smart city monitoring services~\cite{Costa2020CMSS,wang2017joint,cheng2017situation} such as surveillance camera systems, vehicle tracking, face detection, traffic control, environment monitoring, object motion detection, and more. Furthermore, some specific deployments, such
as those employing Low-Power Wide-Area Network~(LPWAN)-based technologies\nomenclature{LPWAN}{Low-Power Wide-Area Network}~\cite{augustin2016study}, suffer from major limitations in data transmission capacity, throughput, and supported packet length. Therefore, integration of multipath strategies with IoT routing leads to improvements in bandwidth aggregation and consequently latency reduction and increasing the lifetime of IoT devices in the network. However, to have an appropriate multipath solution, several factors should be considered regarding routing and path selection due to the decentralized and dynamic nature of these types of networks~\cite{medjiah2012streaming}. For example,  the routing decisions need to be made in a distributed manner and in real-time since the offline routing processes cannot react to topology variations and result in forwarding packets to disconnected routes. Moreover, routing processes should assign the shortest paths as the primary route for delivery of delay-sensitive traffic types.  Also, the selection of the neighboring IoT device for the routing process must take into account the energy of the devices, Their distances from the destination and QoS requirements.}\\

\noindent \textbf{Energy considerations.}~Power efficiency is an essential requirement. The work~\cite{li2012greentube} shows high power consumption by LTE when video {\color{black}streams} over HTTP. Energy consumption even increases more by using multiple network interfaces{\color{black}~\cite{wu2018energy}}. Therefore, optimizing power consumption needs further attention in the proposed approaches.\\

\noindent \textbf{Security.}~Multipath delivery could mitigate some security threats inherently through the use of alternative paths throughout the network. There is little work in the scope of multipath multimedia streaming security. At the same time,  Digital Rights Management~(DRM)\nomenclature{DRM}{Digital Rights Management} and the license issues are also security related issues critical   for some video services.\\

\noindent \textbf{Mobility and Internet of Vehicles~(IoV).}~Although terminal mobility, velocity, motion degree and related mobile aspects are factors affecting video quality, they are rarely discussed in the literature. This type of considerations are key in the delivery of wireless video in mobile environments in {\color{black}the} scope of Intelligent Transport Systems~(ITS) \nomenclature{ITS}{Intelligent Transport Systems}~\cite{8167677} and Vehicle-to-everything~(V2X)\nomenclature{V2X}{Vehicle-to-everything} communications~\cite{7859348}.\\

\noindent \textbf{Machine Learning~(ML) and Artificial Intelligence~(AI).}~{\color{black} Leveraging artificial intelligence and machine learning methods are increasingly becoming key tools for network and service optimization~\cite{latah2018artificial} and can be used for advanced scheduling and adaptive coding decisions~\cite{Pensieve}. The importance of machine learning approaches to improve video quality has been recognized by  Netflix, which proposed a new video quality assessment method named Video Multimethod Assessment  Fusion ~(VMAF)\nomenclature{VMAF}{Video Multimethod Assessment  Fusion}. VMAF is a  machine learning-based model that is trained and tested using the results of a subjective experiment to deliver the best video quality to the user~\cite{trestian2018seamless}. Besides, there are also several machine learning-based efforts to learn QoS measurements~\cite{al2018learnqos} or QoE from user reactions~\cite{vasilev2018predicting,alzahrani2018use} to solve various optimization and control problems for a single path video streaming. In our state-of-art, there are some approaches utilizing machine learning systems learning QoS from {\color{black}the} user device and using it for multipath scheduling decisions~\cite{xing2012rate,xing2014real}. Thus, similarly, an interesting solution could be utilizing machine learning systems learning QoE from user reactions and using it for multipath scheduling decisions.}

\section{Concluding Remarks}{\label{sec: Conclusions}}
One promising approach to improve QoE for wireless video streaming is multipath delivery, which increases available bandwidth, resilience and load balancing. From the industry perspective,  several companies have implemented
their own multipath approaches, such as {\color{black}AVAYA\footnote{{\color{black}https://downloads.avaya.com/css/P8/documents/100134063}}  and Cisco\footnote{{\color{black}https://www.cisco.com/c/en/us/td/docs/switches/lan/catalyst4500/12-2/31sg/configuration/guide/conf/channel.pdf}}}. Apple and {\color{black}Samsung}  have also started to support multipath on {\color{black}smartphones~\cite{bonaventure2016multipath}} for different services like voice recognition, or to increase the download speed of specific software packages.   Therefore, we expect a growth in multipath video streaming in the near future. However, there are still many issues to be solved, especially for  solutions which are not compatible with each other or that require  changes in servers and/or clients, {\color{black}or network equipment software/hardware.}

In this work, we have provided an in-depth survey of multipath {\color{black}wireless} video streaming proposals, covering over forty relevant pieces of work.
We have categorized and explained the surveyed works based on the layer in the protocol stack and the {\color{black} dominating protocols/features. Network equipment compatibility has been also  discussed. In addition, scheduling, resilience and  path selection techniques are presented.  Finally, we have studied different key}  methods, such as packet loss differentiation, video compression, error concealment, etc.

 {\color{black}Several key aspects} should be highlighted  when designing a multipath video streaming approach. We observe that in order to overcome these challenges, packet scheduling strategies should consider several factors.

The first one is the layer dependency that is discussed in Section~\ref{sec: MultiPath Mobile Video Streaming Approaches}, and the surveyed works are summarized and categorized based on it in Table~\ref{layered classification}.
Research shows that the scheduler has better {\color{black}decision capabilities} when it has complete and accurate information about video contents, packet delivery deadlines, playback buffer, RTT, available bandwidth, and other network information. Implementing scheduling functions on a specific layer could access only a part of this information. Therefore, {\color{black}cross-layer approaches benefits from} their ability of gathering information of different layers for improved scheduling.

Another important factor to design a scheduler is client and network equipment compatibility. This topic is also discussed in Section~\ref{sec: MultiPath Mobile Video Streaming Approaches} and summarized in Table~\ref{layered classification}.  While the most flexible case to implement is when only client modification is required, some approaches require changing the server, or both server and client, or also the network infrastructure. It is also important to note the ability to traverse middleboxes. 

{\color{black}
Fundamental aspects to be considered to improve the performance of scheduling functions  discussed in Section~\ref{sec: Scheduler, Resilience, and Routing Functions in Multipath Video} and summarized in Tables~\ref{table:Scheduler Functions-application layer},~\ref{table:Scheduler Functions-transport layer},~\ref{table:Scheduler Functions-network layer},~\ref{table:Scheduler Functions-cross layer} include which packet should be sent next, through which path, and with which type of error protection.
An adequate scheduling strategy should be content-aware, and path-aware, as well as it should utilize a proper channel or source level packet protection methods. }
Such scheduling approaches improve QoE, bandwidth aggregation, load balancing, HOL blocking and out-of-order packet issues.

Section~\ref{sec: Analysis and Comparison of Candidate Methods and Techniques} and Table~\ref{compression} show some related methods used in the surveyed works and the key performance indicators to evaluate the approaches. One observation is that while calculating video quality metrics is very useful to understand the performance of each approach,  {\color{black}many} of the works only consider network QoS metrics without assessing video performance in terms of QoE as the key performance indicators from an end-user perspective.

The road ahead towards the broad realization of video streaming over {\color{black}multipath wireless} solutions is not without issues. In Section~\ref{sec: Open Research Issues}, we overview a series of open challenges and point to some  research opportunities.

\section*{Acknowledgment}
This work is part of the results obtained through the project ``Hyper Realistic Media'', sponsored by Samsung Eletrônica da Amazônia Ltda., in the framework of law No. 8,248/91. We thank CNPq~(Brazilian National Council for Scientific and  Technological Development) for grants \#141778/2015-6, \#310930/2016-2 and \#311378/2020-0.

\bibliography{elsarticle-template}

\begin{thebibliography}{100}
\expandafter\ifx\csname url\endcsname\relax
  \def\url#1{\texttt{#1}}\fi
\expandafter\ifx\csname urlprefix\endcsname\relax\def\urlprefix{URL }\fi
\expandafter\ifx\csname href\endcsname\relax
  \def\href#1#2{#2} \def\path#1{#1}\fi

\bibitem{jarschel2011evaluation}
M.~Jarschel, D.~Schlosser, S.~Scheuring, T.~Ho{\ss}feld, {An evaluation of QoE
  in cloud gaming based on subjective tests}, in: Innovative mobile and
  internet services in ubiquitous computing (imis), 2011 fifth international
  conference on, IEEE, 2011, pp. 330--335.

\bibitem{index2019cisco}
C.~V.~N. Index, {Cisco Visual Networking Index: Global Mobile Data Traffic
  Forecast Update, 2017–2022}, Tech. RepAccessed:21-JUN-2021.

\bibitem{sani2017adaptive}
Y.~Sani, A.~Mauthe, C.~Edwards, {Adaptive bitrate selection: A survey}, IEEE
  Communications Surveys \& Tutorials 19~(4) (2017) 2985--3014.

\bibitem{qadir2015exploiting}
J.~Qadir, A.~Ali, K.-L.~A. Yau, A.~Sathiaseelan, J.~Crowcroft, {Exploiting the
  power of multiplicity: a holistic survey of network-layer multipath}, IEEE
  Communications Surveys \& Tutorials 17~(4) (2015) 2176--2213.

\bibitem{singh2015survey}
S.~K. Singh, T.~Das, A.~Jukan, {A survey on internet multipath routing and
  provisioning}, IEEE Communications Surveys \& Tutorials 17~(4) (2015)
  2157--2175.

\bibitem{li2016multipath}
M.~Li, A.~Lukyanenko, Z.~Ou, A.~Yla-Jaaski, S.~Tarkoma, M.~Coudron, S.~Secci,
  {Multipath transmission for the internet: A survey}, IEEE Communications
  Surveys Tutorials, vol. PP~(99) (2016) 1--41.

\bibitem{domzal2015survey}
J.~Domzal, Z.~Dulinski, M.~Kantor, J.~Rzkasa, R.~Stankiewicz, K.~Wajda,
  R.~Wojcik, {A survey on methods to provide multipath transmission in wired
  packet networks}, Computer Networks 77 (2015) 18--41.

\bibitem{addepalli2013heterogeneous}
S.~Addepalli, H.~G. Schulzrinne, A.~Singh, G.~Ormazabal, {Heterogeneous access:
  Survey and design considerations}, Tech. rep., Department of Computer
  Science, Columbia University (2013).

\bibitem{habak2015bandwidth}
K.~Habak, K.~A. Harras, M.~Youssef, {Bandwidth aggregation techniques in
  heterogeneous multi-homed devices: A survey}, Computer Networks 92 (2015)
  168--188.

\bibitem{barakabitze2018qualitysdn}
A.~A. Barakabitze, I.-H. Mkwawa, L.~Sun, E.~Ifeachor, {QualitySDN: Improving
  Video Quality using MPTCP and Segment Routing in SDN/NFV}, in: IEEE
  Conference on Network Softwarization and Workshops (NetSoft), 2018, pp.
  182--186.

\bibitem{herguner2017towards}
K.~Herguner, R.~S. Kalan, C.~Cetinkaya, M.~Sayit, {Towards QoS-aware routing
  for DASH utilizing MPTCP over SDN}, in: Network Function Virtualization and
  Software Defined Networks (NFV-SDN), IEEE Conference on, 2017, pp. 1--6.

\bibitem{kreutz2015software}
D.~Kreutz, F.~M. Ramos, P.~E. Verissimo, C.~E. Rothenberg, S.~Azodolmolky,
  S.~Uhlig, {Software-defined networking: A comprehensive survey}, Proceedings
  of the IEEE 103~(1) (2015) 14--76.

\bibitem{hasan2017survey}
M.~Z. Hasan, H.~Al-Rizzo, F.~Al-Turjman, {A survey on multipath routing
  protocols for QoS assurances in real-time wireless multimedia sensor
  networks}, IEEE Communications Surveys \& Tutorials 19~(3) (2017) 1424--1456.

\bibitem{sha2013multipath}
K.~Sha, J.~Gehlot, R.~Greve, {Mutipath routing techniques in wireless sensor
  networks: A survey}, Wireless personal communications 70~(2) (2013) 807--829.

\bibitem{Zhang:2012:SPL:2365364.2365643}
X.~Zhang, H.~Hassanein, A survey of peer-to-peer live video streaming schemes -
  an algorithmic perspective, Comput. Netw. 56~(15) (2012) 3548--3579.

\bibitem{liu2008survey}
Y.~Liu, Y.~Guo, C.~Liang, {A survey on peer-to-peer video streaming systems},
  Peer-to-peer Networking and Applications 1~(1) (2008) 18--28.

\bibitem{hodroj2021survey}
A.~Hodroj, M.~Ibrahim, Y.~Hadjadj-Aoul, A survey on video streaming in
  multipath and multihomed overlay networks, IEEE Access 9 (2021) 66816--66828.

\bibitem{mao2006mrtp}
S.~Mao, D.~Bushmitch, S.~Narayanan, S.~S. Panwar, {MRTP: a multiflow real-time
  transport protocol for ad hoc networks}, IEEE Transactions on Multimedia
  8~(2) (2006) 356--369.

\bibitem{singh2013mprtp}
V.~Singh, S.~Ahsan, J.~Ott, {MPRTP: multipath considerations for real-time
  media}, Proceedings of the 4th ACM Multimedia Systems Conference (2013)
  190--201.

\bibitem{xing2014real}
{M. Xing, S. Xiang, and L. Cai}, {A real-time adaptive algorithm for video
  streaming over multiple wireless access networks}, IEEE Journal on Selected
  Areas in communications 32~(4) (2014) 795--805.

\bibitem{sharma2008mplot}
V.~Sharma, S.~Kalyanaraman, K.~Kar, K.~Ramakrishnan, V.~Subramanian, {MPLOT: A
  transport protocol exploiting multipath diversity using erasure codes}, 27th
  Conference on Computer Communications, IEEE INFOCOM (2008) 121--125.

\bibitem{apostolopoulos2000reliable}
J.~G. Apostolopoulos, {Reliable video communication over lossy packet networks
  using multiple state encoding and path diversity}, Photonics West
  2001-Electronic Imaging (2000) 392--409.

\bibitem{chaudhari2021survey}
S.~Chaudhari, A survey on multipath routing techniques in wireless sensor
  networks, International Journal of Networking and Virtual Organisations
  24~(3) (2021) 267--328.

\bibitem{kaur2020survey}
T.~Kaur, D.~Kumar, A survey on qos mechanisms in wsn for computational
  intelligence based routing protocols, Wireless Networks 26~(4) (2020)
  2465--2486.

\bibitem{more2020analytical}
S.~More, U.~Naik, Analytical review and study on multipath routing protocols,
  Control \& Cybernetics 46~(4) (2020) .

\bibitem{trestian2018seamless}
R.~Trestian, I.-S. Comsa, M.~F. Tuysuz, {Seamless multimedia delivery within a
  heterogeneous wireless networks environment: Are we there yet?}, IEEE
  Communications Surveys \& Tutorials 20~(2) (2018) 945--977.

\bibitem{singhmultipath2012}
V.~Singh, T.~Karkkainen, J.~Ott, S.~Ahsan, {Multipath RTP (MPRTP)}, IETF Draft,
  draft-singh-avtcore-mprtp (2012) .

\bibitem{deconinck-quic-multipath-07}
Q.~D. Coninck, O.~Bonaventure,
  \href{https://datatracker.ietf.org/doc/html/draft-deconinck-quic-multipath-07}{{Multipath
  Extensions for QUIC (MP-QUIC)}}, Internet-Draft
  draft-deconinck-quic-multipath-07, Internet Engineering Task Force, work in
  Progress (May 2021).
\newline\urlprefix\url{https://datatracker.ietf.org/doc/html/draft-deconinck-quic-multipath-07}

\bibitem{yuste2015understanding}
L.~B. Yuste, F.~B. Segui, M.~A.~M. Climent, H.~Melvin, {Understanding timelines
  within MPEG standards}, in: Communications Surveys and Tutorials, IEEE
  Communications Society, Vol.~18, Institute of Electrical and Electronics
  Engineers (IEEE), 2015, pp. 368--400.

\bibitem{Schulzrinne1992}
H.~Schulzrinne, {A Transport Protocol for Real-Time Applications} (1992).

\bibitem{lim2014new}
Y.~Lim, S.~Aoki, I.~Bouazizi, J.~Song, {New MPEG transport standard for next
  generation hybrid broadcasting system with IP}, IEEE Transactions on
  Broadcasting 60~(2) (2014) 160--169.

\bibitem{Information2007Part1}
{Information technology -- Generic coding of moving pictures and associated
  audio information: Part 1 Systems}, ISO/IEC 13818-1 (2007) .

\bibitem{yie2016method}
C.~K. Yie, Y.~J. Lee, {Method for hybrid delivery of MMT package and content
  and method for receiving content} (2016) .{US Patent 9414123B2}.

\bibitem{lim2013mmt}
Y.~Lim, K.~Park, J.~Y. Lee, S.~Aoki, G.~Fernando, {MMT: An emerging MPEG
  standard for multimedia delivery over the internet}, IEEE MultiMedia 20~(1)
  (2013) 80--85.

\bibitem{parmar2012real}
H.~Parmar, M.~Thornburgh, Real-time messaging protocol (rtmp) specification,
  Adobe specifications, December (2012) .

\bibitem{hevc}
{High Efficiency Video Coding}, ITU-T Rec. H.265 and ISO/IEC 23008-2 (2018) .

\bibitem{BrueckMark2010Apparatus}
D.~F. Brueck, M.~B. Hurst, {Apparatus, system, and method for multi-bitrate
  content streaming}, {US Patent 7818444B2} (2010).

\bibitem{fielding1999hypertext}
R.~Fielding, J.~Gettys, J.~Mogul, H.~Frystyk, L.~Masinter, P.~Leach,
  T.~Berners-Lee, {Hypertext transfer protocol--HTTP/1.1} (1999).

\bibitem{seufert2015survey}
M.~Seufert, S.~Egger, M.~Slanina, T.~Zinner, T.~Hobfeld, P.~Tran-Gia, {A survey
  on quality of experience of HTTP adaptive streaming}, IEEE Communications
  Surveys \& Tutorials 17~(1) (2015) 469--492.

\bibitem{sodagar2011mpeg}
I.~Sodagar, {The mpeg-dash standard for multimedia streaming over the
  internet}, IEEE MultiMedia 18~(4) (2011) 62--67.

\bibitem{james2016multipath}
C.~James, E.~Halepovic, M.~Wang, R.~Jana, N.~Shankaranarayanan, {Is Multipath
  TCP (MPTCP) Beneficial for Video Streaming over DASH?}, 24th IEEE
  International Symposium on Modeling, Analysis and Simulation of Computer and
  Telecommunication Systems (MASCOTS) (2016) 331--336.

\bibitem{rfc7540}
M.~Belshe, R.~Peon, M.~Thomson,
  \href{https://rfc-editor.org/rfc/rfc7540.txt}{{Hypertext Transfer Protocol
  Version 2 (HTTP/2)}}, RFC 7540 (May 2015).
\newblock \href {http://dx.doi.org/10.17487/RFC7540}
  {\path{doi:10.17487/RFC7540}}.
\newline\urlprefix\url{https://rfc-editor.org/rfc/rfc7540.txt}

\bibitem{rfc8216}
R.~Pantos, W.~May, {HTTP Live Streaming}, RFC 8216 (aug 2017).

\bibitem{DASH2011Part6}
{MPEG systems technologies -- Part 6: Dynamic adaptive streaming over HTTP
  (DASH)}, ISO/IEC FCD 23001-6.

\bibitem{MPEGDASHYouTubeNetflix}
{Why YouTube \& Netflix use MPEG-DASH in HTML5},
  \url{https://bitmovin.com/status-mpeg-dash-today-youtube-netflix-use-html5-beyond/},
  accessed:06-JUN-2021.

\bibitem{yahiahttp}
M.~B. Yahia, Y.~Le~Louedec, L.~Nuaymi, G.~Simon, {When HTTP/2 Rescues DASH:
  Video Frame Multiplexing}, 3rd Communication and Networking Techniques for
  Contemporary Video Workshop (CNTCV), IEEE INFOCOM (2017) .

\bibitem{MPEG-H2014part1}
{Information technology -- High efficiency coding and media delivery in
  heterogeneous environments -- Part 1: MPEG media transport (MMT)}, ISO/IEC
  23008-1 (2014) .

\bibitem{Coding2017}
{MMT Enhancements for mobile environments}, ISO/IEC 23008-1:2017/DAmd 2 (2017)
  .

\bibitem{ye2016shvc}
Y.~Ye, Y.~He, Y.-K. Wang, et~al., {SHVC the Scalable Extensions of HEVC and Its
  Applications}, ZTE Communications 14~(1) (2016) .

\bibitem{karya2018rtp}
O.~Karya, S.~Saesaria, S.~Budiyanto, {RTP analysis for the video transmission
  process on WhatsApp and Skype against signal strength variations in 802.11
  network environments}, in: IOP Conference Series: Materials Science and
  Engineering, Vol. 453, IOP Publishing, 2018, p. 012062.

\bibitem{yap2012making}
K.-K. Yap, T.-Y. Huang, M.~Kobayashi, Y.~Yiakoumis, N.~McKeown, S.~Katti,
  G.~Parulkar, {Making use of all the networks around us: a case study in
  android}, Proceedings of the 2012 ACM SIGCOMM workshop on Cellular networks:
  operations, challenges, and future design (2012) 19--24.

\bibitem{gutterman2019requet}
C.~Gutterman, K.~Guo, S.~Arora, X.~Wang, L.~Wu, E.~Katz-Bassett, G.~Zussman,
  {Requet: Real-Time QoE Detection for Encrypted YouTube Traffic}, MMSys,
  Amherst, MA, US (2019) .

\bibitem{bui2013greenbag}
D.~H. Bui, K.~Lee, S.~Oh, I.~Shin, H.~Shin, H.~Woo, D.~Ban, {Greenbag:
  Energy-efficient bandwidth aggregation for real-time streaming in
  heterogeneous mobile wireless networks}, Real-Time Systems Symposium (RTSS),
  2013 IEEE 34th (2013) 57--67.

\bibitem{han2011end}
S.~Han, H.~Joo, D.~Lee, H.~Song, {An end-to-end virtual path construction
  system for stable live video streaming over heterogeneous wireless networks},
  IEEE Journal on Selected Areas in Communications 29~(5) (2011) 1032--1041.

\bibitem{luo2011digital}
F.-L. Luo, {Digital Front-End in Wireless Communications and Broadcasting:
  circuits and signal processing}, Cambridge University Press, 2011.

\bibitem{chen2013measurement}
Y.-C. Chen, Y.-s. Lim, R.~J. Gibbens, E.~M. Nahum, R.~Khalili, D.~Towsley, {A
  measurement-based study of multipath tcp performance over wireless networks},
  Proceedings of the 2013 conference on Internet measurement conference (2013)
  455--468.

\bibitem{yoon2012muvi}
J.~Yoon, H.~Zhang, S.~Banerjee, S.~Rangarajan, {MuVi: A multicast video
  delivery scheme for 4G cellular networks}, Proceedings of the 18th annual
  international conference on Mobile computing and networking (2012) 209--220.

\bibitem{xu2013cmt}
C.~Xu, T.~Liu, J.~Guan, H.~Zhang, G.-M. Muntean, {CMT-QA: Quality-aware
  adaptive concurrent multipath data transfer in heterogeneous wireless
  networks}, IEEE Transactions on Mobile Computing 12~(11) (2013) 2193--2205.

\bibitem{chan2005tcp}
M.~C. Chan, R.~Ramjee, {TCP/IP performance over 3G wireless links with rate and
  delay variation}, Wireless Networks 11~(1-2) (2005) 81--97.

\bibitem{corbillon2016cross}
X.~Corbillon, R.~Aparicio-Pardo, N.~Kuhn, G.~Texier, G.~Simon, {Cross-layer
  scheduler for video streaming over MPTCP}, Proceedings of the 7th
  International Conference on Multimedia Systems (2016) 7.

\bibitem{ferlin2014tackling}
S.~Ferlin-Oliveira, T.~Dreibholz, {\"O}.~Alay, {Tackling the challenge of
  bufferbloat in multi-path transport over heterogeneous wireless networks},
  in: Quality of Service (IWQoS), 22nd International Symposium of, IEEE, 2014,
  pp. 123--128.

\bibitem{chen2014bufferbloat}
Y.-C. Chen, D.~Towsley, {On bufferbloat and delay analysis of multipath TCP in
  wireless networks}, in: Networking Conference, IFIP, IEEE, 2014, pp. 1--9.

\bibitem{sun2016mars}
G.~Sun, K.~Eng, S.~Yin, G.~Liu, G.~Min, Mars: multiple access radio scheduling
  for a multi-homed mobile device in soft-ran, KSII Transactions on Internet
  and Information Systems (TIIS) 10~(1) (2016) 79--95.

\bibitem{brosh2010delay}
E.~Brosh, S.~A. Baset, V.~Misra, D.~Rubenstein, H.~Schulzrinne, {The
  delay-friendliness of TCP for real-time traffic}, IEEE/ACM Transactions On
  Networking 18~(5) (2010) 1478--1491.

\bibitem{barakovic2013survey}
S.~Barakovi{\'c}, L.~Skorin-Kapov, {Survey and challenges of QoE management
  issues in wireless networks}, Journal of Computer Networks and Communications
  (2013) .

\bibitem{van2011multimedia}
M.~Van~der Schaar, P.~A. Chou, {Multimedia over IP and wireless networks:
  compression, networking, and systems}, Elsevier, 2011.

\bibitem{recommendation20011010}
G.~Recommendation, {1010 End-user multimedia QoS categories}, ITU-T, November
  (2001) .

\bibitem{wu2016bandwidth}
J.~Wu, C.~Yuen, B.~Cheng, Y.~Yang, M.~Wang, J.~Chen, {Bandwidth-efficient
  multipath transport protocol for quality-guaranteed real-time video over
  heterogeneous wireless networks}, IEEE Transactions on Communications 64~(6)
  (2016) 2477--2493.

\bibitem{austerberry2005technology}
D.~Austerberry, {The technology of video and audio streaming}, Taylor \&
  Francis, 2005.

\bibitem{chow2009ems}
A.~L. Chow, H.~Yang, C.~H. Xia, M.~Kim, Z.~Liu, H.~Lei, {EMS: Encoded multipath
  streaming for real-time live streaming applications}, 17th IEEE International
  Conference on Network Protocols (ICNP) (2009) 233--243.

\bibitem{mwela2010impact}
J.~S. Mwela, {Impact of packet loss on the quality of video stream
  transmission}, Ph.D. thesis, Blekinge Institute of Technology (2010).

\bibitem{leiwm-avtcore-mprtp-ar-09}
W.~Lei, W.~Zhang, S.~Liu,
  \href{https://datatracker.ietf.org/doc/html/draft-leiwm-avtcore-mprtp-ar-09}{{Multipath
  Real-Time Transport Protocol Based on Application-Level Relay (MPRTP-AR)}},
  Internet-Draft draft-leiwm-avtcore-mprtp-ar-09, Internet Engineering Task
  Force, work in Progress (Feb. 2018).
\newline\urlprefix\url{https://datatracker.ietf.org/doc/html/draft-leiwm-avtcore-mprtp-ar-09}

\bibitem{xing2012rate}
M.~Xing, S.~Xiang, L.~Cai, {Rate adaptation strategy for video streaming over
  multiple wireless access networks}, IEEE Global Communications Conference
  (GLOBECOM) (2012) 5745--5750.

\bibitem{chowrikoppalu2013multipath}
Y.~Chowrikoppalu, P.~Gowda, {Multipath Adaptive Video Streaming over Multipath
  TCP}, Ph.D. thesis, Intel (2013).

\bibitem{houze2016applicative}
P.~Houz{\'e}, E.~Mory, G.~Texier, G.~Simon, {Applicative-layer multipath for
  low-latency adaptive live streaming}, IEEE International Conference on
  Communications (ICC) (2016) 1--7.

\bibitem{go2019hybrid}
Y.~Go, H.~Noh, G.~Park, H.~Song, {Hybrid TCP/UDP-Based Enhanced HTTP Adaptive
  Streaming System With Multi-Homed Mobile Terminal}, IEEE Transactions on
  Vehicular Technology 68~(5) (2019) 5114--5128.

\bibitem{kolan2016method}
P.~Kolan, I.~Bouazizi, {Method and apparatus for multipath media delivery}, {US
  Patent 10069719B2} (2018).

\bibitem{afzal2018novel}
S.~Afzal, V.~Testoni, J.~F.~F. de~Oliveira, C.~E. Rothenberg, P.~Kolan,
  I.~Bouazizif, {A Novel Scheduling Strategy for MMT-based Multipath Video
  Streaming}, in: IEEE Global Communications Conference (GLOBECOM), 2018, pp.
  206--212.

\bibitem{afzal2021multipath}
S.~Afzal, C.~E. Rothenberg, V.~Testoni, P.~Kolan, I.~Bouazizi, Multipath
  mmt-based approach for streaming high quality video over multiple wireless
  access networks, Computer Networks 185 (2021) 107638.

\bibitem{sohn2015synchronization}
Y.~Sohn, M.~Cho, M.~Seo, J.~Paik, {A synchronization scheme for hierarchical
  video streams over heterogeneous networks}, KSII Transactions on Internet and
  Information Systems (TIIS) 9~(8) (2015) 3121--3135.

\bibitem{michel2018adding}
F.~Michel, Q.~De~Coninck, O.~Bonaventure, {Adding forward erasure correction to
  quic}, arXiv preprint arXiv:1809.04822 (2018) --.

\bibitem{evensen2010quality}
K.~Evensen, T.~Kupka, D.~Kaspar, P.~Halvorsen, C.~Griwodz, {Quality-adaptive
  scheduling for live streaming over multiple access networks}, Proceedings of
  the 20th international workshop on Network and operating systems support for
  digital audio and video (2010) 21--26.

\bibitem{evensen2011improving}
K.~Evensen, D.~Kaspar, C.~Griwodz, P.~Halvorsen, A.~Hansen, P.~Engelstad,
  {Improving the performance of quality-adaptive video streaming over multiple
  heterogeneous access networks}, Proceedings of the second annual ACM
  conference on Multimedia systems (2011) 57--68.

\bibitem{evensen2012using}
K.~Evensen, D.~Kaspar, C.~Griwodz, P.~Halvorsen, A.~F. Hansen, P.~Engelstad,
  {Using bandwidth aggregation to improve the performance of quality-adaptive
  streaming}, Signal Processing: Image Communication 27~(4) (2012) 312--328.

\bibitem{nikravesh2019MP-H2}
A.~Nikravesh, Y.~Guo, X.~Zhu, F.~Qian, Z.~M. Mao, {MP-H2: A Client-only
  Multipath Solution for HTTP/2}, MobiCom. ACM (2019) 1--16.

\bibitem{freris2013distortion}
N.~M. Freris, C.-H. Hsu, J.~P. Singh, X.~Zhu, {Distortion-aware scalable video
  streaming to multinetwork clients}, IEEE/ACM Transactions on Networking
  21~(2) (2013) 469--481.

\bibitem{correia2012optimal}
P.~Correia, L.~Ferreira, P.~A.~A. Assuncao, L.~Cruz, V.~Silva, {Optimal
  priority mdc video streaming for networks with path diversity}, IEEE
  International Conference on Telecommunications and Multimedia (TEMU) (2012)
  54--59.

\bibitem{huang2012qos}
C.-M. Huang, Y.-C. Chen, S.-Y. Lin, {The QoS-Aware Order Prediction Scheduling
  (QOPS) Scheme for Video Streaming Using the Multi-path Datagram Congestion
  Control Protocol (MP-DCCP)}, IEEE 5th International Conference on
  Network-Based Information Systems (NBiS) (2012) 276--283.

\bibitem{wu2016streaming}
J.~Wu, C.~Yuen, B.~Cheng, M.~Wang, J.~Chen, {Streaming high-quality mobile
  video with multipath TCP in heterogeneous wireless networks}, IEEE
  Transactions on Mobile Computing 15~(9) (2016) 2345--2361.

\bibitem{Wu2019Energy}
J.~{Wu}, R.~{Tan}, M.~{Wang}, {Energy-Efficient Multipath TCP for
  Quality-Guaranteed Video Over Heterogeneous Wireless Networks}, IEEE
  Transactions on Multimedia 21~(6) (2019) 1593--1608.

\bibitem{7937943}
J.~Wu, B.~Cheng, M.~Wang, J.~Chen, Quality-aware energy optimization in
  wireless video communication with multipath tcp, IEEE/ACM Transactions on
  Networking 25~(5) (2017) 2701--2718.
\newblock \href {http://dx.doi.org/10.1109/TNET.2017.2701153}
  {\path{doi:10.1109/TNET.2017.2701153}}.

\bibitem{diop2012qos}
C.~Diop, G.~Dugue, C.~Chassot, E.~Exposito, {QoS-oriented MPTCP extensions for
  multimedia multi-homed systems}, IEEE 26th International Conference on
  Advanced Information Networking and Applications Workshops (WAINA) (2012)
  1119--1124.

\bibitem{zhang2015multipath}
C.~Xu, H.~Huang, H.~Zhang, C.~Xiong, L.~Zhu, {Multipath Transmission Control
  Protocol (MPTCP) Partial Reliability Extension} (2016) --.

\bibitem{cao2016pr}
Y.~Cao, Q.~Liu, G.~Luo, Y.~Yi, M.~Huang, {PR-MPTCP+: Context-aware QoE-oriented
  multipath TCP partial reliability extension for real-time multimedia
  applications}, Visual Communications and Image Processing (VCIP), 2016 (2016)
  1--4.

\bibitem{kelly2004delay}
A.~Kelly, G.~Muntean, P.~Perry, J.~Murphy, {Delay-centric handover in SCTP over
  WLAN}, Transactions on Automatic Control and Computer Science 49~(63) (2004)
  1--6.

\bibitem{okamoto2014performance}
K.~Okamoto, N.~Yamai, K.~Okayama, K.~Kawano, M.~Nakamura, T.~Yokohira,
  {Performance improvement of SCTP communication using selective bicasting on
  lossy multihoming environment}, IEEE 38th Annual Computer Software and
  Applications Conference (COMPSAC) (2014) 551--557.

\bibitem{da2016preventing}
C.~A.~G. da~Silva, E.~P. Ribeiro, C.~M. Pedroso, {Preventing quality
  degradation of video streaming using selective redundancy}, Computer
  Communications 91 (2016) 120--132.

\bibitem{sanson2010pr}
H.~Sanson, A.~Neira, L.~Loyola, M.~Matsumoto, {PR-SCTP for real time H. 264/AVC
  video streaming}, IEEE 12th International Conference on Advanced
  Communication Technology (ICACT) 1 (2010) 59--63.

\bibitem{wu2015distortion}
J.~Wu, B.~Cheng, C.~Yuen, Y.~Shang, J.~Chen, {Distortion-aware concurrent
  multipath transfer for mobile video streaming in heterogeneous wireless
  networks}, IEEE Transactions on Mobile Computing 14~(4) (2015) 688--701.

\bibitem{wu2016content}
J.~Wu, C.~Yuen, M.~Wang, J.~Chen, {Content-aware concurrent multipath transfer
  for high-definition video streaming over heterogeneous wireless networks},
  IEEE Transactions on Parallel and Distributed Systems 27~(3) (2016) 710--723.

\bibitem{chebrolu2006bandwidth}
K.~Chebrolu, R.~R. Rao, {Bandwidth aggregation for real-time applications in
  heterogeneous wireless networks}, IEEE Transactions on Mobile Computing 5~(4)
  (2006) 388--403.

\bibitem{ojanpera2016network}
T.~Ojanper{\"a}, J.~Vehkaper{\"a}, {Network-assisted multipath DASH using the
  distributed decision engine}, IEEE International Conference on Computing,
  Networking and Communications (ICNC) (2016) 1--6.

\bibitem{wu2015goodput}
J.~Wu, C.~Yuen, B.~Cheng, Y.~Shang, J.~Chen, {Goodput-aware load distribution
  for real-time traffic over multipath networks}, IEEE Transactions on Parallel
  and Distributed Systems 26~(8) (2015) 2286--2299.

\bibitem{wu2013joint}
J.~Wu, Y.~Shang, J.~Huang, X.~Zhang, B.~Cheng, J.~Chen, {Joint source-channel
  coding and optimization for mobile video streaming in heterogeneous wireless
  networks}, European Association for Signal Processing (EURASIP) Journal on
  Wireless Communications and Networking 2013~(1) (2013) 283.

\bibitem{deng2021cross}
Z.~Deng, Y.~Liu, J.~Liu, A.~Argyriou, Cross-layer dash-based multipath video
  streaming over lte and 802.11 ac networks, Multimedia Tools and Applications
  80~(10) (2021) 16007--16026.

\bibitem{han2016mp}
B.~Han, F.~Qian, L.~Ji, V.~Gopalakrishnan, N.~Bedminster, {MP-DASH: Adaptive
  Video Streaming Over Preference-Aware Multipath}, Proceedings of the 12th
  International on Conference on emerging Networking EXperiments and
  Technologies (2016) 129--143.

\bibitem{nam2016towards}
H.~Nam, D.~Calin, H.~Schulzrinne, {Towards dynamic MPTCP Path control using
  SDN}, IEEE NetSoft Conference and Workshops (NetSoft) (2016) 286--294.

\bibitem{xu2015cross}
C.~Xu, Z.~Li, J.~Li, H.~Zhang, G.-M. Muntean, {Cross-layer fairness-driven
  concurrent multipath video delivery over heterogeneous wireless networks},
  IEEE Transactions on Circuits and Systems for Video Technology 25~(7) (2015)
  1175--1189.

\bibitem{xing2021low}
Y.~Xing, K.~Xue, Y.~Zhang, J.~Han, J.~Li, J.~Liu, R.~Li, A low-latency mptcp
  scheduler for live video streaming in mobile networks, IEEE Transactions on
  Wireless Communications (2021) .

\bibitem{schulzrinne1998real}
H.~Schulzrinne, {Real time streaming protocol (RTSP)}, RFC 2326 (1998).

\bibitem{singh2012performance}
A.~Singh, C.~Goerg, A.~Timm-Giel, M.~Scharf, T.-R. Banniza, {Performance
  comparison of scheduling algorithms for multipath transfer} (2012)
  2653--2658.

\bibitem{xiao2016evaluating}
M.~Xiao, V.~Swaminathan, S.~Wei, S.~Chen, {Evaluating and improving push based
  video streaming with HTTP/2}, Proceedings of the 26th International Workshop
  on Network and Operating Systems Support for Digital Audio and Video (2016)
  3.

\bibitem{ojanpera2016wireless}
T.~Ojanper{\"a}, H.~Kokkoniemi-Tarkkanen, {Wireless bandwidth management for
  multiple video clients through network-assisted DASH}, in: World of Wireless,
  Mobile and Multimedia Networks (WoWMoM), IEEE, 2016, pp. 1--3.

\bibitem{kleinrouweler2016delivering}
J.~W. Kleinrouweler, S.~Cabrero, P.~Cesar, {Delivering stable high-quality
  video: An SDN architecture with DASH assisting network elements}, in:
  Proceedings of the 7th International Conference on Multimedia Systems, ACM,
  2016, p.~4.

\bibitem{cofano2017design}
G.~Cofano, L.~D. Cicco, T.~Zinner, A.~Nguyen-Ngoc, P.~Tran-Gia, S.~Mascolo,
  {Design and performance evaluation of network-assisted control strategies for
  HTTP adaptive streaming}, ACM Transactions on Multimedia Computing,
  Communications, and Applications (TOMM) 13~(3s) (2017) 42.

\bibitem{Information2014Part5}
{Information technology -- Dynamic adaptive streaming over HTTP (DASH) -- part
  5: Server and network assisted DASH (SAND)}, ISO/IEC CD 23009-5 (2014) .

\bibitem{thomas2016applications}
E.~Thomas, M.~van Deventer, T.~Stockhammer, A.~C. Begen, M.-L. Champel,
  O.~Oyman, {Applications and deployments of server and network assisted DASH
  (SAND)} (2016) .

\bibitem{luoto2015distributed}
M.~Luoto, T.~Rautio, T.~Ojanper{\"a}, J.~M{\"a}kel{\"a}, {Distributed decision
  engine—An information management architecture for autonomie wireless
  networking}, IFIP/IEEE International Symposium on Integrated Network
  Management (IM) (2015) 713--719.

\bibitem{habib2016past}
S.~Habib, J.~Qadir, A.~Ali, D.~Habib, M.~Li, A.~Sathiaseelan, {The past,
  present, and future of transport-layer multipath}, Journal of Network and
  Computer Applications 75 (2016) 236--258.

\bibitem{bokani2013http}
A.~Bokani, M.~Hassan, S.~Kanhere, {Http-based adaptive streaming for mobile
  clients using markov decision process}, IEEE 20th International Packet Video
  Workshop (PV) (2013) 1--8.

\bibitem{floyd2000equation}
S.~Floyd, M.~Handley, J.~Padhye, J.~Widmer, {Equation-based congestion control
  for unicast applications}, ACM SIGCOMM Computer Communication Review 30~(4)
  (2000) 43--56.

\bibitem{cen2003end}
S.~Cen, P.~C. Cosman, G.~M. Voelker, {End-to-end differentiation of congestion
  and wireless losses}, IEEE/ACM Transactions on networking 11~(5) (2003)
  703--717.

\bibitem{bae2013method}
S.~BAE, {Method for configuring and transmitting m-unit}, {US Patent
  20130094594A1} (2013).

\bibitem{aoki2017Emerging}
S.~Aoki, {Emerging 8K services and their applications towards 2020}, ITU-T 2nd
  mini-Workshop on Immersive Live Experience.

\bibitem{ITU-R-MMT}
R.~I.-R. BT.2074-1, {Service configuration, media transport protocol, and
  signalling information for MMT-based broadcasting systems} (2017) .

\bibitem{I-D.bouazizi-mmtp}
I.~Bouazizi,
  \href{https://datatracker.ietf.org/doc/html/draft-bouazizi-mmtp-01}{{MPEG
  Media Transport Protocol (MMTP)}}, Internet-Draft draft-bouazizi-mmtp-01,
  Internet Engineering Task Force, work in Progress (Sep. 2014).
\newline\urlprefix\url{https://datatracker.ietf.org/doc/html/draft-bouazizi-mmtp-01}

\bibitem{jung2015overview}
T.-J. Jung, H.-r. Lee, K.-d. Seo, {Overview on MPEG MMT Technology and Its
  Application to Hybrid Media Delivery over Heterogeneous Networks}, Pacific
  Rim Conference on Multimedia (2015) 660--669.

\bibitem{MPEG-H2013Part13}
{Information technology — High efficiency coding and media delivery in
  heterogeneous environments — Part 13: 3rd Edition MPEG Media Transport
  Implementation Guidelines}, ISO/IEC 23008-13 (2013) .

\bibitem{de2021method}
J.~F.~F. de~Oliveira, S.~Afzal, V.~Testoni, C.~R.~E. Rothenberg, Method for
  digital video transmission adopting packaging forwarding strategies with path
  and content monitoring in heterogeneous networks using mmt protocol, method
  for reception and communication system, {US Patent 10887151} (2021).

\bibitem{de2021methodINPI}
J.~F.~F. de~Oliveira, S.~Afzal, V.~Testoni, {M}Étodo para transmissÃo de
  vÍdeo digital adotando estratÉgias de encaminhamento de pacotes com
  monitoramento de percurso e conteÚdo em redes heterogÊnas usando protocolo
  {MMT}, {INPI Patent 1020180769529} (2020).

\bibitem{langley2017quic}
A.~Langley, A.~Riddoch, A.~Wilk, A.~Vicente, C.~Krasic, D.~Zhang, F.~Yang,
  F.~Kouranov, I.~Swett, J.~Iyengar, et~al., The quic transport protocol:
  Design and internet-scale deployment, in: Proceedings of the Conference of
  the ACM Special Interest Group on Data Communication, ACM, 2017, pp.
  183--196.

\bibitem{IETFQUICworkinggroup}
{IETF QUIC working group.}, \url{https://datatracker.ietf.org/wg/quic/},
  accessed:06-JUN-2021.

\bibitem{kakhki2017rigorous}
A.~M. Kakhki, Rigorous evaluation of performance and policy impacts of
  transport protocols and in-network devices, Ph.D. thesis, Northeastern
  University (2017).

\bibitem{choi2017streaming}
J.~Choi, C.~Jung, I.~Yeom, Y.~Kim, {Streaming Service Enhancement on QUIC
  Protocol}, in: Proceedings on the International Conference on Internet
  Computing (ICOMP), The Steering Committee of The World Congress in Computer
  Science, Computer~…, 2017, pp. 123--124.

\bibitem{li2016mmt}
B.~Li, C.~Wang, Y.~Xu, Z.~Ma, {An MMT based heterogeneous multimedia system
  using QUIC}, IEEE 2nd International Conference on Cloud Computing and
  Internet of Things (CCIOT) (2016) 129--133.

\bibitem{de2017multipath}
Q.~De~Coninck, O.~Bonaventure, Multipath quic: Design and evaluation, in:
  Proceedings of the 13th International Conference on emerging Networking
  EXperiments and Technologies, ACM, 2017, pp. 160--166.

\bibitem{Khalili2013MPTCP}
R.~{Khalili}, N.~{Gast}, M.~{Popovic}, J.~{Le Boudec}, {MPTCP Is Not
  Pareto-Optimal: Performance Issues and a Possible Solution}, IEEE/ACM
  Transactions on Networking 21~(5) (2013) 1651--1665.

\bibitem{przylucki2019simulation}
S.~Przylucki, D.~Czerwinski, The simulation study on the multipath adaptive
  video transmission, in: MATEC Web of Conferences, Vol. 252, EDP Sciences,
  2019, p. 05018.

\bibitem{johansen2009davvi}
D.~Johansen, H.~Johansen, T.~Aarflot, J.~Hurley, {\AA}.~Kvalnes, C.~Gurrin,
  S.~Zav, B.~Olstad, E.~Aaberg, T.~Endestad, et~al., {DAVVI: A prototype for
  the next generation multimedia entertainment platform}, Proceedings of the
  17th ACM international conference on Multimedia (2009) 989--990.

\bibitem{kaspar2010using}
D.~Kaspar, K.~Evensen, P.~Engelstad, A.~F. Hansen, {Using HTTP pipelining to
  improve progressive download over multiple heterogeneous interfaces}, IEEE
  International Conference on Communications (ICC) (2010) 1--5.

\bibitem{lederer2012dynamic}
S.~Lederer, C.~M{\"u}ller, C.~Timmerer, {Dynamic adaptive streaming over HTTP
  dataset}, Proceedings of the 3rd Multimedia Systems Conference (2012) 89--94.

\bibitem{raiciu2012hard}
C.~Raiciu, C.~Paasch, S.~Barre, A.~Ford, M.~Honda, F.~Duchene, O.~Bonaventure,
  M.~Handley, {How hard can it be? designing and implementing a deployable
  multipath TCP}, Proceedings of the 9th USENIX conference on Networked Systems
  Design and Implementation (2012) 29--29.

\bibitem{postel1980rfc}
J.~Postel, {User Datagram Protocol}, RFC 768 (1980).

\bibitem{fairhurst2017services}
G.~Fairhurst, B.~Trammell, M.~Kuehlewind, {Services provided by IETF transport
  protocols and congestion control mechanisms}, RFC 8095 (2017).

\bibitem{hossfeld2014qoe}
T.~Ho{\ss}feld, R.~Schatz, U.~R. Krieger, {QoE of YouTube video streaming for
  current Internet transport protocols}, in: Measurement, Modelling, and
  Evaluation of Computing Systems and Dependability and Fault Tolerance,
  Springer, 2014, pp. 136--150.

\bibitem{postel1981rfc}
J.~Postel, {Transmission control protocol}, RFC 793 (1981).

\bibitem{floyd2000extension}
S.~Floyd, J.~Mahdavi, M.~Mathis, M.~Podolsky, {An extension to the selective
  acknowledgement (SACK) option for TCP}, RFC 2883 (2000).

\bibitem{black2018rfc}
D.~Black, {Relaxing Restrictions on Explicit Congestion Notification (ECN)
  Experimentation}, RFC 8311 (2018).

\bibitem{li2014tolerating}
M.~Li, A.~Lukyanenko, S.~Tarkoma, Y.~Cui, A.~Yl{\"a}-J{\"a}{\"a}ski,
  {Tolerating path heterogeneity in multipath TCP with bounded receive
  buffers}, Computer Networks 64 (2014) 1--14.

\bibitem{kohler2006datagram}
E.~Kohler, M.~Handley, S.~Floyd, {Datagram Congestion Control Protocol (DCCP)},
  RFC 4340 (2006).

\bibitem{floyd2006rfcTCP-like}
S.~Floyd, E.~Kohler, {Profile for Datagram Congestion Control Protocol (DCCP)
  Congestion Control ID 2: TCP-like Congestion Control}, RFC 4341 (2006).

\bibitem{floyd2006rfcTCP-Friendly}
S.~Floyd, E.~Kohler, J.~Padhye, {Profile for Datagram Congestion Control
  Protocol (DCCP) Congestion Control ID 3: TCP-Friendly Rate Control (TFRC)},
  RFC 4342 (2006).

\bibitem{azad2009comparative}
M.~A. Azad, R.~Mahmood, T.~Mehmood, {A comparative analysis of DCCP variants
  (CCID2, CCID3), TCP and UDP for MPEG4 video applications}, International
  Conference on Information and Communication Technologies. ICICT (2009)
  40--45.

\bibitem{ford2009mptcp}
A.~Ford, C.~Raiciu, S.~Barre, Louvain, {TCP Extensions for Multipath Operation
  with Multiple Addresses}, RFC 6824, subsequent updates (2009).

\bibitem{Ford2011Architectural}
J.~Iyengar, C.~Raiciu, S.~Barre, M.~J. Handley, A.~Ford,
  \href{https://rfc-editor.org/rfc/rfc6182.txt}{{Architectural Guidelines for
  Multipath TCP Development}}, RFC 6182 (Mar. 2011).
\newblock \href {http://dx.doi.org/10.17487/RFC6182}
  {\path{doi:10.17487/RFC6182}}.
\newline\urlprefix\url{https://rfc-editor.org/rfc/rfc6182.txt}

\bibitem{bonaventure2016multipath}
O.~Bonaventure, S.~Seo, {Multipath TCP deployments}, IETF Journal 12~(2) (2016)
  24--27.

\bibitem{wischik2011design}
D.~Wischik, C.~Raiciu, A.~Greenhalgh, M.~Handley, {Design, Implementation and
  Evaluation of Congestion Control for Multipath TCP}, Networked Systems Design
  and Implementation (NSDI) 11 (2011) 8--8.

\bibitem{raiciu2011rfc}
C.~Raiciu, M.~Handley, D.~Wischik, {Coupled Congestion Control for Multipath
  Transport Protocols}, RFC 6356 (2011).

\bibitem{paasch2014experimental}
C.~Paasch, S.~Ferlin, O.~Alay, O.~Bonaventure, {Experimental evaluation of
  multipath TCP schedulers}, Proceedings of the ACM SIGCOMM workshop on
  Capacity sharing workshop (2014) 27--32.

\bibitem{barre2011multipath}
S.~Barr{\'e}, C.~Paasch, O.~Bonaventure, {Multipath TCP: from theory to
  practice}, International Conference on Research in Networking (2011)
  444--457.

\bibitem{hesmans2013tcp}
B.~Hesmans, F.~Duchene, C.~Paasch, G.~Detal, O.~Bonaventure, {Are TCP
  extensions middlebox-proof?}, Proceedings of the workshop on Hot topics in
  middleboxes and network function virtualization (2013) 37--42.

\bibitem{deng2014wifi}
S.~Deng, R.~Netravali, A.~Sivaraman, H.~Balakrishnan, {Wifi, lte, or both?:
  Measuring multi-homed wireless internet performance}, Proceedings of the
  Conference on Internet Measurement Conference (2014) 181--194.

\bibitem{lim2014cross}
Y.-s. Lim, Y.-C. Chen, E.~M. Nahum, D.~Towsley, K.-W. Lee, {Cross-layer path
  management in multi-path transport protocol for mobile devices} (2014)
  1815--1823.

\bibitem{exposito2009building}
E.~Exposito, M.~Gineste, L.~Dairaine, C.~Chassot, {Building self-optimized
  communication systems based on applicative cross-layer information}, Computer
  Standards \& Interfaces 31~(2) (2009) 354--361.

\bibitem{stewart2000rfc}
R.~Stewart, Q.~Xie, K.~Morneault, C.~Sharp, H.~Schwarzbauer, T.~Taylor,
  I.~Rytina, M.~Kalia, L.~Zhang, V.~Paxson, {Stream control transmission
  protocol}, RFC 2960 (2000).

\bibitem{rfc49602007stream}
R.~R. Stewart, M.~Tüxen, K.~Nielsen,
  \href{https://datatracker.ietf.org/doc/html/draft-ietf-tsvwg-rfc4960-bis-12}{{Stream
  Control Transmission Protocol}}, Internet-Draft
  draft-ietf-tsvwg-rfc4960-bis-12, Internet Engineering Task Force, work in
  Progress (Jul. 2021).
\newline\urlprefix\url{https://datatracker.ietf.org/doc/html/draft-ietf-tsvwg-rfc4960-bis-12}

\bibitem{fu2004sctp}
S.~Fu, M.~Atiquzzaman, {SCTP: State of the art in research, products, and
  technical challenges}, IEEE Communications Magazine 42~(4) (2004) 64--76.

\bibitem{Stewart2004rfc3758}
R.~Stewart, M.~Ramalho, Q.~Xie, M.~Tuexen, P.~Conrad, {Stream Control
  Transmission Protocol (SCTP) Partial Reliability Extension}, RFC 3758 (2004).

\bibitem{tuexen2015additionalrfc}
M.~Tuexen, R.~Seggelmann, R.~Stewart, S.~Loreto, {Additional Policies for the
  Partially Reliable Stream Control Transmission Protocol Extension}, RFC 7496
  (2015).

\bibitem{wang2003performance}
H.~Wang, Y.~Jin, W.~Wang, J.~Ma, D.~Zhang, {The performance comparison of
  PRSCTP, TCP and UDP for MPEG-4 multimedia traffic in mobile network}, IEEE
  International Conference on Communication Technology Proceedings (ICCT) 1
  (2003) 403--406.

\bibitem{iyengar2006concurrent}
J.~R. Iyengar, P.~D. Amer, R.~Stewart, {Concurrent multipath transfer using
  SCTP multihoming over independent end-to-end paths}, IEEE/ACM Transactions on
  networking 14~(5) (2006) 951--964.

\bibitem{chen2019buffer}
F.~Chen, J.~Zhang, Z.~Chen, J.~Wu, N.~Ling, {Buffer-Driven Rate Control and
  Packet Distribution for Real-Time Videos in Heterogeneous Wireless Networks},
  IEEE Access 7 (2019) 27401--27415.

\bibitem{wu2013csi}
K.~Wu, J.~Xiao, Y.~Yi, D.~Chen, X.~Luo, L.~M. Ni, {CSI-based indoor
  localization}, IEEE Transactions on Parallel and Distributed Systems 24~(7)
  (2013) 1300--1309.

\bibitem{le1991mpeg}
D.~Le~Gall, {MPEG: A video compression standard for multimedia applications},
  Communications of the ACM 34~(4) (1991) 46--58.

\bibitem{zhai2004rate}
F.~Zhai, Y.~Eisenberg, T.~N. Pappas, R.~Berry, A.~K. Katsaggelos,
  {Rate-distortion optimized product code forward error correction for video
  transmission over IP-based wireless networks}, in: IEEE International
  Conference on Acoustics, Speech, and Signal Processing (ICASSP), Vol.~5,
  2004, pp. V--857.

\bibitem{frossard2001fec}
P.~Frossard, {FEC performance in multimedia streaming}, IEEE Communications
  Letters 5~(3) (2001) 122--124.

\bibitem{nakachi2013next}
T.~Nakachi, T.~Yamaguchi, Y.~Tonomura, T.~Fujii, {Next-generation media
  transport MMT for 4K/8K video transmission}, NTT Technical Review 12~(5)
  (2013) 1--7.

\bibitem{shokrollahi2006raptor}
A.~Shokrollahi, {Raptor codes}, IEEE transactions on information theory 52~(6)
  (2006) 2551--2567.

\bibitem{RLC-16}
V.~Roca, B.~Teibi,
  \href{https://datatracker.ietf.org/doc/html/draft-ietf-tsvwg-rlc-fec-scheme-16.}{{Sliding
  Window Random Linear Code (RLC) Forward Erasure Correction (FEC) Schemes for
  FECFRAME}}, Internet-Draft draft-ietf-tsvwg-rlc-fec-scheme-16, Internet
  Engineering Task Force, work in Progress (2019).
\newline\urlprefix\url{https://datatracker.ietf.org/doc/html/draft-ietf-tsvwg-rlc-fec-scheme-16.}

\bibitem{kuo2014modeling}
C.-I. Kuo, C.-H. Shih, C.-K. Shieh, W.-S. Hwang, C.-H. Ke, {Modeling and
  analysis of frame-level forward error correction for MPEG video over
  burst-loss channels}, Applied Mathematics \& Information Sciences 8~(4)
  (2014) 1845.

\bibitem{wallace2012review}
T.~D. Wallace, A.~Shami, {A review of multihoming issues using the stream
  control transmission protocol}, IEEE Communications Surveys \& Tutorials
  14~(2) (2012) 565--578.

\bibitem{kazemi2014review}
M.~Kazemi, S.~Shirmohammadi, K.~H. Sadeghi, {A review of multiple description
  coding techniques for error-resilient video delivery}, Multimedia Systems
  20~(3) (2014) 283--309.

\bibitem{kobayashi2009robust}
M.~Kobayashi, H.~Nakayama, N.~Ansari, N.~Kato, {Robust and efficient stream
  delivery for application layer multicasting in heterogeneous networks}, IEEE
  Transactions on Multimedia 11~(1) (2009) 166--176.

\bibitem{mitzenmacher2001power}
M.~Mitzenmacher, {The power of two choices in randomized load balancing}, IEEE
  Transactions on Parallel and Distributed Systems 12~(10) (2001) 1094--1104.

\bibitem{ribeiro2006minimum}
E.~P. Ribeiro, V.~C. Leung, {Minimum delay path selection in multi-homed
  systems with path asymmetry}, IEEE Communications Letters 10~(3) (2006)
  135--137.

\bibitem{bentaleb2019bandwidth}
A.~Bentaleb, C.~Timmerer, A.~C. Begen, R.~Zimmermann, {Bandwidth prediction in
  low-latency chunked streaming}, in: Proceedings of the 29th ACM Workshop on
  Network and Operating Systems Support for Digital Audio and Video, ACM, 2019,
  pp. 7--13.

\bibitem{paul2016enhanced}
A.~K. Paul, A.~Tachibana, T.~Hasegawa, {An enhanced available bandwidth
  estimation technique for an end-to-end network path}, IEEE Transactions on
  Network and Service Management 13~(4) (2016) 768--781.

\bibitem{zhou2008new}
A.~Zhou, M.~Liu, Y.~Song, Z.~Li, H.~Deng, Y.~Ma, {A new method for end-to-end
  available bandwidth estimation}, in: IEEE Global Communications Conference
  (GLOBECOM), 2008, pp. 1--5.

\bibitem{wu2016delay}
J.~Wu, C.~Yuen, N.-M. Cheung, J.~Chen, {Delay-constrained high definition video
  transmission in heterogeneous wireless networks with multi-homed terminals},
  IEEE Transactions on Mobile Computing 15~(3) (2016) 641--655.

\bibitem{ribeiro2003pathchirp}
V.~J. Ribeiro, R.~H. Riedi, R.~G. Baraniuk, J.~Navratil, L.~Cottrell,
  {pathchirp: Efficient available bandwidth estimation for network paths}, in:
  Passive and active measurement workshop, 2003.

\bibitem{ramakrishnan2001rfc}
K.~Ramakrishnan, S.~Floyd, D.~Black, {The Addition of Explicit Congestion
  Notification (ECN) to IP}, RFC 3168 (2001).

\bibitem{rijkse1996h}
K.~Rijkse, {H. 263: video coding for low-bit-rate communication}, IEEE
  Communications magazine 34~(12) (1996) 42--45.

\bibitem{h264avc}
{Advanced Video Coding for Generic Audio-Visual Services}, ITU-T Rec. H.264 and
  ISO/IEC 14496-10 (MPEG-4 AVC) (2017) .

\bibitem{h264svc}
H.~Schwarz, D.~Marpe, T.~Wiegand, {Overview of the scalable video coding
  extension of the H. 264/AVC standard}, IEEE Transactions on circuits and
  systems for video technology 17~(9) (2007) 1103--1120.

\bibitem{shvc}
J.~Boyce, J.~Chen, Y.~Chen, D.~Flynn, M.~Hannuksela, M.~Naccari, C.~Rosewarne,
  K.~Sharman, J.~Sole, G.~Sullivan, et~al., {Edition 2 Draft Text of High
  Efficiency Video Coding (HEVC), Including Format Range (RExt), Scalability
  (SHVC), and Multi-View (MV-HEVC) Extensions}, document JCTVC-R1013 (2014) .

\bibitem{chang2012network}
Y.-L. Chang, T.-L. Lin, P.~C. Cosman, {Network-based H. 264/AVC whole-frame
  loss visibility model and frame dropping methods}, IEEE Transactions on Image
  Processing 21~(8) (2012) 3353--3363.

\bibitem{mn-wifi}
R.~R. Fontes, S.~Afzal, S.~H.~B. Brito, M.~A.~S. Santos, C.~E. Rothenberg,
  Mininet-wifi: Emulating software-defined wireless networks, in: 2015 11th
  International Conference on Network and Service Management (CNSM), 2015, pp.
  384--389.

\bibitem{wang2004image}
Z.~Wang, A.~C. Bovik, H.~R. Sheikh, E.~P. Simoncelli, {Image quality
  assessment: from error visibility to structural similarity}, IEEE
  transactions on image processing 13~(4) (2004) 600--612.

\bibitem{wang2003multiscale}
Z.~Wang, E.~P. Simoncelli, A.~C. Bovik, {Multiscale structural similarity for
  image quality assessment}, IEEE Thirty-Seventh Asilomar Conference on
  Signals, Systems and Computers 2 (2003) 1398--1402.

\bibitem{blin2006new}
J.-L. Blin, {New quality evaluation method suited to multimedia context:
  SAMVIQ}, Proceedings of the Second International Workshop on Video Processing
  and Quality Metrics, VPQM 6 (2006) .

\bibitem{frommgen2017programming}
A.~Fr{\"o}mmgen, A.~Rizk, T.~Erbsh{\"a}u{\ss}er, M.~Weller, B.~Koldehofe,
  A.~Buchmann, R.~Steinmetz, {A programming model for application-defined
  multipath TCP scheduling}, in: Proceedings of the 18th ACM/IFIP/USENIX
  Middleware Conference, ACM, 2017, pp. 134--146.

\bibitem{7437151}
M.~{Yan}, J.~{Casey}, P.~{Shome}, A.~{Sprintson}, A.~{Sutton}, Ætherflow:
  Principled wireless support in sdn, in: 2015 IEEE 23rd International
  Conference on Network Protocols (ICNP), 2015, pp. 432--437.

\bibitem{Lee:2014:MME:2609908.2609948}
J.~Lee, M.~Uddin, J.~Tourrilhes, S.~Sen, S.~Banerjee, M.~Arndt, K.-H. Kim,
  T.~Nadeem, mesdn: Mobile extension of sdn, in: Proceedings of the Fifth
  International Workshop on Mobile Cloud Computing \&\#38; Services, MCS '14,
  ACM, New York, NY, USA, 2014, pp. 7--14.

\bibitem{SARAIVADESOUSA201969}
N.~F.~S. de~Sousa, D.~A.~L. Perez, R.~V. Rosa, M.~A. Santos, C.~E. Rothenberg,
  Network service orchestration: A survey, Computer Communications 142-143
  (2019) 69 -- 94.

\bibitem{dandachi2017multihoming}
G.~Dandachi, {Multihoming in heterogeneous wireless networks}, Ph.D. thesis,
  Evry, Institut national des t{\'e}l{\'e}communications (2017).

\bibitem{ibnalfakih2016multi}
S.~Ibnalfakih, E.~Sabir, M.~Sadik, {Multi-homing as an Enabler for 5G Networks:
  Survey and Open Challenges}, in: International Symposium on Ubiquitous
  Networking, Springer, 2016, pp. 347--356.

\bibitem{karimi2017evaluating}
P.~Karimi, M.~Sherman, F.~Bronzino, I.~Seskar, D.~Raychaudhuri, A.~Gosain,
  Evaluating 5g multihoming services in the mobilityfirst future internet
  architecture, in: IEEE Vehicular Technology Conference (VTC Spring), IEEE,
  2017, pp. 1--5.

\bibitem{I-D.purkayastha-mptcp-considerations-for-nextgen}
D.~Purkayastha, M.~Perras, A.~Rahman,
  \href{http://www.ietf.org/internet-drafts/draft-purkayastha-mptcp-considerations-for-nextgen-00.txt}{{Considerations
  for MPTCP operation in 5G}}, Internet-Draft
  draft-purkayastha-mptcp-considerations-for-nextgen-00, IETF Secretariat
  (October 2017).
\newline\urlprefix\url{http://www.ietf.org/internet-drafts/draft-purkayastha-mptcp-considerations-for-nextgen-00.txt}

\bibitem{lei2018ndn}
K.~Lei, S.~Zhong, F.~Zhu, K.~Xu, H.~Zhang, {An NDN IoT Content Distribution
  Model With Network Coding Enhanced Forwarding Strategy for 5G}, IEEE
  Transactions on Industrial Informatics 14~(6) (2018) 2725--2735.

\bibitem{barakabitze2018novel}
A.~A. Barakabitze, L.~Sun, I.-H. Mkwawa, E.~Ifeachor, {A Novel QoE-Centric
  SDN-based Multipath Routing Approach for Multimedia Services over 5G
  Networks}, in: IEEE International Conference on Communications (ICC), 2018,
  pp. 1--7.

\bibitem{ghasempour2017ieee}
Y.~Ghasempour, C.~R. da~Silva, C.~Cordeiro, E.~W. Knightly, {IEEE 802.11 ay:
  Next-generation 60 GHz communication for 100 Gb/s Wi-Fi}, IEEE Communications
  Magazine 55~(12) (2017) 186--192.

\bibitem{oughton2021revisiting}
E.~J. Oughton, W.~Lehr, K.~Katsaros, I.~Selinis, D.~Bubley, J.~Kusuma,
  Revisiting wireless internet connectivity: 5g vs wi-fi 6, Telecommunications
  Policy 45~(5) (2021) 102127.

\bibitem{khorov2020current}
E.~Khorov, I.~Levitsky, I.~F. Akyildiz, Current status and directions of ieee
  802.11 be, the future wi-fi 7, IEEE Access 8 (2020) 88664--88688.

\bibitem{sultana2019choice}
T.~Sultana, K.~A. Wahid, Choice of application layer protocols for next
  generation video surveillance using internet of video things, IEEE Access 7
  (2019) 41607--41624.

\bibitem{Costa2020CMSS}
L.~Costa, S.~Afzal, P.~Calcina-Ccori, G.~Fedrescheski, M.~Zuffo, {Collaborative
  Mobile Surveillance System for Smart Cities}, in: accepted in 7th Annual
  Conf. on Computational Science \& Computational Intelligence {(CSCI'20)},
  2020.

\bibitem{wang2017joint}
M.~Wang, B.~Cheng, C.~Yuen, Joint coding-transmission optimization for a video
  surveillance system with multiple cameras, IEEE Transactions on Multimedia
  20~(3) (2017) 620--633.

\bibitem{cheng2017situation}
B.~Cheng, M.~Wang, S.~Zhao, Z.~Zhai, D.~Zhu, J.~Chen, Situation-aware dynamic
  service coordination in an iot environment, IEEE/ACM Transactions On
  Networking 25~(4) (2017) 2082--2095.

\bibitem{augustin2016study}
A.~Augustin, J.~Yi, T.~Clausen, W.~M. Townsley, A study of lora: Long range \&
  low power networks for the internet of things, Sensors 16~(9) (2016) 1466.

\bibitem{medjiah2012streaming}
S.~Medjiah, T.~Ahmed, A.~H. Asgari, Streaming multimedia over wmsns: an online
  multipath routing protocol, International Journal of Sensor Networks 11~(1)
  (2012) 10--21.

\bibitem{li2012greentube}
X.~Li, M.~Dong, Z.~Ma, F.~C. Fernandes, {Greentube: power optimization for
  mobile videostreaming via dynamic cache management}, Proceedings of the 20th
  ACM international conference on Multimedia (2012) 279--288.

\bibitem{wu2018energy}
J.~Wu, B.~Cheng, M.~Wang, J.~Chen, {Energy-aware concurrent multipath transfer
  for real-time video streaming over heterogeneous wireless networks}, IEEE
  Transactions on Circuits and Systems for Video Technology 28~(8) (2018)
  2007--2023.

\bibitem{8167677}
A.~{Kaul}, K.~{Obraczka}, M.~A.~S. {Santos}, C.~E. {Rothenberg}, T.~{Turletti},
  Dynamically distributed network control for message dissemination in its, in:
  2017 IEEE/ACM 21st International Symposium on Distributed Simulation and Real
  Time Applications (DS-RT), 2017, pp. 1--9.

\bibitem{7859348}
R.~D.~R. Fontes, C.~Campolo, C.~E. Rothenberg, A.~Molinaro, From theory to
  experimental evaluation: Resource management in software-defined vehicular
  networks, IEEE Access 5 (2017) 3069--3076.

\bibitem{latah2018artificial}
M.~Latah, L.~Toker, {Artificial Intelligence Enabled Software Defined
  Networking: A Comprehensive Overview}, arXiv preprint arXiv:1803.06818.

\bibitem{Pensieve}
H.~Mao, R.~Netravali, M.~Alizadeh, Neural adaptive video streaming with
  pensieve, in: Proceedings of the Conference of the ACM Special Interest Group
  on Data Communication, SIGCOMM '17, ACM, New York, NY, USA, 2017, pp.
  197--210.

\bibitem{al2018learnqos}
A.~Al-Jawad, P.~Shah, O.~Gemikonakli, R.~Trestian, {LearnQoS: a learning
  approach for optimizing QoS over multimedia-based SDNs}, in: IEEE
  International Symposium on Broadband Multimedia Systems and Broadcasting
  (BMSB), 2018, pp. 1--6.

\bibitem{vasilev2018predicting}
V.~Vasilev, J.~Leguay, S.~Paris, L.~Maggi, M.~Debbah, {Predicting QoE factors
  with machine learning}, in: IEEE International Conference on Communications
  (ICC), 2018, pp. 1--6.

\bibitem{alzahrani2018use}
I.~R. Alzahrani, N.~Ramzan, S.~Katsigiannis, A.~Amira, {Use of Machine Learning
  for Rate Adaptation in MPEG-DASH for Quality of Experience Improvement}, in:
  5th International Symposium on Data Mining Applications, Springer, 2018, pp.
  3--11.

\end{thebibliography}
\begin{footnotesize}
\printnomenclature
\end{footnotesize}
\end{document}